\documentclass[sn-mathphys-num]{sn-jnl}

\usepackage{adjustbox}
\usepackage{graphicx}%
\usepackage{multirow}%
\usepackage{amsmath,amssymb,amsfonts}%
\usepackage{amsthm}%
\usepackage{mathrsfs}%
\usepackage[title]{appendix}%
\usepackage{xcolor,colortbl}
\usepackage[most]{tcolorbox}
\usepackage{textcomp}%
\usepackage{manyfoot}%
\usepackage{booktabs}%
\usepackage[ruled,linesnumbered]{algorithm2e}
\usepackage{listings}%
\usepackage{subfigure}
\SetKwInput{KwInput}{Input}                
\SetKwInput{KwOutput}{Output} 

\definecolor{mycolor_box}{HTML}{F5FFFA} 
\definecolor{mycolor_title}{HTML}{FFEBCD} 

\theoremstyle{thmstyleone}%
\theoremstyle{thmstyletwo}%

\theoremstyle{thmstylethree}%

\begin{document}

\title[Investigating Adversarial Attacks in Software Analytics via Machine Learning Explainability]{Investigating Adversarial Attacks in Software Analytics via Machine Learning Explainability}


\author*[1]{\fnm{Md. Abdul} \sur{Awal}}\email{abdul.awal@usask.ca}
\author[1]{\fnm{Mrigank} \sur{Rochan}}\email{mrochan@cs.usask.ca}
\author[1]{\fnm{Chanchal K.} \sur{Roy}}\email{chanchal.roy@usask.ca}

\affil[1]{
  \orgdiv{Department of Computer Science}, 
  \orgname{University of Saskatchewan}, 
  \orgaddress{\city{Saskatoon}, \state{Saskatchewan}, \country{Canada}}
}









\abstract{With the recent advancements in machine learning (ML), numerous ML-based approaches have been extensively applied in software analytics tasks to streamline software development and maintenance processes. Nevertheless, studies indicate that despite their potential usefulness, ML models are vulnerable to adversarial attacks, which may result in significant monetary losses in these processes. As a result, the ML models' robustness against adversarial attacks must be assessed before they are deployed in software analytics tasks. Despite several techniques being available for adversarial attacks in software analytics tasks, exploring adversarial attacks using ML explainability is largely unexplored. Therefore, this study aims to investigate the relationship between ML explainability and adversarial attacks to measure the robustness of ML models in software analytics tasks. In addition, unlike most existing attacks that directly perturb input-space, our attack approach focuses on perturbing feature-space. Our extensive experiments, involving six datasets, three ML explainability techniques, and seven ML models, demonstrate that ML explainability can be used to conduct successful adversarial attacks on ML models in software analytics tasks. This is achieved by modifying only the top \textbf{1--3} important features identified by ML explainability techniques. Consequently, the ML models under attack fail to accurately predict up to \textbf{86.6\%} of instances that were correctly predicted before adversarial attacks, indicating the models' low robustness against such attacks. Finally, our proposed technique demonstrates promising results compared to four state-of-the-art adversarial attack techniques targeting tabular data.}




\maketitle

\section{Introduction}
\label{sec1}
In recent years, machine learning (ML) models have been extensively applied in software analytics tasks such as code clone detection \cite{nafi2019clcdsa, feng2024machine, hu2022treecen, wu2022detecting}, API recommendation \cite{moreno2015can, nguyen2019focus}, automatic code summarization \cite{mcburney2015automatic, zhu2019automatic}, malware detection \cite{sahs2012machine, gavriluct2009malware}, code completion \cite{das2015contextual, svyatkovskiy2021fast}, Just-in-Time (JIT) defect prediction \cite{catolino2019cross, yatish2019mining, kamei2012large}, and source code authorship attribution \cite{alsulami2017source, ullah2019source}. However, research shows that despite the effectiveness of the state-of-the-art ML models, they are vulnerable to adversarial attacks \cite{szegedy2013intriguing, liu2021practical, quiring2019misleading, chen2017adversarial, bielik2020adversarial, springer2020strata, jha2023codeattack, yang2022natural, zhou2022adversarial, zhang2020generating, severi2021explanation}. For example, carefully crafted data from large open-source software repositories in an API recommender system can lead to malicious API calls by developers \cite{nguyen2021adversarial}. As a result, researchers have proposed various techniques for generating adversarial examples to evaluate the robustness of the ML models in software analytics tasks \cite{zeng2022extensive, du2023extensive, nguyen2021adversarial, t2021adversarial, suciu2019exploring, liu2019atmpa, liu2021practical, quiring2019misleading, bielik2020adversarial, springer2020strata, jha2023codeattack, yang2022natural, zhou2022adversarial, zhang2020generating}.

Adversarial attacks generate perturbations to deceive and manipulate ML model predictions, potentially leading to system failures in real-life applications. For example, failing to predict JIT defect introducing commit using ML models under adversarial attacks can significantly impact software development and maintenance processes and optimize the allocation of limited software quality assurance (SQA) resources \cite{pornprasit2021pyexplainer, jiarpakdee2020empirical}. It can also affect the quality maintenance and execution of exhaustive code review activities for all incoming commits with limited SQA resources in modern code review \cite{mcintosh2014impact, bosu2015characteristics}. Consequently, the overall software development and maintenance process may experience significant delays, resulting in financial losses and wasted time, as the underlying ML models struggle to make accurate decisions in the face of adversarial attacks \cite{kamei2012large, catolino2019cross}. Furthermore, attacks on ML models used in code review comment classification tasks can affect automatic support for improving non-useful comments in code review \cite{rahman2017predicting}. Mondal et al. \cite{mondal2019empirical} empirically showed that code clones are directly related to bugs, and the intensity of bug propagation through code cloning is significant. In addition, as software systems evolve and the demand for code grows, the occurrence of code clones may pose a significant threat to vulnerability propagation within these systems \cite{feng2024machine}. Thus, failure to detect code clones accurately under adversarial attacks can lead to the development of buggy software, resulting in technical debt and a huge loss ($\$3.61$ per line of code) in monetary value \cite{guo2016exploring, technicaldebtcost}. Therefore, having a thorough understanding of the performance and resilience of ML models in various software analytics tasks regarding input manipulation is crucial for developing robust ML models.

Most research on adversarial attacks in software analytics tasks focuses on deep learning (DL) models trained on the source code \cite{zeng2022extensive, du2023extensive, nguyen2021adversarial, t2021adversarial, liu2019atmpa, liu2021practical, quiring2019misleading, bielik2020adversarial, springer2020strata, jha2023codeattack, yang2022natural, zhou2022adversarial, zhang2020generating, awal2024large}. Furthermore, the attack approach targets the input-space (e.g., source code) when generating adversarial examples. However, in other software analytics tasks such as JIT defect prediction, clone detection, and code review comment classification, many classical \footnote{In this study, unless otherwise stated, ML models refer to classical machine learning models, not deep learning models.} ML models (not DL) have been extensively used. For example, Feng et al. \cite{feng2024machine}, Hu et al. \cite{hu2022treecen}, and Wu et al. \cite{wu2022detecting} demonstrated that  ML models (e.g., Random Forest) outperformed several state-of-the-art DL-based approaches (e.g., \textit{ASTNN} \cite{zhang2019novel}, \textit{SCDetector} \cite{wu2020scdetector}, \textit{DeepSIM \cite{zhao2018deepsim}}, \textit{FCCA} \cite{hua2020fcca}, \textit{CDLH} \cite{wei2017supervised}, and \textit{TBCNN} \cite{mou2014tbcnn}) in software clone detection on the BigCloneBench \cite{svajlenko2014towards} dataset. Moreover, in these tasks, features from the input data are extracted using a pre-trained machine learning model or some other metrics first \cite{kamei2012large, catolino2019cross, nafi2019clcdsa} and then represented in tabular format. These tabular data are later used to tackle the tasks of interest. Ravid et al. \cite{shwartz2022tabular} demonstrated that  ML models trained on tabular data outperformed DL models trained on the same data in various tasks. Therefore, it is essential to assess the robustness of  ML models trained on tabular data through feature-space manipulation in software analytics tasks. Adversarial attacks in feature-space have been studied in computer vision \cite{xu2020towards, inkawhich2019feature, simonetto2021unified}. However, to the best of our knowledge, this area remains largely unexplored in software analytics tasks. Therefore, our study focuses on bypassing the input-space and instead examines attacks on the feature-space for  ML models in software analytics tasks.

Despite significant efforts to advance adversarial attacks targeting DL models trained on source code \cite{zeng2022extensive, du2023extensive}, to the best of our knowledge, adversarial attacks on classical ML models trained on tabular data for various software analytics tasks have not been explored. Moreover, due to fundamental differences in the characteristics of the training datasets and the model's architecture, the methods proposed for adversarial attacks on DL models cannot be directly applied to ML models trained on tabular data. Therefore, there is an urgent need to propose novel techniques to perform adversarial attacks on  ML models trained specifically on tabular data in software analytics tasks. Furthermore, investigating adversarial attacks using machine learning explainability in software analytics tasks remains largely unexplored. Recently, Severi et al. \cite{severi2021explanation} proposed explanation-guided backdoor poisoning attacks to assess the robustness of malware classifiers. The primary distinction between Severi et al. \cite{severi2021explanation} and our work is that they focused on influencing the ML training process, while our approach is specifically designed to attack the ML inference process. Similarly, Amich et al. \cite{amich2021explanation, amich2022eg} employ ML explainability to boost and diagnose evasion attacks on ML models, particularly those trained on image classifier datasets. However, our approach diverges from previous studies \cite{severi2021explanation, amich2021explanation, amich2022eg} in terms of how we select important features, modify them, and attack feature-space while generating adversarial examples, making our work \textit{novel}.

This study aims to assess the robustness of ML models in software analytics tasks by generating adversarial examples using ML explainability techniques such as SHapley Additive exPlanation (SHAP) \cite{lundberg2017unified}, Local Interpretable Model-agnostic Explanations (LIME) \cite{ribeiro2016should}, and PyExplainer \cite{pornprasit2021pyexplainer}. The key contributions of our study are listed below:
\begin{itemize}
    \item We apply existing ML explainability techniques (e.g., SHAP, LIME, and PyExplainer) to figure out important features of an instance that influence ML models toward decision-making. We also determine the impacts of each feature on the model's prediction for individual instances.
    \item Our explanation-guided adversarial example generation creates transformed instances, significantly reducing the accuracy of the ML models in software analytics tasks.
    \item We comprehensively evaluate the proposed adversarial example generation technique using six datasets and seven ML models. Additionally, we compare our approach with four state-of-the-art adversarial attack techniques (e.g., \textit{Zoo} \cite{chen2017zoo}, \textit{Boundary attack} \cite{brendel2017decision}, \textit{PermuteAttack} \cite{hashemi2020permuteattack}, and \textit{HopSkipJump} \cite{chen2020hopskipjumpattack}) designed for ML models in a model-agnostic way. We also compare our approach with two makeshift tools. Our proposed approach demonstrates promising results compared to the baselines and makeshift tools, showcasing its effectiveness in generating adversarial examples to evaluate the robustness of ML models.
    \item Our code and the corresponding dataset are publicly available to enhance further research\footnote{\href{https://zenodo.org/doi/10.5281/zenodo.7865487}{Replication-package}}.
    
\end{itemize}

\section{Background}
\label{back}



\textbf{Adversarial Attacks}: 
Szegedy et al. \cite{szegedy2013intriguing} were the first to introduce the concept of adversarial attacks within the domain of computer vision. Their work demonstrated that minimal, pixel-level perturbations—imperceptible to the human eye—could successfully deceive even state-of-the-art image classifiers. Extending this idea to discrete domains, such as clone detection, a classifier $f: X \rightarrow Y$ aims to determine whether two code snippets are similar. In this context, an adversarial attack may apply small, semantic-preserving transformations—such as variable renaming, dead code insertion, or statement reordering—to produce an adversarial example $x'$ from an original input $x$, such that $f(x') \neq f(x)$, misleading the model without altering the code's functionality. Similarly, in just-in-time defect prediction, models classify code changes as either buggy or clean. Adversarial examples in this setting can be created by subtly modifying the code—e.g., inserting dead code or removing redundant lines—while preserving the model's functionality, thereby causing it to change its prediction.

An adversarial example is defined as a slightly modified version of an original input that remains visually indistinguishable from the original to a human observer. Despite these minimal changes, the modified input is crafted with the specific intent to deceive a model into making an incorrect prediction. Formally, given an input $x$ and a small perturbation $\delta$, the adversarial example $x' = x + \delta$ is constructed such that $f(x') \neq f(x)$, where $f$ denotes the ML model's prediction function. When such a perturbation successfully alters the model’s output without altering the input’s actual semantics or functionality, the phenomenon is referred to as an adversarial attack \cite{szegedy2013intriguing}.

In practice, the implications of such attacks become clearer when we examine how they can subvert real-world software analytics pipelines. To illustrate this, consider a real-world scenario involving a large software company that employs a machine learning–based just-in-time (JIT) defect prediction model integrated into its continuous integration (CI) pipeline. This model automatically analyzes every submitted code commit and predicts whether it is likely to introduce a bug. Commits predicted as \textit{buggy} are flagged and blocked for further review, while \textit{clean} commits are automatically merged. By doing so, the model influences key software engineering practices. This enables software developers to better allocate their limited software quality assurance (SQA) resources by reviewing the highest-risk commits first \cite{catolino2019cross, kamei2012large}. It also supports the development of proactive measures aimed at enhancing software quality and preventing the recurrence of issues that may lead to bugs in future releases \cite{mcintosh2014impact}.

Yet, the same automation that improves efficiency may open the door to adversarial vulnerabilities. For example, consider an adversary—such as a malicious developer or a third party—who is aware of this automated screening process and intercepts the input to the machine learning model before committing code changes. The adversary subtly modifies the input in a way that preserves the functionality of the buggy code but disrupts the feature patterns learned by the model. As a result, the model misclassifies the \textit{buggy} commit as \textit{clean}, allowing it to bypass the code review process. Consequently, the concurrency bug is deployed to production and remains undetected until it causes a critical system failure under high load. Over time, such adversarial evasions accumulate, contributing to technical debt, complicating future maintenance, increasing debugging costs, and degrading overall software quality \cite{mondal2019empirical, guo2016exploring, technicaldebtcost}. Such attacks are not merely technical exploits—they can serve for deliberate strategic purposes. By causing critical bugs to slip into production, an adversary can significantly damage the reputation of the software company, erode user trust, or even trigger substantial financial losses due to system outages, regulatory penalties, or gain an unfair market advantage. This scenario illustrates how adversarial attacks can have far-reaching consequences that extend beyond model accuracy, affecting software reliability, business operations, and stakeholder trust. Thus, it underscores the importance of assessing model robustness against such attacks prior to deployment in real-world software engineering tasks.

\textbf{Adversarial Robustness}: 
Adversarial robustness refers to a model's ability to maintain its predictive performance when subjected to adversarial perturbations—small, intentionally crafted modifications to input data designed to mislead the model into making incorrect predictions. A model is considered adversarially robust if its output remains stable and consistent despite such adversarial manipulations. In clone detection, a robust model should consistently identify clones even after minor, semantic-preserving transformations, such as variable renaming or control structure rewriting. In just-in-time defect prediction, robustness implies that modifications—such as adding dead code or removing redundant lines—should not cause the model's prediction to flip from clean to buggy, or vice versa. Formally, for a model $f: X \rightarrow Y$, adversarial robustness implies that for any small perturbation $\delta$ bounded by a constraint (e.g., $|\delta| \leq \epsilon$ for some small $\epsilon > 0$), the model satisfies $f(x) = f(x+\delta)$ for most inputs $x \in X$ \cite{szegedy2013intriguing, goodfellow2014explaining, carlini2017towards}.



\textbf{Risk Score}: The Risk Score represents the likelihood of an instance being classified into a predefined class. For instance, let's consider a ML model trained on the JIT defect prediction dataset, such as cross-project mobile apps. In Figure \ref{pyexp_a}, the selected instance is predicted as a \textit{clean} commit with a Risk Score of $33\%$. This indicates a higher likelihood of the instance being classified as a \textit{clean} commit, as the Risk Score is below $50\%$. Conversely, a higher Risk Score implies a greater probability of the instance being classified as a \textit{buggy} commit. 


\textbf{Feature Importance Rank}: Feature importance rank refers to the assessment of how much each input feature contributes to the prediction made by a ML model for any given input instance. The rank helps to identify the most significant features that affect the model's output. For example, Figure \ref{shap_a} demonstrates the name of each feature in the $y$-axis and the contribution of each feature to the magnitude of the model output in the $x$-axis. This figure also shows that the most important feature is \textit{ndev} (number of developers).

\textbf{Machine Learning Explainability}: 
Machine Learning (ML) explainability refers to understanding, interpreting, and articulating a machine learning model's internal mechanisms, reasoning processes, and outputs. In other words, explainability seeks to provide human-interpretable insights into why a model arrives at a particular prediction or decision \cite{ribeiro2016should}. Common approaches to ML explainability include feature importance analysis, surrogate modeling (e.g., LIME, SHAP), attention mechanisms, and rule extraction. These methods help practitioners and researchers interpret model outputs and identify biases, errors, and potential vulnerabilities in the models, including their susceptibility to adversarial attacks.


\section{Research Methodology}
\label{method}
In this section, we discuss the overall methodology of our explanation-guided adversarial attack approach. Our objective is to assess the robustness of ML models through an extensive study by generating adversarial examples based on altering the top-$k$ important features identified in ML explainability. Figure \ref{workflow} depicts the workflow of our research methodology, which is divided into five distinct steps described below:

\begin{figure}[htbp] 
\centerline{
\includegraphics[width=12cm, height = 6.5cm]{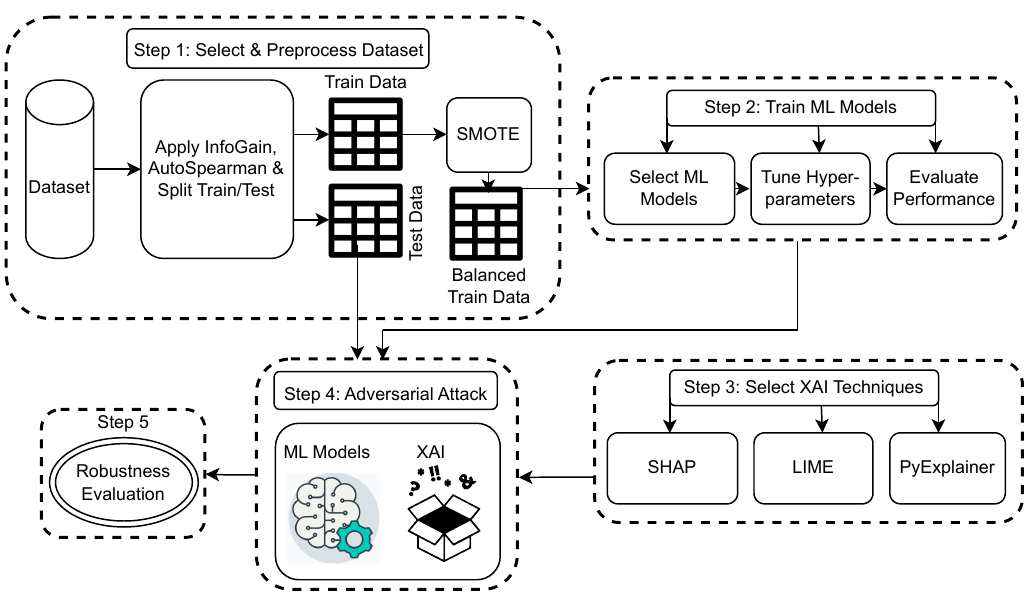}} 
\caption{Our proposed workflow for evaluating the robustness of ML models under adversarial attacks. Step 1 includes the dataset selection and preprocessing methods. Step 2 involves training ML models, while Step 3 involves selecting ML explainability techniques. Step 4 then describes how we perform adversarial attacks using ML explainability, and Step 5 investigates the robustness of ML models under adversarial attacks. 
} 
\label{workflow} 
\end{figure}

\subsection{Dataset Selection and Preprocessing}
\label{DC}
This study aims to evaluate the robustness of ML models by generating adversarial examples through feature-space manipulation. We meticulously select and process the datasets as follows:


\subsubsection{Dataset Selection}
\label{DS}
This study investigates the robustness of ML models against adversarial attacks across various software analytics tasks, including defect prediction, clone detection, and code review comment classification. Additionally, our study shifts the focus of attack from the input-space to the feature-space when generating adversarial examples. Thus, we aim to select software analytics tasks where features extracted from the input (e.g., source code) are utilized to perform the tasks of interest. For example, in cross-language clone detection, various features such as the \textit{`Number of Variables Declared,' `Total Number of Operators,' `Number of Arguments,' `Number of Expressions,' `Total Number of Operands,' `Number of Loops (for, while),' `Number of Exceptions Thrown,' `Number of Exceptions Referenced,' and `McCabe's Cyclomatic Complexity'} were extracted first, and ML models were then trained based on these extracted features \cite{nafi2019clcdsa}. Consequently, we select six different datasets: three for defect prediction, two for clone detection, and one for code review comment classification, as the targets for adversarial attacks, where features are extracted from the input (e.g., source code) and stored in tabular format.

\begin{table}[htbp]
\centering
\caption{Summary of datasets used in this study}
\label{datasetsummary}
\begin{tabular}{l|l|l|l}
\hline
\textbf{Task}                            & \textbf{Dataset} & \textbf{\#   Data Points} & \textbf{\#   Features} \\ \hline
\multirow{3}{*}{JIT-defect   Prediction} & Cross   Project  & 30,000                    & 14                     \\ \cmidrule{2-4} 
                                         & Java   Project   & 25,000                    & 27                     \\ \cmidrule{2-4} 
                                         & Postgres         & 20,431                    & 14                     \\ \hline
\multirow{2}{*}{Clone   Detection}       & CLCDSA           & 30,000                    & 18                     \\ \cmidrule{2-4} 
                                         & BigCloneBench    & 5,46,000                  & 6                      \\ \hline
Code   Review Comments Classification    & Code   Review    & 1,100                     & 15                     \\ \hline
\end{tabular}
\end{table}

Table \ref{datasetsummary} summarizes the key characteristics of the datasets used in this study. For JIT defect prediction, we utilize a dataset for cross-project mobile apps \cite{catolino2019cross} and ten defect prediction datasets from various Java projects \cite{yatish2019mining}. Additionally, we select the Postgres dataset \cite{kamei2012large} from six subject systems for the Desktop application. The code review comment dataset \cite{rahman2017predicting} comprises 15 features and approximately 1,100 data points for detecting useful comments. Finally, we extend our experimental analysis by incorporating the cross-language clone dataset used by Nafi et al. \cite{nafi2019clcdsa} and the BigCloneBench (BCB) dataset \cite{svajlenko2014towards}. The selected cross-project mobile apps dataset contains 14 features and approximately 30,000 data points, while different Java projects have datasets with 65 features and approximately 25,000 data points. The CLCDSA dataset consists of 18 features and approximately 30,000 data points. Feng et al. \cite{feng2024machine} extracted six features from the BCB dataset based on some similarity metrics. In our study, we also use these six features to train the ML models. The preprocessed dataset ended up with six features and approximately 546K samples. Please refer to these articles \cite{svajlenko2014towards, kamei2012large, catolino2019cross, nafi2019clcdsa, rahman2017predicting} for details about the features of each dataset.

\subsubsection{Feature Engineering}
The rationale for selecting six datasets in this paper is described in Section \ref{DS}. Therefore, to optimize the performance of ML models in our experiment, we adhere to the same feature engineering techniques previously applied to these datasets by other studies. In the study conducted by Catolino et al. \cite{catolino2019cross}, the \textit{InfoGain} method was employed to select a subset of features. From the original 14 features, this technique identified 6 as particularly relevant, which we incorporated into our model. Similarly, Nafi et al. \cite{nafi2019clcdsa} conducted feature selection for their code fragments, retaining 9 features from an initial pool of 24. This process resulted in the derivation of 18 feature values, each associated with a single class label for every clone pair. We utilized these filtered features to train ML models in our study. When dealing with the Java project dataset, we encounter the issue of collinearity among features. To address this challenge, we apply the \textit{AutoSpearman} technique, which has been utilized by Yatish et al. \cite{yatish2019mining} on the same dataset. Applying this technique allows us to identify and retain 27 pertinent features from an initial set of 65. In our investigations on the Postgres dataset, we also encounter the challenge of the multi-collinearity problem. To address this issue effectively, we implement the same feature selection and normalization methodologies outlined in Kamei et al. \cite{kamei2012large} on this dataset.

Feng et al. \cite{feng2024machine} proposed a code clone detection method, namely \textit{Toma}, using the BCB dataset and classical machine learning models. They converted the source code into type sequences and applied six similarity metrics, such as the \textit{Jaccard similarity coefficient}, \textit{Dice similarity coefficient}, \textit{Levenshtein distance}, \textit{Levenshtein ratio}, \textit{Jaro similarity}, and \textit{Jaro-Winkler similarity}, to extract six features from the type sequences. Thus, the BCB dataset has six features and is used in our study to train the ML models.

\subsubsection{Dataset Balancing}
The JIT defect prediction task faces the challenge of imbalanced binary classification, with one class (e.g., \textit{clean} commit) having significantly more examples than the other (e.g., \textit{buggy} commit). To tackle this issue, we employ the \textit{Synthetic Minority Over-sampling Technique (SMOTE)}, a technique previously utilized in previous studies \cite{kamei2012large, roy2022don, pornprasit2021pyexplainer, awal2024evaluatexai} on the defect prediction datasets selected for our study. Additionally, we apply \textit{SMOTE} to rectify imbalanced datasets within the cross-language clone dataset. We must emphasize that we exclusively use this technique for the training data, ensuring that the test dataset remains unaffected by the oversampling process. Finally, the BCB dataset has more clone pairs than non-clone pairs. Since the BCB dataset contains only 278,838 non-clone pairs, we randomly select approximately 270,000 clone pairs from eight million clones, following the study by Feng et al. \cite{feng2024machine}, ensuring that the BCB dataset used in our study is balanced.

\subsection{Machine Learning Model Selection and Training}

Selecting and training ML models is crucial for the success of any ML project. This process entails choosing the appropriate algorithm or model architecture, training it on the preprocessed dataset, and tuning hyperparameter values to enhance generalization while avoiding overfitting.

\subsubsection{Machine Learning Models}
Ravid et al. \cite{shwartz2022tabular} demonstrated that classical ML models trained on tabular data outperformed DL models trained on the same data in various tasks. In addition, Feng et al. \cite{feng2024machine}, Hu et al. \cite{hu2022treecen}, and Wu et al. \cite{wu2022detecting} demonstrated that  ML models (e.g., Random Forest) outperformed several state-of-the-art deep learning models (e.g., \textit{ASTNN} \cite{zhang2019novel}, \textit{SCDetector} \cite{wu2020scdetector}, \textit{DeeSIM \cite{zhao2018deepsim}}, \textit{FCCA} \cite{hua2020fcca}, \textit{CDLH} \cite{wei2017supervised}, and \textit{TBCNN} \cite{mou2014tbcnn}) in software clone detection on the BCB dataset. Since the datasets selected for this study contain tabular data, we aim to evaluate the robustness of classical  ML models trained on features extracted from the original input (e.g., source code) presented in a tabular format. Consequently, we focus on  ML models commonly used for tabular data in various software analytics tasks, including defect prediction, clone detection, and classification of code review comments. For example, Logistic Regression and Random Forest are commonly used classification techniques in software defect prediction \cite{yang2024cfexplainer, yu2024formal}. Thus, we select widely used ML models for our experiments, encompassing three standard ML models (Logistic Regression (LR), Multi-Layered Perceptron (MLP), and Decision Tree (DT)), as well as four ensemble methods (Gradient Boosting Classifier (GBC), Random Forest (RF), AdaBoost (ADA), and Bagging (BAG)). It is worth noting that all of these models have been successfully employed in previous software analytics studies \cite{awal2024evaluatexai, kamei2012large, pornprasit2021pyexplainer, catolino2019cross, roy2022don, rahman2017predicting, feng2024machine, hu2022treecen, wu2022detecting, yang2024cfexplainer, yu2024formal}.


\subsubsection{Hyperparameters Tuning}
\label{tune_param}

In the realm of model optimization, the tuning of hyperparameters plays a pivotal role in facilitating effective generalization and mitigating the risk of overfitting. To maximize model performance, we explore various hyperparameter combinations, following established standard settings for different ML models \cite{kamei2012large, pornprasit2021pyexplainer, catolino2019cross, roy2022don, rahman2017predicting, tantithamthavorn2016automated, tantithamthavorn2018impact, feng2024machine}. For example, Feng et al. \cite{feng2024machine} observed that setting the depth parameters to 32, 32, 64, and 16 for RF, DT, Adaboost, and GBDT models, respectively, achieves the highest F1-scores for clone detection on the BCB dataset. In addition, our approach leverages various optimization techniques, including grid search \cite{liashchynskyi2019grid}, random search \cite{bergstra2012random}, and Bayesian optimization \cite{wu2019hyperparameter}, to meticulously fine-tune hyperparameters, ultimately selecting the optimal combinations for our specific objectives. This rigorous optimization process is a cornerstone of our ML model training, ensuring that our models are finely calibrated to yield the best possible performance. It is important to note that we only tune hyperparameters while leaving the model's parameters unaltered.

\subsubsection{Evaluate Model Performance}
Model validation is a critical step in ML, ensuring the model's ability to perform well on new, unseen data. Since PyExplainer often requires excessive time to generate an explanation for a single instance, we adopt a dataset-splitting approach, dividing it into training (90\%) and testing (10\%) subsets to expedite the overall process. To validate the ML models using the 90\% training data, we employ a 10-fold cross-validation technique, which was applied in other software analytics studies \cite{awal2024evaluatexai, feng2024machine, hu2022treecen, wu2022detecting, roy2022don, kamei2012large, yatish2019mining}. Finally, we conduct a thorough evaluation of model performance, utilizing key metrics, including accuracy, F1-score, and the Area Under the Receiver Operating Characteristic Curve (AUC), as recommended by recent studies \cite{yatish2019mining, catolino2019cross, roy2022don}.

The AUC assesses the discriminatory power of models by measuring the true negative rate (coverage of the negative class) on the $x$-axis and the true positive rate (coverage of the positive class) on the $y$-axis. AUC values range from $0$ (indicating the worst performance) to $0.5$ (no better than random guessing) up to $1$ (representing the best performance) \cite{hanley1982meaning}. Therefore, AUC-ROC is a comprehensive tool for evaluating the effectiveness of binary classification models in distinguishing between positive instances (e.g., \textit{buggy} commit) and negative instances (e.g., \textit{clean} commit). Finally, since we aim to perform explanation-guided adversarial attacks on ML models, a test-AUC value of 0.75 is the minimum threshold for ensuring the reliability of explanations \cite{roy2022don, lyu2021towards}. Thus, we evaluate the performance of ML models trained on the selected datasets for this study.



\subsection{ML Explainability Technique Selection}
\label{exp tech}
This study aims to investigate adversarial attacks in software analytics tasks using the ML explainability technique. Our goal is to modify the fewest features possible when generating adversarial examples \cite{mathov2020not}. Therefore, ML explainability techniques can serve as an effective solution for identifying the important features on which ML models base their decisions. We then modify the top-$k$ features from the important feature list to generate adversarial examples, where $k$ ranges between 1 and half of the total features. The details of adversarial attacks using ML explainability techniques are described in Section \ref{Adv Attack}.

While numerous ML explainability techniques have been proposed in recent years to shed light on the decisions made by ML models \cite{lundberg2017unified, ribeiro2016should, tantithamthavorn2021explainable}, our focus is solely on model-agnostic techniques, which are applied across various software analytics tasks. Jiarpakdee et al. \cite{jiarpakdee2020empirical} empirically demonstrated the effectiveness of model-agnostic ML explainability techniques (e.g., LIME \cite{ribeiro2016should}) in elucidating the outcomes of defect prediction models. Recently, Roy et al. \cite{roy2022don} demonstrated that model-agnostic techniques such as SHAP and LIME could be successfully used for post-hoc analysis of black-box ML models in software analytics tasks. Additionally, Amich demonstrated that LIME \cite{ribeiro2016should} and SHAP \cite{lundberg2017unified} yield more accurate results when compared to other model-agnostic explanation methods, such as DeepLIFT \cite{shrikumar2017learning} and LEMNA \cite{guo2018lemna}. Therefore, we select two highly cited model-agnostic ML explainability techniques, SHAP and LIME. Furthermore, Pornprasit et al. \cite{pornprasit2021pyexplainer} introduced the model-agnostic technique PyExplainer, which builds upon the concept of LIME for JIT defect prediction models and supports all ML models available in the \textit{scikit-learn} library. Hence, we also include PyExplainer in our study due to its model-agnostic nature.

\subsection{Adversarial Attacks}
\label{Adv Attack}

In computer vision \cite{szegedy2013intriguing}, a carefully crafted pixel-level perturbation added to an input image yields an adversarial example that cannot be differentiated from the original input upon visual inspection. The main property of adversarial examples is that they are inputs intentionally modified to cause incorrect predictions while remaining imperceptible to the human eye. However, this is not true for any software analytics task. In software analytics, any transformations to feature values in tabular format or tokens of source code lead to perceptible changes to the human eye. Therefore, adversarial example generation in software analytics tasks differs fundamentally from that in computer vision. Ballet et al. \cite{ballet2019imperceptible}, and Cartella et al. \cite{cartella2021adversarial} introduced a distinct concept of imperceptibility concerning adversarial examples within financial domain data. They altered non-important features based on human judgment while generating these examples. For instance, according to expert analysis, only a subset of features, such as income and age, might be crucial for specific predictions (e.g., fraudulent transaction detection). The attacker should not aim to change these important features; instead, they should target the non-important features so that the generated adversarial examples remain imperceptible. However, these imperceptibility concepts are domain-specific and are not directly applicable to tabular data in software analytics tasks. Therefore, to ensure imperceptibility while attacking, we consider the criterion introduced by Mathov et al. \cite{mathov2020not}, which states that \textit{``an attacker should aim to minimize the number of modified features (e.g., a minimal $\ell_0$ perturbation) when dealing with tabular data."} The $\ell_0$ perturbation is defined as follows:

\begin{equation}
\label{l0_norm}
\|\mathbf{x} - \mathbf{x'}\|_0 = \sum_{i=1}^{n}\mathbf{1} \hspace{0.25cm} (x_i \ne x'_i \text{ } \& \text{ } f(\mathbf{x}) \ne f(\mathbf{x'}))
\end{equation}

Where $\mathbf{x}$ is the original input, $\mathbf{x'}$ is the generated adversarial input obtained by modifying the smallest number of top-$k$ feature values, $f(\mathbf{.})$ is the model's prediction and $n$ is the number of features. The goal of our adversarial attack is to minimize the $\ell_0$ perturbation as much as possible.

To conduct the adversarial attack while maintaining imperceptibility, as defined in equation \ref{l0_norm}, this study aims to address two key aspects: (1) which features in a given instance have the most significant influence on guiding ML models to make specific predictions? and (2) how can these influential features be strategically manipulated to create transformed examples effective in adversarial attacks? To address the first aspect, we present a use case using PyExplainer to generate an explanation for an instance (e.g., a data point of the test dataset), as shown in Table \ref{instance}. This instance includes six distinct features: $nd$ (number of modified directories), $nf$ (number of modified files), $la$ (lines of code added), $ld$ (lines of code deleted), $ndev$ (number of developers), and $nuc$ (number of unique changes to the modified files).

\begin{table}[htbp]
\small
\centering
\caption{An instance that we use to generate an explanation in Fig. \ref{pyexp_a} and calculation of one standard deviation (STD) value of each feature using the whole training dataset}
\begin{tabular}{l|c|c|c|c|c|c}
\hline
Feature name                   & nd & nf & la    & ld   & ndev & nuc \\ \hline
Feature value      & 1  & 4  & 25    & 0    & 4    & 1   \\ \hline
One standard deviation & 14 & 72 & 8377 & 5592 & 10   & 229 \\ \hline
\end{tabular}
\label{instance}
\end{table}

\begin{figure}[htbp]
\centering     
\subfigure[SHAP]{\label{shap_a}\includegraphics[width=5.75cm, height=2.5cm]{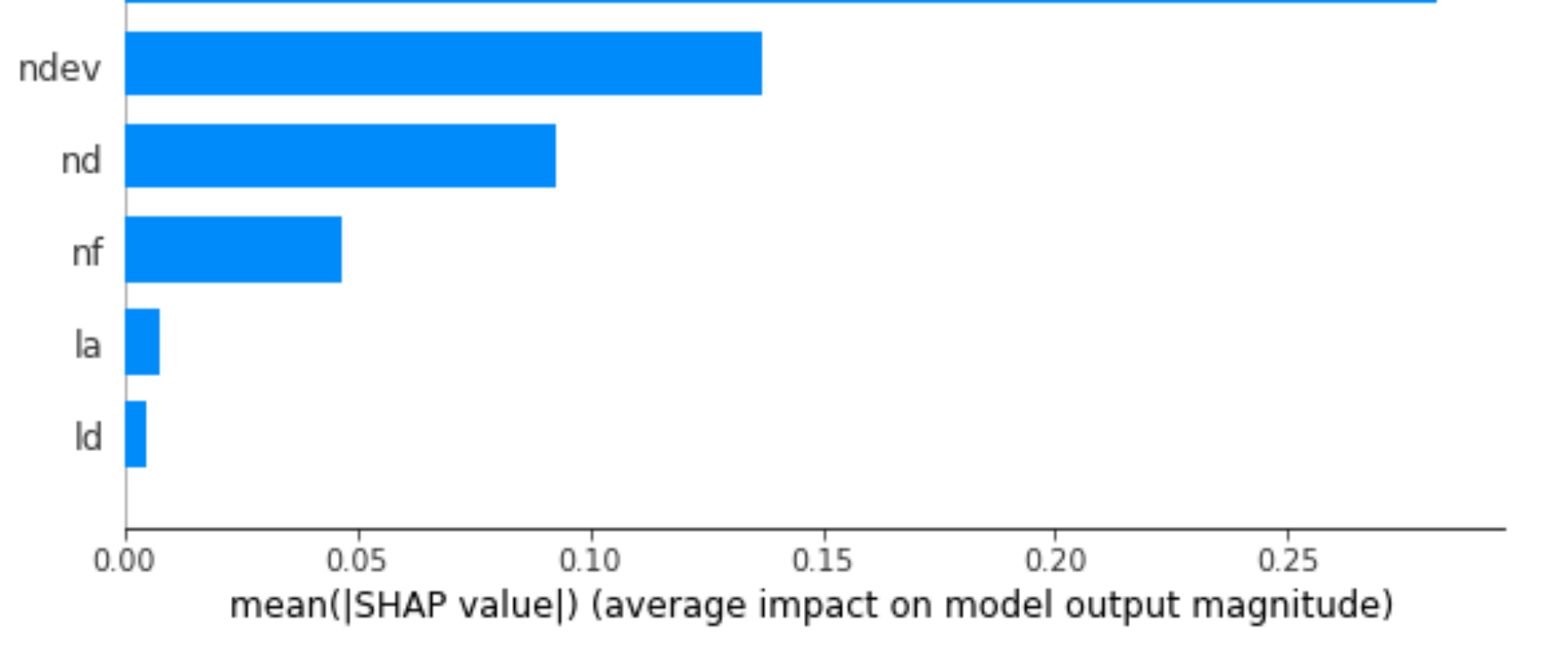}}
\vspace{-0.75em}
\subfigure[LIME]{\label{lime_a}\includegraphics[width=5.75cm, height=2.5cm]{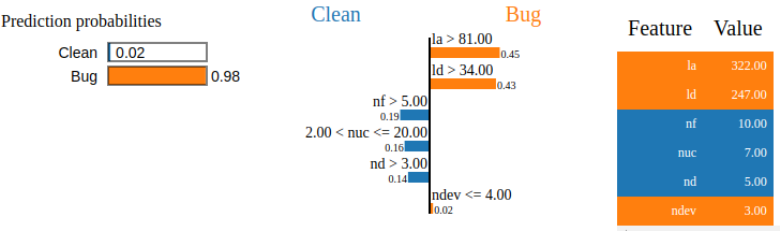}}
\subfigure[PyExplainer]{\label{pyexp_a}\includegraphics[width=5.75cm, height=2.8cm]{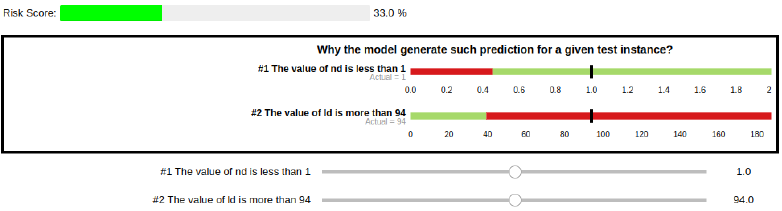}} 
\vspace{-0.75em}
\subfigure[PyExplainer]{\label{pyexp_b}\includegraphics[width=5.75cm, height=2.8cm]{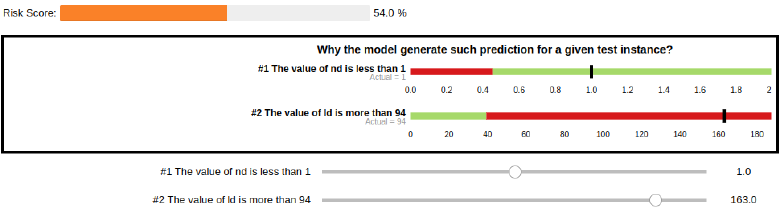}}
\caption{Important features found by SHAP (Fig. \ref{shap_a}), LIME (Fig. \ref{lime_a}), and PyExplainer (Fig. \ref{pyexp_a}) on the cross-project mobile apps dataset. Figure \ref{pyexp_a} depicts the explanation and Risk Score when PyExplainer works on the original input. Figure \ref{pyexp_b} shows the explanation and Risk Score when PyExplainer works on the modified features in the guided direction.}
\label{feat_imp}
\end{figure}

Figures \ref{pyexp_a} and \ref{pyexp_b} illustrate how the features \textit{nd} and \textit{ld} influence the LR model's decision-making process. A comparison between Figures \ref{pyexp_a} and \ref{pyexp_b} reveals an increase in the \textbf{Risk Score} from $33.0\%$ to $54.0\%$ after changing the feature values in the \textit{red} zone, effectively flipping the ML model's prediction. These two figures collectively demonstrate that increasing the feature values in the \textit{red} zone will elevate the probability of an instance being predicted as a positive class (e.g., \textit{buggy} commit). Conversely, raising the feature values in the \textit{green} zone will reduce the probability of an instance being categorized as a positive class, implying that the given instance will be predicted as a negative class (e.g., \textit{clean} commit). Our manual investigation typically found that PyExplainer identified the number of important features for the selected datasets to be within the range of 1--3.

While PyExplainer directly illustrates which features are responsible for guiding ML models toward a decision, SHAP and LIME provide us with a feature importance rank. In adversarial example generation, the goal is to identify minimal perturbations or changes (e.g., a minimal $\ell_0$ perturbation) to the input that will cause the model to misclassify the transformed instance. For example, in the case of ML models trained on tabular data, our objective is to modify the minimum number (e.g., top-\textit{k}) of feature values from the feature importance rank to generate adversarial examples. However, a fundamental question arises: how do we determine the optimal value of \textit{k} for the top-\textit{k} features?

To determine the value of $k$ for the top-\textit{k} important features, we leveraged the concept of the \textit{Elbow} method in the $K$-means clustering algorithm \cite{kodinariya2013review}, which we refer to as the \textit{Reverse Elbow Method}. Initially, we modify the top-$1$ feature, proceed to the top-$2$ features, and so forth to calculate the \%ASR (definition of \%ASR\footnote{In this study, \%ASR and ASR represent the same metric.} is provided in Section \ref{robust eval}.) metric value until we reach half the total number of features in the feature importance rank. For example, in Figure \ref{shap_a}, the feature importance rank of the cross-project mobile apps dataset is displayed from top to bottom. As a result, when considering half of the features (e.g., $3$ out of $6$ features), we obtain three feature combinations (Line $7$, Algorithm \ref{alg:one}): \{\textit{ndev}\}, \{\textit{ndev, nd}\}, and \{\textit{ndev, nd, nf}\}, as shown in Table \ref{topk}.

\begin{table}[htbp]
\small
\centering
\caption{Determining optimal top-$k$ features for adversarial attacks using ASR metric values}
\begin{tabular}{c|c|c|c}
\hline
\multirow{3}{*}{ASR} & Top-1                      & Top-2                      & Top-3                      \\ \cmidrule{2-4} 
                     & \{ndev\}                   & \{ndev, nd\}               & \{ndev, nd, nf\}           \\ \cmidrule{2-4} 
                     & \multicolumn{1}{c|}{49.9} & \multicolumn{1}{c|}{58.4} & \multicolumn{1}{c}{62.4} \\ \hline
\end{tabular}
\label{topk}
\vspace{-1.05em}
\end{table}

In each iteration, we select a combination to modify feature values when generating adversarial examples, allowing us to calculate ASR metric (Lines 7--13, as outlined in Algorithm \ref{alg:one}). If, in consecutive iterations, the difference in ASR values does not increase, we can halt the process and set the value of \textit{k} equal to the length of the previous feature combination (Lines 15--34, Algorithm \ref{alg:one}). Additionally, if the ASR value decreases after the first iteration, we set the value of $k$ to $1$ (Lines 21--24, Algorithm \ref{alg:one}). Let us consider a concrete example to illustrate this process. Table \ref{topk} presents the ASR values for different feature combinations in each iteration, considering the DT model trained on the cross-project mobile apps dataset. The ASR difference between the $2^{nd}$ and $1^{st}$ iterations is $58.4 - 49.9 = 8.5$. In contrast, the ASR difference between the $3^{rd}$ and $2^{nd}$ iterations is $62.4 - 58.4 = 4$. It is evident that the ASR value does not significantly increase after the $2^{nd}$ iteration. Therefore, we can set the value of $k=2$ to determine the top-$k$ features.

\newlength{\commentWidth}
\setlength{\commentWidth}{5cm}
\newcommand{\atcp}[1]{\tcp*[r]{\makebox[\commentWidth]{#1\hfill}}}
\begin{algorithm}[hbt!]
\small
\DontPrintSemicolon
\SetAlgoLined
\KwInput{ML Model ($M$), ML Explainability Techniques (XAI) (\textit{SHAP, LIME, PyExplainer}), Train ($X_{train}$), and Test ($X_{test}$) Data.}
\KwOutput{Adversarial Examples ($X_{test}'$), and \textit{ASR} Metric Value.}
$ASR \gets 0$ \tcp*[r]{ASR for all test samples}
$ASR_{topk} \gets 0$ \tcp*[r]{ASR for changing top-$k$ features}
$Top_k \gets 0$ \tcp*[r]{Store $k$ as top-$k$ important features}
$FC_{ASR} \gets []$ \tcp*[r]{Store ASR for changing top-$k$ features}
\BlankLine
$FC_{comb} \gets getFeatureCombination(getFeatureImportanceRank(XAI, X_{test}'))$ 
\BlankLine
\tcp{Calculate the ASR and top-$k$ values for different feature combinations by optimizing the multi-objective optimization function}
\ForEach{$FC \in FC_{comb}$}
{
\tcp{$X_{test}'$ contains the correctly predicted test instances}
\ForEach{$x \in X_{test}'$} 
{
    $x' \gets getTransformedInstance(x,FC)$ \\ 
    $y_{pred} = getPrediction(M, x')$\\
    \If{$y_{pred} \neq y_{test}$}
    {
            $ASR_{topk} \gets ASR_{topk} + 1$
        }
}
$FC_{ASR}[FC] \gets ASR_{topk}$ \\
\For{$i$ in range($len(FC_{ASR})$)}{
$ASR_{diff} \gets -\infty$ \tcp*[r]{Initialize and store ASR difference between iteration}
        \If{$i == 0$}
        {
        $ASR_{diff} \gets FC_{ASR}[i]$ \\
        $ASR \gets FC_{ASR}[0]$
        }
        \ElseIf{$FC_{ASR}[0] > FC_{ASR}[1]$}
        {
            $Top_k = 1$ \\
            $ASR \gets FC_{ASR}[0]$ \\
            break
        }
        \ElseIf{$FC_{ASR}[i+1] - FC_{ASR}[i] > ASR_{diff}$}{
            $ASR_{diff} \gets FC_{ASR}[i+1] - FC_{ASR}[i]$ \\
            $ASR \gets FC_{ASR}[i]$
        }
        \Else{
            $Top_k \gets len(FC_{ASR}[i])$ \\ 
            break
        }               
}
}

\caption{Explanation-guided Adversarial Attacks.}\label{alg:one}
\end{algorithm}

    



We leverage the technique introduced in the previous study by Pornprasit et al. \cite{pornprasit2021pyexplainer} to address the second aspect and perform the adversarial attacks (Lines 8--12, Algorithm \ref{alg:one}), and this is one of the many ways to change the feature values \cite{hashemi2020permuteattack}. PyExplainer explicitly identifies the features that influence ML models' decision-making processes. It also offers threshold values for each feature, enabling us to make additions or subtractions of one STD with the threshold values in alignment with the guided directions for generating adversarial examples. We calculate STD from the training set as mentioned in the previous study \cite{pornprasit2021pyexplainer}. As SHAP and LIME do not provide specific threshold values with their explanations, we perform one STD addition and subtraction operation relative to the original feature values. It is crucial to note that we discard changes that would result in a negative value after subtracting one STD from the current value. This precaution ensures that the transformed instances remain valid, do not violate the fundamental properties of the features, and do not generate out-of-distribution samples. For example, the feature \textit{number of modified directories (nd)} can't have a negative value after its alteration (e.g., $1 - 14 = -13$).

Our objective is to generate adversarial examples such that the feature value changes satisfy the $\ell_0$ perturbations, causing the ML models under attack to fail to accurately predict the maximum number of these adversarial examples, which were predicted correctly before the adversarial attack. To achieve this, we use a multi-objective optimization function as an aggregation of (1) Finding the minimum number of modified features (e.g., $\ell_0$ perturbations) using the \textit{Reverse Elbow Method}; and (2) Maximizing the ASR metric values. Thus, the multi-objective optimization function can be defined as follows:

\[
\label{optim_eqn}
\text{min}\|\mathbf{x} - \mathbf{x'}\|_0 \text{ } (f(\mathbf{x}) \ne  f(\mathbf{x'})) - \lambda \cdot \text{ASR}
\]

In the context of the above multi-objective optimization function, the minus sign ($-$) indicates that we are simultaneously minimizing the number of modified features and maximizing the ASR metric value. Additionally, $\lambda$ is a weighting factor that balances the two objectives by denoting the number of modified features. Thus, we generate transformed instances to conduct adversarial attacks on the ML models.

\subsection{Robustness Evaluation}
\label{robust eval}
We evaluate the robustness of the selected ML models using the generated adversarial examples. To assess the quality of these adversarial examples and evaluate the robustness of the ML models, we employ the Attack Success Rate (ASR) metric, as previously used by Yang et al. \cite{yang2022natural}. The ASR quantifies the percentage of instances that were correctly predicted but are no longer predicted correctly by the ML models following adversarial attacks. The ASR is defined as follows:

\begin{equation}
\label{eqn}
\%ASR = \frac{|\{x|x \in X \wedge M(x') \neq M(x)\}|}{|X|}
\end{equation}

where $X$ is a dataset, $x \in X$ represents an instance, $x'$ denotes the generated adversarial example, and $M$ denotes the ML model under adversarial attack. A higher ASR value indicates that the generated examples possess sufficient quality to challenge the robustness of the ML models. Conversely, a lower ASR value demonstrates the models' resistance to adversarial attacks.

\section{Experimental Results and Analysis}
\label{exp_res}
We implement and evaluate our experiment using the scikit-learn \cite{sklearn_api}, SHAP \cite{lundberg2017unified}, LIME \cite{ribeiro2016should}, PyExplainer \cite{pornprasit2021pyexplainer}, and Adversarial Robustness Toolbox (ART) \cite{art2018} libraries. Our study involves six datasets, three ML explainability techniques, and seven classical ML models, resulting in 126 ($6 \times 3 \times 7 = 126$) experimental combinations. Similarly to prior studies \cite{yang2022natural, hashemi2020permuteattack, ballet2019imperceptible, zhang2020generating}, we focus exclusively on the proportion of correctly predicted instances in the test dataset. Thus, we start by evaluating the performance of our chosen ML models, and then, we assess the effectiveness of our proposed approach through extensive experiments by addressing the following two research questions:

\textbf{RQ1:} \textit{Do changes in important feature values found from ML explainability techniques affect ML model's prediction probability?}

\textbf{RQ2:} \textit{Can we use ML explainability techniques to generate adversarial examples to assess the robustness of machine learning models in software analytics tasks?}

The study proceeds with a comprehensive analysis of the performance of selected ML models. Subsequently, we investigate how altering the top-$k$ important features influences the prediction probabilities of these models (\textbf{RQ1}). We then delve into examining the impact of adversarial attacks on ML model accuracy, utilizing the \textit{ASR} metric (\textbf{RQ2-a}). Finally, we assess the effectiveness of our approach in comparison to baselines in terms of imperceptibility and the ASR metric (\textbf{RQ2-b}).


\textbf{Baseline Attacks:}
To the best of our knowledge, adversarial attacks targeting classical ML models trained on tabular data for various software analytics tasks, such as JIT defect prediction, code review comment classification, and clone detection, have not been explored. Since our proposed attack technique, similar to existing studies \cite{cartella2021adversarial, mathov2020not, du2023extensive, zhang2021advdoor}, operates under a black-box setting (i.e., \textit{it does not require access to model internals such as gradients or parameters}), we compare it against four state-of-the-art black-box attack techniques: \textit{Zoo} \cite{chen2017zoo}, \textit{Boundary Attack} \cite{brendel2017decision}, \textit{PermuteAttack} \cite{hashemi2020permuteattack}, and \textit{HopSkipJump} \cite{chen2020hopskipjumpattack}, all specifically designed for ML models trained on tabular data. We deliberately selected these baselines due to their black-box nature and the availability of their public implementations. A brief description of each baseline attack technique is provided below.


\begin{itemize}
    \item \textbf{Zeroth Order Optimization (ZOO) Attack:} The ZOO attack is a black-box adversarial attack method used to generate adversarial examples for machine learning models. Unlike gradient-based attacks, ZOO does not require direct access to the model’s gradients. Instead, it estimates the gradients by querying the model and using zeroth-order optimization techniques. This makes ZOO particularly useful for attacking black-box models where only input-output pairs are accessible. The key steps involve: (1) \textit{Initialization}: Start with a clean input sample, initialize the perturbation vector and define parameters such as learning rate, number of iterations, and batch size for gradient estimation; (2) \textit{Gradient Estimation}: Estimate the gradient of the loss function with respect to the input using finite difference methods. This involves querying the model with slightly perturbed versions of the input and observing the change in the output; (3) \textit{Gradient Perturbation}: Use the estimated gradients to update the perturbation vector. Apply gradient descent or a similar optimization technique to minimize the loss function with respect to the input; (4) \textit{Projection}: Project the perturbed input back to the valid input-space to ensure it remains within allowable bounds (e.g., valid pixel values for images); (5) \textit{Repeat}: Iterate over steps 2, 3, and 4 for a predefined number of iterations or until the adversarial example successfully fools the model.

    \item \textbf{Boundary Attack:} The Boundary Attack is an iterative, gradient-free method for finding adversarial examples. It starts with an initial adversarial example or a sample near the decision boundary and refines the perturbation through multiple iterations to find one that successfully fools the model. The steps involve: (1) \textit{Initialization}: Start with a set of valid inputs close to the decision boundary of the model; (2) \textit{Boundary Perturbation}: Perturb the input along the direction of the decision boundary, making minor adjustments; (3) \textit{Optimization}: Adjust the perturbation iteratively to find the minimal perturbation that causes the model to misclassify the input. This method is computationally efficient and can be applied to a wide range of models.

    \item \textbf{HopSkipJump Attack:} The HopSkipJumpAttack is a hyperparameter-free and query-efficient adversarial attack technique designed to generate adversarial examples by perturbing the input data to maximize the model's misclassification while remaining within a specified distance from the original input. It works in three steps: (1) \textit{Hop}: Perturb the input data using small, incremental changes; (2) \textit{Skip}: Skip intermediate steps that do not show significant progress towards generating a successful adversarial example; (3) \textit{Jump}: Apply more significant perturbations if needed to achieve the desired adversarial example. HopSkipJump does not rely on gradients to generate adversarial examples, making it suitable for models where gradients are unavailable. Additionally, HopSkipJump is computationally efficient, as it skips unnecessary steps and focuses on promising perturbations.

    \item \textbf{PermuteAttack:} It is a black-box adversarial attack technique that generates counterfactual examples to evaluate the robustness of machine learning models trained on tabular data, including discrete and categorical variables. PermuteAttack uses gradient-free optimization based on a genetic algorithm to generate adversarial examples. The goal of PermuteAttack is to find the $x_{perm}$ where the number of altered features is minimized to $\delta_{0, max}$, and the change in feature values is minimized within an $\ell_2$-ball of $\delta_{2, max}$. The permuted sample that meets these two conditions is considered the counterfactual $x_{cnt}$. PermuteAttack solves equation \ref{permute_eqn} by randomly selecting features and changing their values based on the fitness function defined in the equation.

    \[
    \text{arg} \max_{c \in C} f(x_{\text{cnt}}) = t \text{ such that}
    \]

    \begin{equation}
    \label{permute_eqn}
    \|x_{orig} - x_{cnt}\|_0 \leq \delta_{0, max} \text{ and } \|x_{orig} - x_{cnt}\|_2 \leq \delta_{2, max}
    \end{equation}

    \begin{equation}
    \label{fitnexx_eqn}
    \text{ComputeFitness}(\text{X}) = \|f(x)_t - f(x_{orig})_t\|_2 - \rho_0\|x_{orig} - x\|_0 - \rho_1\|x_{orig} - x\|_2
    \end{equation}

    Where $f(x)_t$ is the outcome for the target class $t$, samples $x$' with higher fitness values are selected for the next iteration of the genetic algorithm. The set $\rho = \{\rho_0, \rho_1\}$ denotes the two conditions mentioned in equation \ref{permute_eqn}.

\end{itemize}

\subsection{Analyse Model Performance}
\label{AnalyzeModelPerformance}
The hyperparameter tuning process for ML models is described in Section \ref{tune_param}. The hyperparameters and their optimized values for the ML models trained on the cross-project mobile app dataset are shown in Table \ref{hyper_val}. Please also refer to our replication package for the hyperparameter settings of other datasets. After determining the optimized hyperparameters, we trained the ML models.

\begin{table*}[htbp]
\centering
\caption{Optimized hyperparameters for different ML models on the CLCDSA dataset after tuning}
\begin{adjustbox}{width=1.0\textwidth}
\begin{tabular}{l|l}
\hline
\textbf{ML Model} & \multicolumn{1}{c}{\textbf{Hyperparameters}}                                                                                                 \\ \hline
LR                 & C: 1.0, dual: False, max\_iter: 140, penalty: `l2', solver:   `lbfgs'                                                           \\ \hline
DT                 & criterion: `entropy', max\_depth: 15, min\_samples\_leaf: 20,   min\_samples\_split: 8                                            \\ \hline
RF                 & bootstrap: False, max\_depth: None, max\_features: `sqrt',   min\_samples\_leaf: 1, min\_samples\_split: 2, n\_estimators: 50 \\ \hline
MLP                & solver: `adam', learning\_rate: `adaptive', hidden\_layer\_sizes:   (10, 30, 10), alpha: 0.05, activation: `relu'               \\ \hline
ADA                & algorithm: `SAMME.R', learning\_rate: 1.02, n\_estimators: 20                                                                       \\ \hline
BAG                & max\_features: 13, max\_samples: 100, n\_estimators: 800                                                                            \\ \hline
GBC                & learning\_rate: 1, max\_depth: 7, min\_samples\_leaf: 0.1,   min\_samples\_split: 0.1, n\_estimators: 200                       \\ \hline
\end{tabular}
\end{adjustbox}
\label{hyper_val}
\end{table*}

Table \ref{dataset} presents the accuracy, F1-score, and AUC values for our ML models trained on various datasets without adversarial attacks. Notably, when considering the cross-project mobile apps dataset, the LR model displays the lowest AUC value (0.73), while the BAG model achieves the highest AUC value (0.85). On the other hand, ML models trained on the CLCDSA dataset consistently exhibit high AUC values, ranging from 0.82 to 0.96. In contrast, for the code review dataset, Table \ref{dataset} reveals relatively lower AUC values; however, the AUC values range from 0.63 to 0.96 for all the ML models. A closer look at Table \ref{dataset} emphasizes that all the ML models maintain acceptable accuracy, F1-scores, and AUC values, indicating their high accuracy and non-overfitting characteristics. Furthermore, all test-AUCs (with the exception of two cases and the code review dataset) exceed 0.75, a benchmark recommended by previous studies as the minimum required to guarantee the reliability of explanations \cite{roy2022don, lyu2021towards}.

\begin{table*}[htbp]
\centering
\caption{Performance of the ML models on different datasets without adversarial attacks considering different evaluation metrics such as Accuracy (Acc), F1-score (F1), and AUC values}
\begin{adjustbox}{max width=\textwidth}
\begin{tabular}{l|llllllllllllllllll}
\hline
                                                                              & \multicolumn{18}{c}{\textbf{Dataset}}                                                                                                                                                                                                                                                                                                                                                                                                                                                                                                                                                              \\ \cmidrule{2-19} 
                                                                              & \multicolumn{3}{c|}{\textbf{Cross Project}}                                                                        & \multicolumn{3}{c|}{\textbf{Java Project}}                                        & \multicolumn{3}{c|}{\textbf{Postgres}}                                                                             & \multicolumn{3}{c|}{\textbf{CLCDSA}}                                              & \multicolumn{3}{c|}{\textbf{Code Review}}                                                                          & \multicolumn{3}{c}{\textbf{BigCloneBench}}                  \\ \cmidrule{2-19} 
\multirow{-3}{*}{\textbf{\begin{tabular}[c]{@{}l@{}}ML\\ Model\end{tabular}}} & \multicolumn{1}{l|}{Acc}  & \multicolumn{1}{l|}{F-1}  & \multicolumn{1}{l|}{AUC}                                   & \multicolumn{1}{l|}{Acc}  & \multicolumn{1}{l|}{F-1}  & \multicolumn{1}{l|}{AUC}  & \multicolumn{1}{l|}{Acc}  & \multicolumn{1}{l|}{F-1}  & \multicolumn{1}{l|}{AUC}                                   & \multicolumn{1}{l|}{Acc}  & \multicolumn{1}{l|}{F-1}  & \multicolumn{1}{l|}{AUC}  & \multicolumn{1}{l|}{Acc}  & \multicolumn{1}{l|}{F-1}  & \multicolumn{1}{l|}{AUC}                                   & \multicolumn{1}{l|}{Acc}  & \multicolumn{1}{l|}{F-1}  & AUC  \\ \hline
LR                                                                            & \multicolumn{1}{l|}{0.68} & \multicolumn{1}{l|}{0.49} & \multicolumn{1}{l|}{\cellcolor[HTML]{C0C0C0}\textbf{0.73}} & \multicolumn{1}{l|}{0.75} & \multicolumn{1}{l|}{0.37} & \multicolumn{1}{l|}{0.76} & \multicolumn{1}{l|}{0.75} & \multicolumn{1}{l|}{0.57} & \multicolumn{1}{l|}{0.76}                                  & \multicolumn{1}{l|}{0.76} & \multicolumn{1}{l|}{0.81} & \multicolumn{1}{l|}{0.82} & \multicolumn{1}{l|}{0.62} & \multicolumn{1}{l|}{0.69} & \multicolumn{1}{l|}{\cellcolor[HTML]{C0C0C0}\textbf{0.63}} & \multicolumn{1}{l|}{0.75} & \multicolumn{1}{l|}{0.74} & 0.76 \\ \hline
DT                                                                            & \multicolumn{1}{l|}{0.79} & \multicolumn{1}{l|}{0.60} & \multicolumn{1}{l|}{0.83}                                  & \multicolumn{1}{l|}{0.80} & \multicolumn{1}{l|}{0.49} & \multicolumn{1}{l|}{0.80} & \multicolumn{1}{l|}{0.78} & \multicolumn{1}{l|}{0.59} & \multicolumn{1}{l|}{0.79}                                  & \multicolumn{1}{l|}{0.86} & \multicolumn{1}{l|}{0.88} & \multicolumn{1}{l|}{0.93} & \multicolumn{1}{l|}{0.67} & \multicolumn{1}{l|}{0.73} & \multicolumn{1}{l|}{\cellcolor[HTML]{C0C0C0}\textbf{0.68}} & \multicolumn{1}{l|}{0.84} & \multicolumn{1}{l|}{0.84} & 0.85 \\ \hline
RF                                                                            & \multicolumn{1}{l|}{0.81} & \multicolumn{1}{l|}{0.58} & \multicolumn{1}{l|}{0.84}                                  & \multicolumn{1}{l|}{0.88} & \multicolumn{1}{l|}{0.55} & \multicolumn{1}{l|}{0.85} & \multicolumn{1}{l|}{0.80} & \multicolumn{1}{l|}{0.59} & \multicolumn{1}{l|}{0.82}                                  & \multicolumn{1}{l|}{0.91} & \multicolumn{1}{l|}{0.93} & \multicolumn{1}{l|}{0.96} & \multicolumn{1}{l|}{0.71} & \multicolumn{1}{l|}{0.76} & \multicolumn{1}{l|}{\cellcolor[HTML]{C0C0C0}\textbf{0.72}} & \multicolumn{1}{l|}{0.88} & \multicolumn{1}{l|}{0.87} & 0.88 \\ \hline
MLP                                                                           & \multicolumn{1}{l|}{0.75} & \multicolumn{1}{l|}{0.57} & \multicolumn{1}{l|}{0.83}                                  & \multicolumn{1}{l|}{0.80} & \multicolumn{1}{l|}{0.48} & \multicolumn{1}{l|}{0.81} & \multicolumn{1}{l|}{0.71} & \multicolumn{1}{l|}{0.51} & \multicolumn{1}{l|}{\cellcolor[HTML]{C0C0C0}\textbf{0.73}} & \multicolumn{1}{l|}{0.83} & \multicolumn{1}{l|}{0.86} & \multicolumn{1}{l|}{0.92} & \multicolumn{1}{l|}{0.60} & \multicolumn{1}{l|}{0.68} & \multicolumn{1}{l|}{\cellcolor[HTML]{C0C0C0}\textbf{0.63}} & \multicolumn{1}{l|}{0.83} & \multicolumn{1}{l|}{0.83} & 0.84 \\ \hline
ADA                                                                           & \multicolumn{1}{l|}{0.73} & \multicolumn{1}{l|}{0.56} & \multicolumn{1}{l|}{0.81}                                  & \multicolumn{1}{l|}{0.79} & \multicolumn{1}{l|}{0.47} & \multicolumn{1}{l|}{0.81} & \multicolumn{1}{l|}{0.78} & \multicolumn{1}{l|}{0.58} & \multicolumn{1}{l|}{0.79}                                  & \multicolumn{1}{l|}{0.81} & \multicolumn{1}{l|}{0.77} & \multicolumn{1}{l|}{0.85} & \multicolumn{1}{l|}{0.63} & \multicolumn{1}{l|}{0.65} & \multicolumn{1}{l|}{\cellcolor[HTML]{C0C0C0}\textbf{0.67}} & \multicolumn{1}{l|}{0.83} & \multicolumn{1}{l|}{0.82} & 0.83 \\ \hline
BAG                                                                           & \multicolumn{1}{l|}{0.77} & \multicolumn{1}{l|}{0.59} & \multicolumn{1}{l|}{0.85}                                  & \multicolumn{1}{l|}{0.78} & \multicolumn{1}{l|}{0.50} & \multicolumn{1}{l|}{0.83} & \multicolumn{1}{l|}{0.78} & \multicolumn{1}{l|}{0.61} & \multicolumn{1}{l|}{0.82}                                  & \multicolumn{1}{l|}{0.81} & \multicolumn{1}{l|}{0.84} & \multicolumn{1}{l|}{0.89} & \multicolumn{1}{l|}{0.65} & \multicolumn{1}{l|}{0.73} & \multicolumn{1}{l|}{\cellcolor[HTML]{C0C0C0}\textbf{0.71}} & \multicolumn{1}{l|}{0.9}  & \multicolumn{1}{l|}{0.89} & 0.9  \\ \hline
GBC                                                                           & \multicolumn{1}{l|}{0.81} & \multicolumn{1}{l|}{0.59} & \multicolumn{1}{l|}{0.84}                                  & \multicolumn{1}{l|}{0.81} & \multicolumn{1}{l|}{0.48} & \multicolumn{1}{l|}{0.80} & \multicolumn{1}{l|}{0.78} & \multicolumn{1}{l|}{0.57} & \multicolumn{1}{l|}{0.80}                                  & \multicolumn{1}{l|}{0.85} & \multicolumn{1}{l|}{0.89} & \multicolumn{1}{l|}{0.94} & \multicolumn{1}{l|}{0.63} & \multicolumn{1}{l|}{0.69} & \multicolumn{1}{l|}{\cellcolor[HTML]{C0C0C0}\textbf{0.65}} & \multicolumn{1}{l|}{0.87} & \multicolumn{1}{l|}{0.86} & 0.87 \\ \hline
\end{tabular}
\end{adjustbox}
\label{dataset}
\end{table*}


\subsection{Present Experimental Results}
\label{PresentExpResutls}
We present the findings of our experimental study regarding our two research questions. The following sections provide a detailed analysis and the corresponding results, with each research question addressed separately.

\textbf{Finding the Top-$k$ Important Features and Their Impact on Model's Predictions when Changed:} 
Our initial step to answering RQ1 involves identifying the top-$k$ important features from the feature importance rank using the \textit{Reverse Elbow Method} and considering the test dataset. Figure \ref{feat_imp} illustrates the features that SHAP, LIME, and PyExplainer identified for the LR model trained on the cross-project mobile apps dataset. SHAP provides a feature importance rank, allowing us to select the top-$k$ important features. Figure \ref{shap_a} presents the feature importance rank identified by SHAP from top to bottom, with each feature's relative importance score on the \textit{x-}axis. For example, the figure highlights that the most important feature is $ndev$ with an approximate relative importance score of $0.13$, followed by $nd$ with an approximate relative importance score of $0.096$. 


Similar to SHAP, LIME also reveals the relative importance of each feature in the model's prediction process. For instance, the explanation in Figure \ref{lime_a} indicates that the ML model predicts the given instance as a \textit{buggy} commit with a $98\%$ probability. Furthermore, we observe that the top-$3$ important features are $la$, $ld$, and $nf$ based on their contribution scores in influencing the model's prediction. In contrast to SHAP and LIME, PyExplainer explicitly identifies the most important features in the explanation that influence the model's prediction. For example, Figure \ref{pyexp_a} displays the important features $nd$ and $ld$ identified by PyExplainer.

\begin{figure}[htbp]
\centering     
\subfigure[Mobile + SHAP]{\label{ASR_Mob_SH}\includegraphics[width=6cm, height=3.5cm]{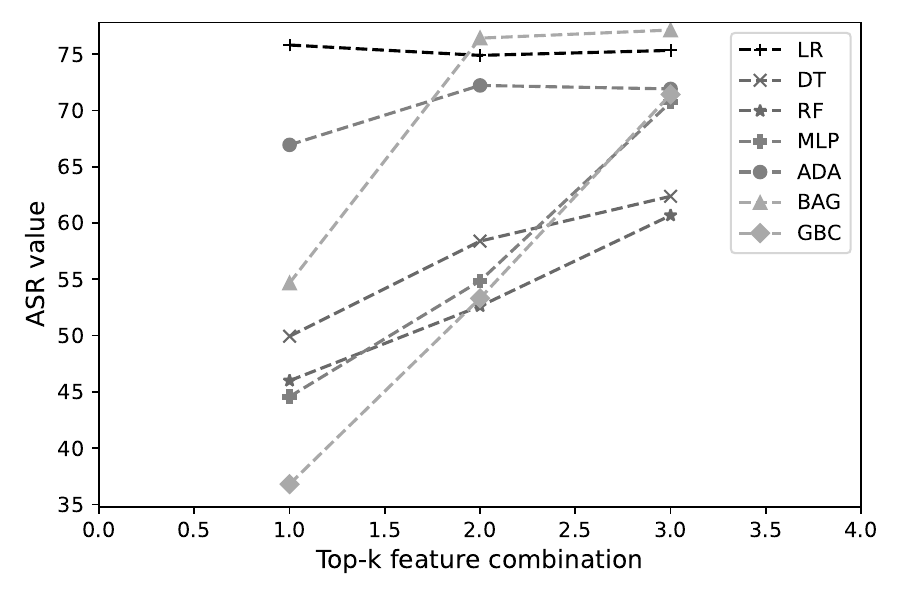}}
\subfigure[Mobile + LIME ]{\label{ASR_Mob_LM}\includegraphics[width=6cm, height=3.5cm]{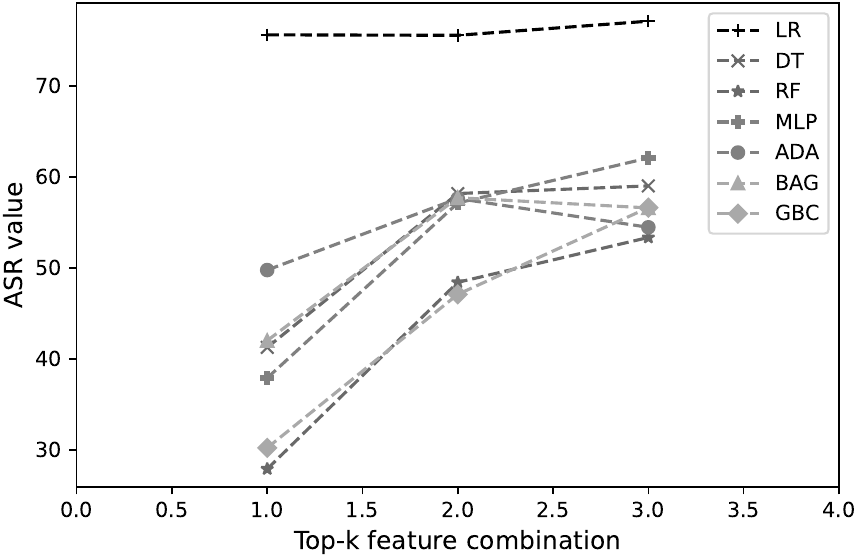}}


\subfigure[Java + SHAP]{\label{ASR_Java_SH}\includegraphics[width=5.75cm, height=3.5cm]{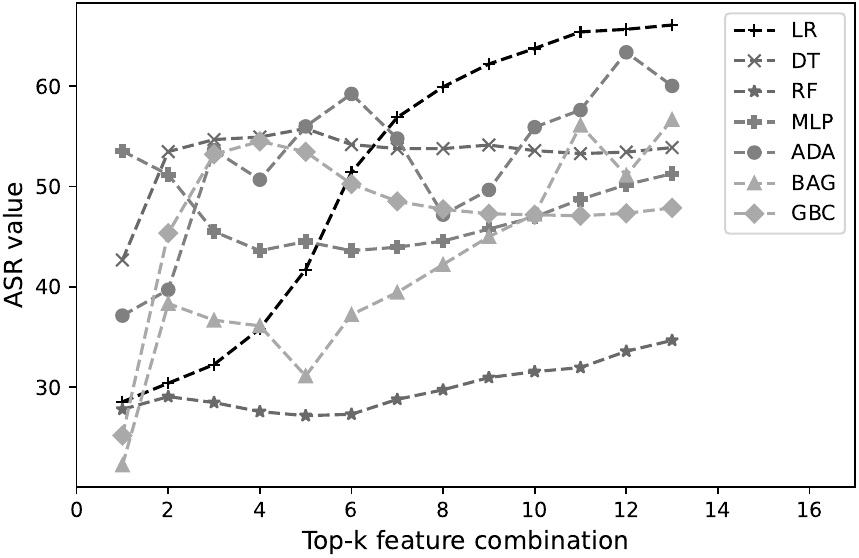}} 
\subfigure[Java + LIME]{\label{ASR_Java_LM}\includegraphics[width=5.75cm, height=3.5cm]{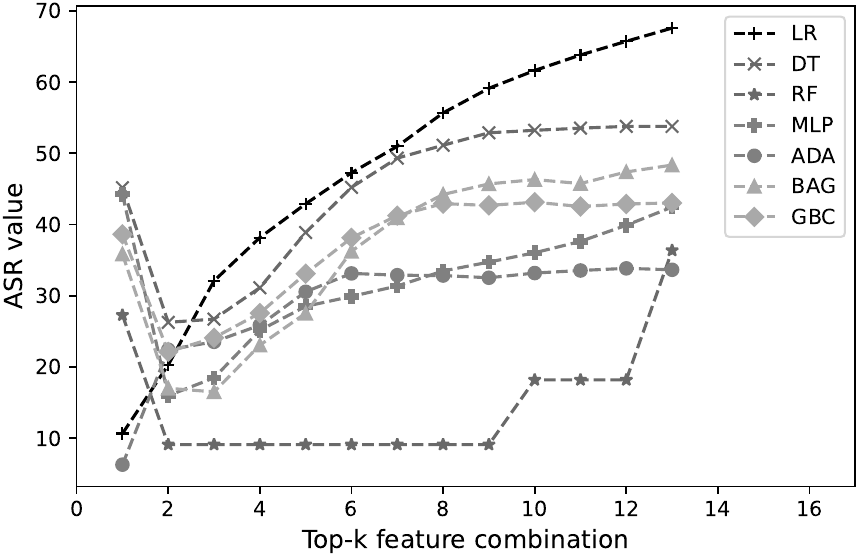}} 

\subfigure[CLCDSA + SHAP]{\label{ASR_CLCDSA_SH}\includegraphics[width=5.75cm, height=3.5cm]{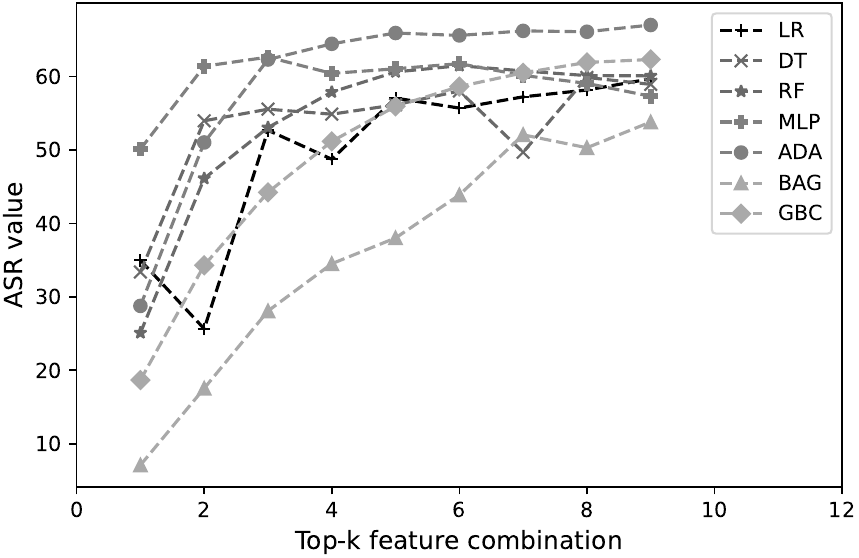}}
\subfigure[CLCDSA + LIME]{\label{ASR_CLCDSA_LM}\includegraphics[width=5.75cm, height=3.5cm]{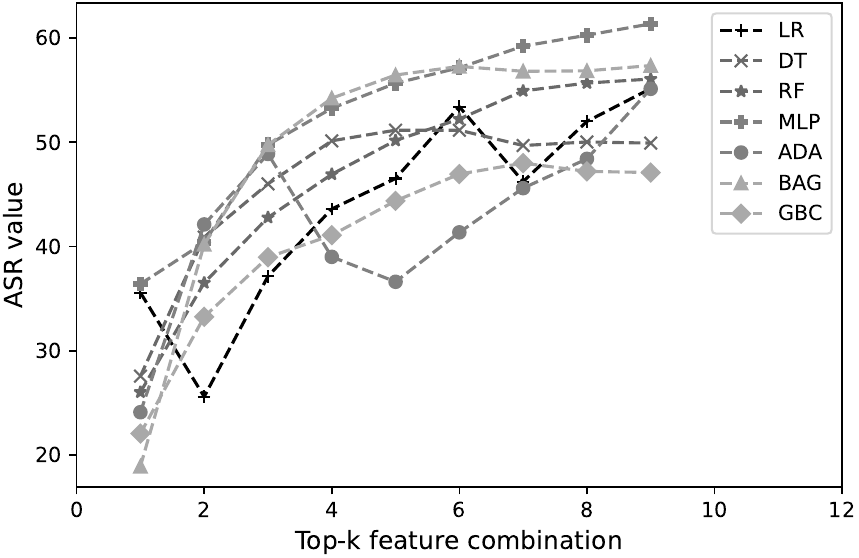}}

\subfigure[BCB + SHAP]{\label{ASR_BCB_SH}\includegraphics[width=5.75cm, height=3.5cm]{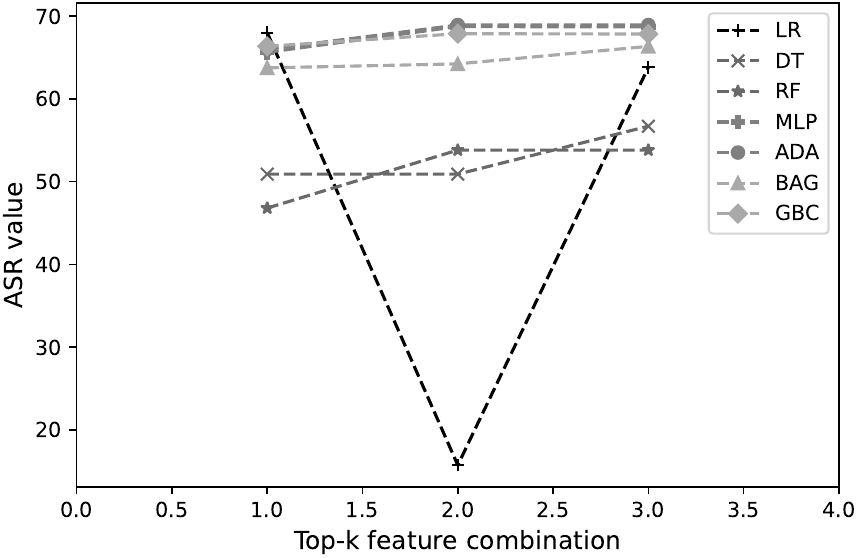}}
\subfigure[BCB + LIME]{\label{ASR_BCB_LM}\includegraphics[width=5.75cm, height=3.5cm]{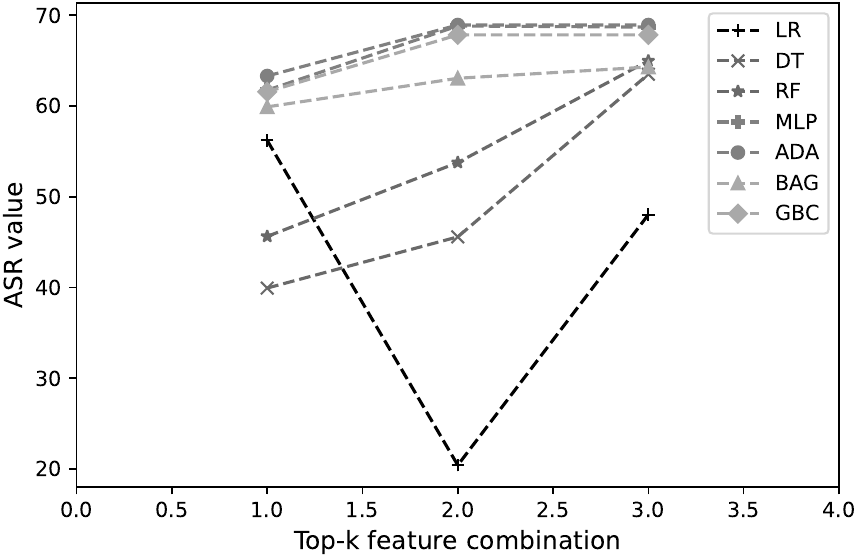}}


\caption{selecting the top-$k$ feature values for different datasets. The first column (e.g., \ref{ASR_Mob_SH} for cross-project mobile apps, \ref{ASR_Java_SH} for Java projects, \ref{ASR_CLCDSA_SH} for CLCDSA, and \ref{ASR_BCB_SH} for BCB) shows how the ASR metric value changes when different combinations of features are selected from the feature importance rank using \textit{SHAP} explainability. The second column (e.g., \ref{ASR_Mob_LM} for cross-project mobile apps, \ref{ASR_Java_LM} for Java projects, \ref{ASR_CLCDSA_LM} for CLCDSA, and \ref{ASR_BCB_LM} for BCB) shows how the ASR metric value changes when different combinations of features are selected from the feature importance rank using LIME explainability.}
\label{ASR_TOP_Feat} 
\end{figure}

Figure \ref{ASR_TOP_Feat} illustrates variations in ASR values resulting from alterations in the values of the top-$k$ feature combinations in SHAP and LIME approaches. We present experimental results for ML models trained on the cross-project mobile apps, Java project, CLCDSA, and BCB datasets. Regarding the SHAP and cross-project mobile apps dataset, the BAG models exhibit a sharp increase in ASR value when changing from top-$1$ to top-$2$ feature values, followed by a flat increase. The GBC model shows a linear increase in the ASR metric value. The LR model displays a decrease in ASR metric value after changing the top-$1$ feature. The ASR metric value remains relatively flat for the DT model after changing the top-$2$ feature values. A similar trend is observed across all ML models trained on the cross-project mobile apps dataset when we employ the LIME explainability to identify the top-$k$ important features.

Figure \ref{ASR_CLCDSA_SH} illustrates that when considering \textit{SHAP}, there is a notable increase in the ASR metric value after changing the top-$1$ or top-$2$ features for all ML models trained on the CLCDSA dataset except the LR model. We observe a sharp decrease in ASR metric value for the LR model after changing top-$2$ features. We observe a very similar result for all the ML models trained on the BCB dataset except the LR model. In the case of the LR model, we observe a sharp decrease after changing the top-$2$ features. In the case of LIME, a sharp decrease in the ASR metric value is observed after altering the top-$1$ feature for all ML models trained on the Java dataset, except for LR. Overall, Figure \ref{ASR_TOP_Feat} visualizes that significant impacts on the ASR metric value occur after changing the top-$2$ or top-$3$ feature values. Thus, we can select the top-$k$ feature values from the feature importance rank using \textit{Reverse Elbow Method}.

\begin{center}
\begin{tcolorbox}[
    enhanced,
    attach boxed title to top left={yshift=-3mm,yshifttext=-1mm}, 
    colback=mycolor_box,                 
    width=0.9\textwidth            
]
    Our \textit{Reverse Elbow Method} effectively identifies the least amount of altered feature values required to impact the model's prediction accuracy to a great extent.
\end{tcolorbox}
\end{center}

Figures \ref{pyexp_a} and \ref{pyexp_b} illustrate how the features \textit{nd} and \textit{ld} influence the LR model's decision-making process. A comparison between these figures reveals that changing the feature values in the guided direction increases the \textbf{Risk Score} from $33.0\%$ to $54.0\%$. The lower Risk Score initially denotes that the given instance is predicted as a \textit{clean} commit. However, changing the values of these important features increases the Risk Score, indicating that the same instance is more likely to be classified as a \textit{buggy} commit. Therefore, our objective is to investigate whether there are differences, and to what extent, in prediction probabilities between the original and transformed instances across all test instances (e.g., considering the entire test dataset) and among all the ML models.

\begin{figure}[htbp]
\centering     
\subfigure[LR]{\label{LR_RQ1}\includegraphics[width=4.25cm, height=3.25cm]{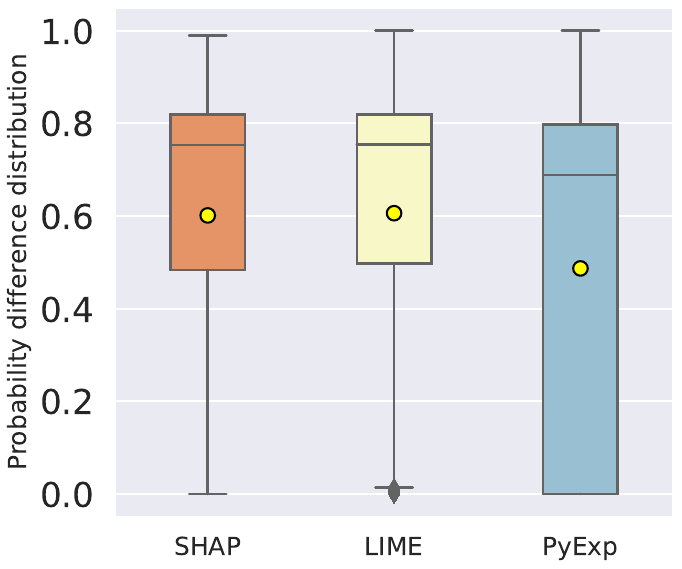}}
\subfigure[DT]{\label{DT_RQ1}\includegraphics[width=4.25cm, height=3.25cm]{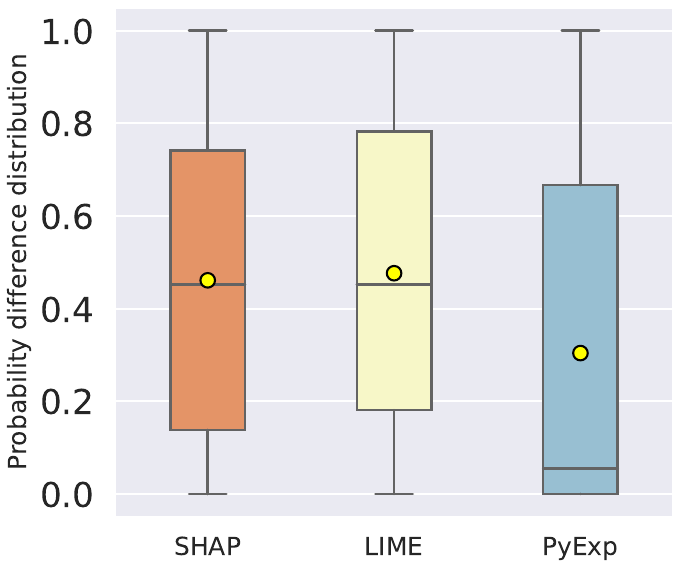}} 
\subfigure[RF]{\label{RF_RQ1}\includegraphics[width=4.25cm, height=3.25cm]{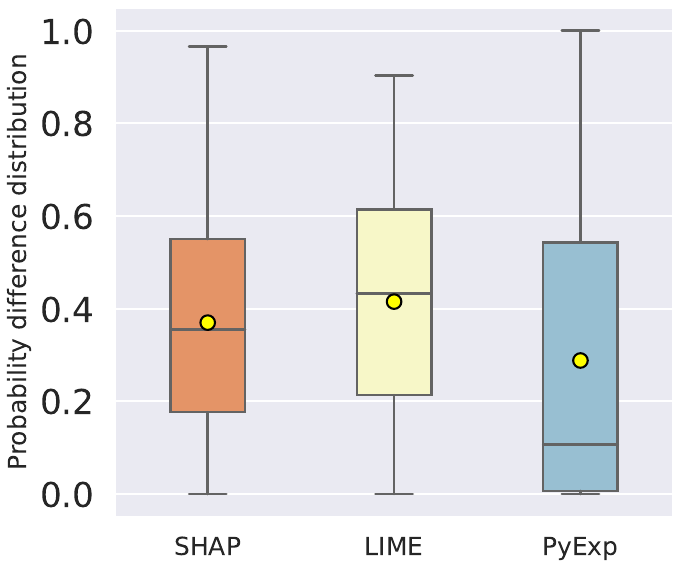}} 
\subfigure[MLP]{\label{MLP_RQ1}\includegraphics[width=4.25cm, height=3.25cm]{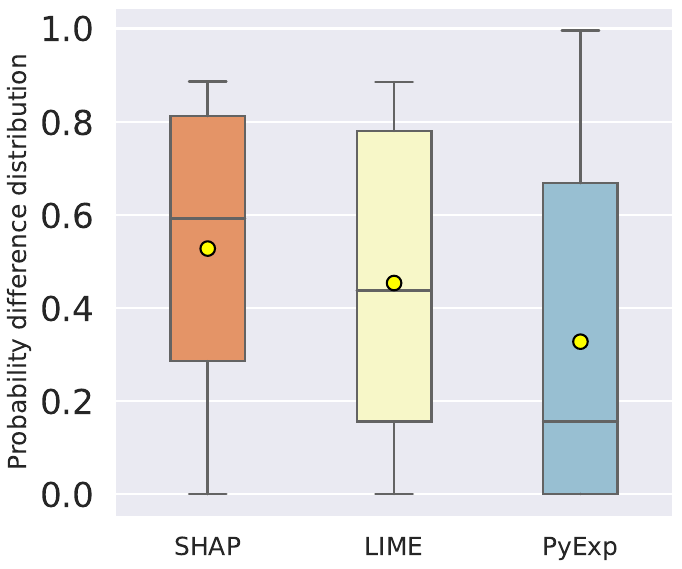}}
\vspace{-0.5em}
\subfigure[ADA]{\label{ADA_RQ1}\includegraphics[width=4.25cm, height=3.25cm]{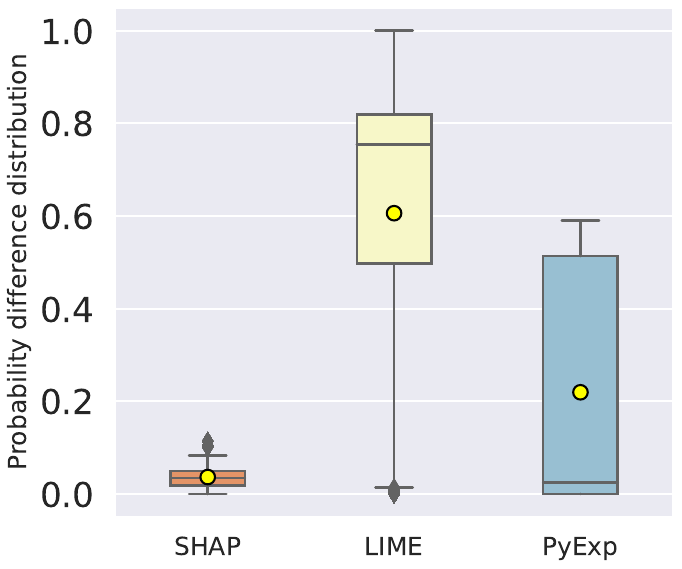}} 
\subfigure[BAG]{\label{BAG_RQ1}\includegraphics[width=4.25cm, height=3.25cm]{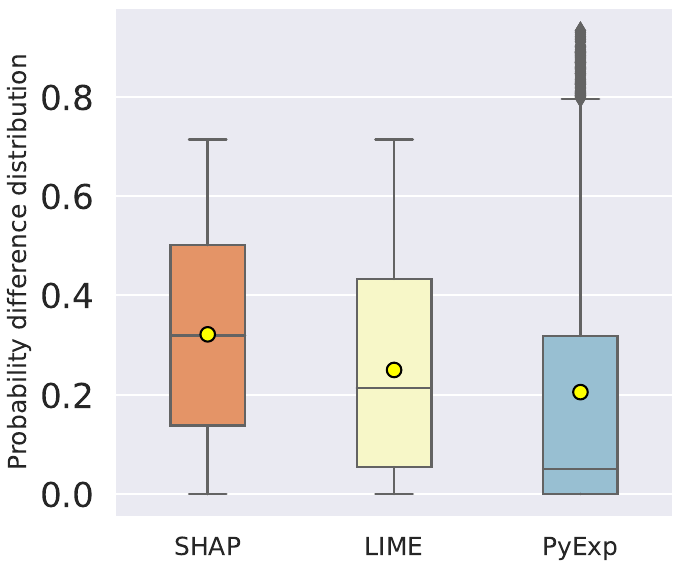}}
\subfigure[GBC]{\label{GBC_RQ1}\includegraphics[width=4.25cm, height=3.25cm]{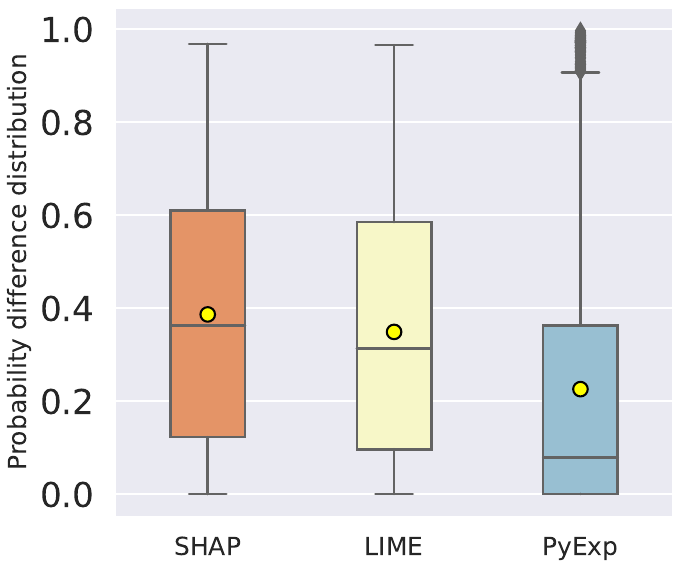}} 

\caption{Distribution of prediction probability differences between the original and transformed instances for various ML models trained on the cross-project mobile apps dataset are shown in Figures \ref{LR_RQ1}, \ref{DT_RQ1}, \ref{RF_RQ1}, \ref{MLP_RQ1}, \ref{ADA_RQ1}, \ref{BAG_RQ1}, and \ref{GBC_RQ1}. These figures represent the distribution for LR, DT, RF, MLP, ADA, BAG, and GBC models, respectively.}
\label{RQ1_Fig}
\end{figure}

\begin{figure}[htbp]
\centering     
\subfigure[LR]{\label{LR_RQ1_BCB}\includegraphics[width=4.25cm, height=3.25cm]{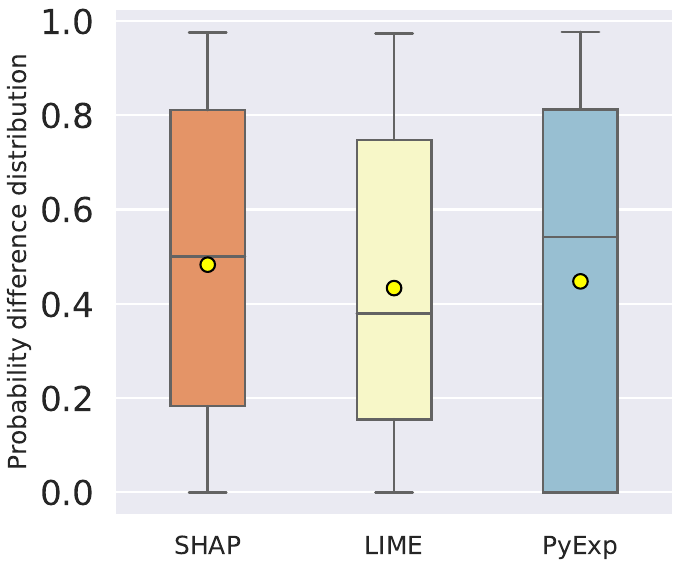}}
\subfigure[DT]{\label{DT_RQ1_BCB}\includegraphics[width=4.25cm, height=3.25cm]{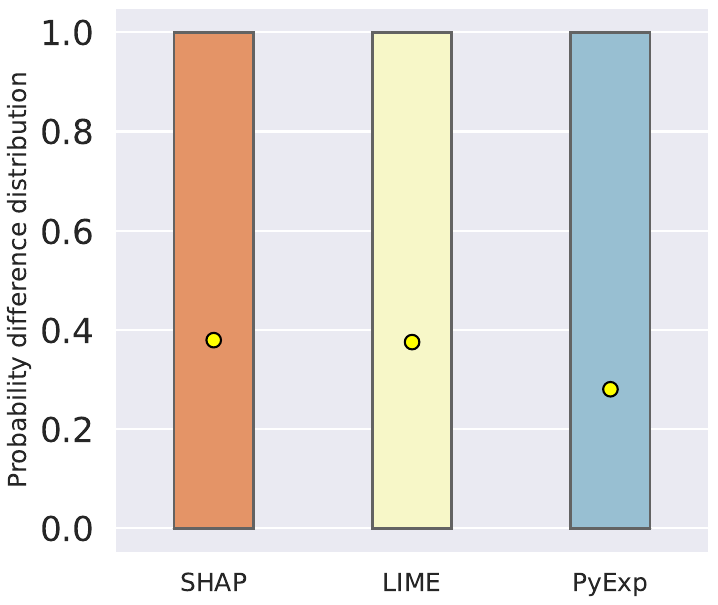}} 
\subfigure[RF]{\label{RF_RQ1_BCB}\includegraphics[width=4.25cm, height=3.25cm]{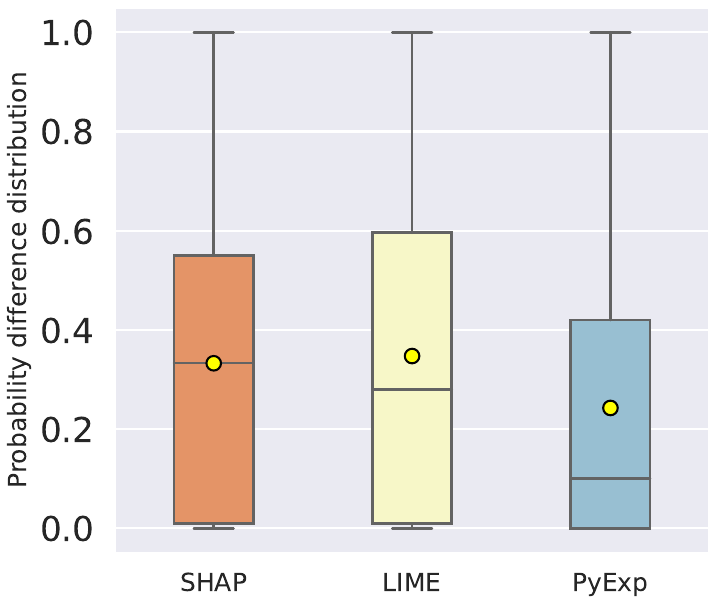}} 
\subfigure[MLP]{\label{MLP_RQ1_BCB}\includegraphics[width=4.25cm, height=3.25cm]{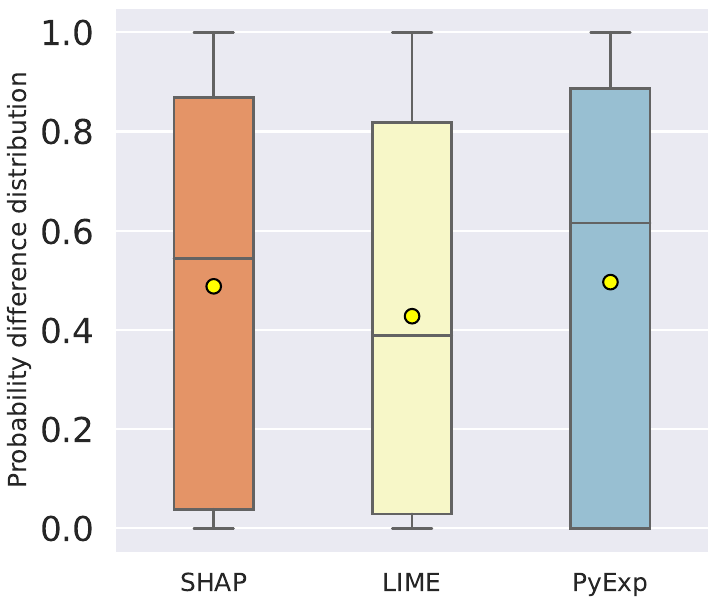}}
\vspace{-0.5em}
\subfigure[ADA]{\label{ADA_RQ1_BCB}\includegraphics[width=4.25cm, height=3.25cm]{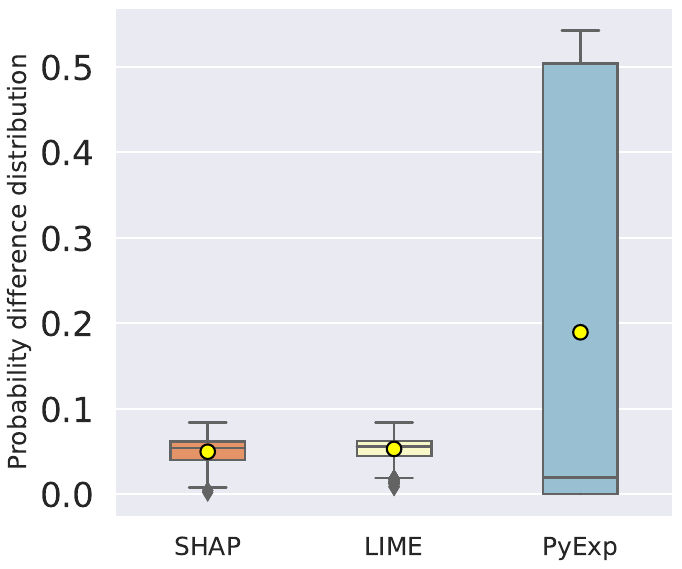}} 
\subfigure[BAG]{\label{BAG_RQ1_BCB}\includegraphics[width=4.25cm, height=3.25cm]{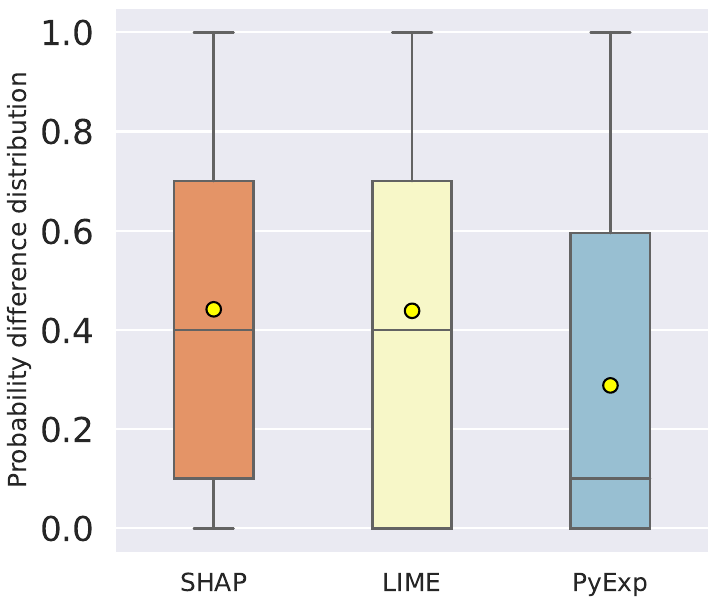}}
\subfigure[GBC]{\label{GBC_RQ1_BCB}\includegraphics[width=4.25cm, height=3.25cm]{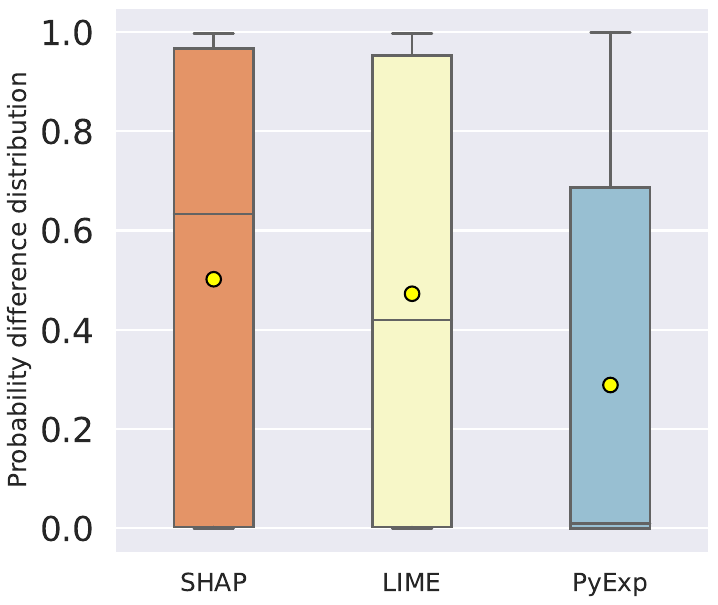}} 

\caption{Distribution of prediction probability differences between the original and transformed instances for various ML models trained on the BigCloneBench dataset are shown in Figures \ref{LR_RQ1_BCB}, \ref{DT_RQ1_BCB}, \ref{RF_RQ1_BCB}, \ref{MLP_RQ1_BCB}, \ref{ADA_RQ1_BCB}, \ref{BAG_RQ1_BCB}, and \ref{GBC_RQ1_BCB}. These figures represent the distribution for LR, DT, RF, MLP, ADA, BAG, and GBC models, respectively.}
\label{RQ1_BCB_Fig}
\end{figure}

Figures \ref{RQ1_Fig} and \ref{RQ1_BCB_Fig} illustrate the distribution of prediction probability differences between the original and transformed instances for the ML models trained on the cross-project mobile apps and BCB datasets. The both figures clearly indicate substantial differences in prediction probabilities between the original and transformed instances, except for the ADA model and SHAP (Figures \ref{ADA_RQ1} and \ref{ADA_RQ1_BCB}). For example, in the case of the \textit{LR} model, the \textit{mean} difference is approximately $0.61$ when changing the top-$k$ important feature values identified by SHAP and LIME. The \textit{mean} probability difference is approximately $0.47$ when modifying the top-$k$ important feature values identified by PyExplainer (PyExp). The \textit{mean} difference consistently exceeds $0.22$ for all other ML models and explainability techniques. This denotes that a significant prediction probability difference exists between the original and the transformed instances when we modify the original instance based on the top-$k$ important features identified by ML explainability techniques. The only exception is in the case of the ADA model, where SHAP shows a comparatively lower prediction probability difference. In contrast, LIME and PyExplainer maintain consistency compared to SHAP for all the ML models.

While Figures \ref{RQ1_Fig} and~\ref{RQ1_BCB_Fig} visually demonstrate the differences in prediction probabilities between the original and transformed instances, we further perform a Wilcoxon Signed-Rank test \cite{wilcoxon1992individual} to assess these differences statistically. We also report Cliff’s $|\delta|$ effect size \cite{cliff1993dominance}, whose interpretation is provided in Table \ref{cliff_delta_table}. Our results show that, in all cases, the $p$-values are below the significance threshold (0.05), and the effect sizes exceed 0.74, except for the ADA model trained on the cross-project mobile apps and BigCloneBench datasets using the SHAP and LIME explainers. These findings provide strong statistical evidence of a significant difference in prediction probabilities between the original and transformed instances.

\begin{table}[htbp]
\centering
\small
\caption{Interpretation of Cliff's $|\delta|$ effect size}
\begin{tabular}{|l|l|}
\hline
\multicolumn{1}{|c|}{\textbf{Interval}} & \multicolumn{1}{c|}{\textbf{Interpret Effect}} \\ \hline
$\delta < 0.147$                        & Negligible effect                                 \\ \hline
$0.147 \leq  0.33$                      & Small effect                                   \\ \hline
$0.33 \leq  0.474$                      & Medium effect                                  \\ \hline
$\delta \geq 0.474$                     & Large effect                                   \\ \hline
\end{tabular}
\label{cliff_delta_table}
\end{table}

The experimental results for the ML models trained on the other four datasets are available in our replication package. However, our experimental findings demonstrate similar results across all other datasets. Therefore, based on our experimental results, we conclude that changing the top-$k$ important feature values identified by ML explainability techniques can be employed to generate adversarial examples to assess the robustness of ML models in software analytics tasks.



\begin{center}
\begin{tcolorbox}[
    enhanced,
    attach boxed title to top left={yshift=-3mm,yshifttext=-1mm}, 
    colback=mycolor_box,                 
    colframe=black,                
    colbacktitle= mycolor_title,            
    coltitle=black,                
    title=Result RQ1,            
    fonttitle=\bfseries,           
    boxed title style={size=small},
    width=0.9\textwidth            
]
    Changing important feature values identified by ML explainability techniques significantly affects the prediction probability of the ML models in software analytics tasks.
\end{tcolorbox}
\end{center}

\begin{table*}[htbp]
\centering
\caption{Attack Success Rate (ASR) metric values and the Top-$k$ modified features (e.g., a minimal $\ell_0$ perturbation) when we apply explanation-guided adversarial attacks and selected baseline attacks on the ML models for various datasets}
\begin{adjustbox}{max width=\textwidth}
\begin{tabular}{|llllllllllllll|}
\hline
\multicolumn{1}{|l|}{}                                                                                    & \multicolumn{13}{c|}{\textbf{Cross-project   Mobile Apps}}                                                                                                                                                                                                                                                                                                                                                                                                                                                                                                                                                                                                                                    \\ \cmidrule{2-14} 
\multicolumn{1}{|l|}{}                                                                                    & \multicolumn{2}{c|}{\textbf{SHAP}}                                        & \multicolumn{2}{c|}{\textbf{LIME}}                                        & \multicolumn{1}{c|}{\textbf{PyExplainer}} & \multicolumn{2}{c|}{\textbf{PermuteAttack}}                                                                              & \multicolumn{2}{c|}{\textbf{Zoo}}                                                                                         & \multicolumn{2}{c|}{\textbf{Boundary}}                                                                                     & \multicolumn{2}{c|}{\textbf{HopSkipJump}}                                                             \\ \cmidrule{2-14} 
\multicolumn{1}{|l|}{\multirow{-3}{*}{\textbf{\begin{tabular}[c]{@{}l@{}}ML\\    \\ Model\end{tabular}}}} & \multicolumn{1}{l|}{ASR}            & \multicolumn{1}{l|}{Top-$k$}        & \multicolumn{1}{l|}{ASR}            & \multicolumn{1}{l|}{Top-$k$}        & \multicolumn{1}{l|}{}                     & \multicolumn{1}{l|}{\cellcolor[HTML]{ECF4FF}ASR}           & \multicolumn{1}{l|}{\cellcolor[HTML]{ECF4FF}Top-$k$}        & \multicolumn{1}{l|}{\cellcolor[HTML]{ECF4FF}ASR}           & \multicolumn{1}{l|}{\cellcolor[HTML]{ECF4FF}Top-$k$}         & \multicolumn{1}{l|}{\cellcolor[HTML]{ECF4FF}ASR}            & \multicolumn{1}{l|}{\cellcolor[HTML]{ECF4FF}Top-$k$}         & \multicolumn{1}{l|}{\cellcolor[HTML]{ECF4FF}ASR}            & \cellcolor[HTML]{ECF4FF}Top-$k$         \\ \hline
\multicolumn{1}{|l|}{LR}                                                                                  & \multicolumn{1}{l|}{\textbf{75.8}}  & \multicolumn{1}{l|}{\textbf{Top-1}} & \multicolumn{1}{l|}{75.6}           & \multicolumn{1}{l|}{Top-1}          & \multicolumn{1}{l|}{56.3}                 & \multicolumn{1}{l|}{\cellcolor[HTML]{ECF4FF}88.1}          & \multicolumn{1}{l|}{\cellcolor[HTML]{ECF4FF}Top-4}          & \multicolumn{1}{l|}{\cellcolor[HTML]{ECF4FF}78.2}          & \multicolumn{1}{l|}{\cellcolor[HTML]{ECF4FF}Top-6}           & \multicolumn{1}{l|}{\cellcolor[HTML]{ECF4FF}\textbf{56.8}}  & \multicolumn{1}{l|}{\cellcolor[HTML]{ECF4FF}\textbf{Top-6}}  & \multicolumn{1}{l|}{\cellcolor[HTML]{ECF4FF}\textbf{55.7}}  & \cellcolor[HTML]{ECF4FF}\textbf{Top-6}  \\ \hline
\multicolumn{1}{|l|}{DT}                                                                                  & \multicolumn{1}{l|}{\textbf{58.4}}  & \multicolumn{1}{l|}{\textbf{Top-2}} & \multicolumn{1}{l|}{58.2}           & \multicolumn{1}{l|}{Top-2}          & \multicolumn{1}{l|}{23.3}                 & \multicolumn{1}{l|}{\cellcolor[HTML]{ECF4FF}99.6}          & \multicolumn{1}{l|}{\cellcolor[HTML]{ECF4FF}Top-3}          & \multicolumn{1}{l|}{\cellcolor[HTML]{ECF4FF}\textbf{25.2}} & \multicolumn{1}{l|}{\cellcolor[HTML]{ECF4FF}\textbf{Top-2}}  & \multicolumn{1}{l|}{\cellcolor[HTML]{ECF4FF}85.7}           & \multicolumn{1}{l|}{\cellcolor[HTML]{ECF4FF}Top-6}           & \multicolumn{1}{l|}{\cellcolor[HTML]{ECF4FF}84.9}           & \cellcolor[HTML]{ECF4FF}Top-6           \\ \hline
\multicolumn{1}{|l|}{RF}                                                                                  & \multicolumn{1}{l|}{\textbf{60.8}}  & \multicolumn{1}{l|}{\textbf{Top-3}} & \multicolumn{1}{l|}{48.2}           & \multicolumn{1}{l|}{Top-2}          & \multicolumn{1}{l|}{32.1}                 & \multicolumn{1}{l|}{\cellcolor[HTML]{ECF4FF}98.1}          & \multicolumn{1}{l|}{\cellcolor[HTML]{ECF4FF}Top-4}          & \multicolumn{1}{l|}{\cellcolor[HTML]{ECF4FF}\textbf{40.8}} & \multicolumn{1}{l|}{\cellcolor[HTML]{ECF4FF}\textbf{Top-3}}  & \multicolumn{1}{l|}{\cellcolor[HTML]{ECF4FF}99.7}           & \multicolumn{1}{l|}{\cellcolor[HTML]{ECF4FF}Top-6}           & \multicolumn{1}{l|}{\cellcolor[HTML]{ECF4FF}99.9}           & \cellcolor[HTML]{ECF4FF}Top-6           \\ \hline
\multicolumn{1}{|l|}{MLP}                                                                                 & \multicolumn{1}{l|}{\textbf{70.7}}  & \multicolumn{1}{l|}{\textbf{Top-3}} & \multicolumn{1}{l|}{54.1}           & \multicolumn{1}{l|}{Top-2}          & \multicolumn{1}{l|}{38.7}                 & \multicolumn{1}{l|}{\cellcolor[HTML]{ECF4FF}92.1}          & \multicolumn{1}{l|}{\cellcolor[HTML]{ECF4FF}Top-4}          & \multicolumn{1}{l|}{\cellcolor[HTML]{ECF4FF}71.5}          & \multicolumn{1}{l|}{\cellcolor[HTML]{ECF4FF}Top-6}           & \multicolumn{1}{l|}{\cellcolor[HTML]{ECF4FF}99.9}           & \multicolumn{1}{l|}{\cellcolor[HTML]{ECF4FF}Top-6}           & \multicolumn{1}{l|}{\cellcolor[HTML]{ECF4FF}99.9}           & \cellcolor[HTML]{ECF4FF}Top-6           \\ \hline
\multicolumn{1}{|l|}{ADA}                                                                                 & \multicolumn{1}{l|}{\textbf{72.2}}  & \multicolumn{1}{l|}{\textbf{Top-2}} & \multicolumn{1}{l|}{57.7}           & \multicolumn{1}{l|}{Top-2}          & \multicolumn{1}{l|}{48.0}                 & \multicolumn{1}{l|}{\cellcolor[HTML]{ECF4FF}11.4}          & \multicolumn{1}{l|}{\cellcolor[HTML]{ECF4FF}Top-2}          & \multicolumn{1}{l|}{\cellcolor[HTML]{ECF4FF}\textbf{5.1}}  & \multicolumn{1}{l|}{\cellcolor[HTML]{ECF4FF}\textbf{Top-2}}  & \multicolumn{1}{l|}{\cellcolor[HTML]{ECF4FF}\textbf{23.3}}  & \multicolumn{1}{l|}{\cellcolor[HTML]{ECF4FF}\textbf{Top-6}}  & \multicolumn{1}{l|}{\cellcolor[HTML]{ECF4FF}\textbf{23.1}}  & \cellcolor[HTML]{ECF4FF}\textbf{Top-6}  \\ \hline
\multicolumn{1}{|l|}{BAG}                                                                                 & \multicolumn{1}{l|}{\textbf{76.4}}  & \multicolumn{1}{l|}{\textbf{Top-2}} & \multicolumn{1}{l|}{57.7}           & \multicolumn{1}{l|}{Top-2}          & \multicolumn{1}{l|}{28.6}                 & \multicolumn{1}{l|}{\cellcolor[HTML]{ECF4FF}74.1}          & \multicolumn{1}{l|}{\cellcolor[HTML]{ECF4FF}Top-4}          & \multicolumn{1}{l|}{\cellcolor[HTML]{ECF4FF}\textbf{31.1}} & \multicolumn{1}{l|}{\cellcolor[HTML]{ECF4FF}\textbf{Top-6}}  & \multicolumn{1}{l|}{\cellcolor[HTML]{ECF4FF}\textbf{22.2}}  & \multicolumn{1}{l|}{\cellcolor[HTML]{ECF4FF}\textbf{Top-6}}  & \multicolumn{1}{l|}{\cellcolor[HTML]{ECF4FF}\textbf{22.2}}  & \cellcolor[HTML]{ECF4FF}\textbf{Top-6}  \\ \hline
\multicolumn{1}{|l|}{GBC}                                                                                 & \multicolumn{1}{l|}{\textbf{71.4}}  & \multicolumn{1}{l|}{\textbf{Top-3}} & \multicolumn{1}{l|}{47.1}           & \multicolumn{1}{l|}{Top-2}          & \multicolumn{1}{l|}{28.4}                 & \multicolumn{1}{l|}{\cellcolor[HTML]{ECF4FF}99.3}          & \multicolumn{1}{l|}{\cellcolor[HTML]{ECF4FF}Top-4}          & \multicolumn{1}{l|}{\cellcolor[HTML]{ECF4FF}\textbf{42.8}} & \multicolumn{1}{l|}{\cellcolor[HTML]{ECF4FF}\textbf{Top-2}}  & \multicolumn{1}{l|}{\cellcolor[HTML]{ECF4FF}84.4}           & \multicolumn{1}{l|}{\cellcolor[HTML]{ECF4FF}Top-6}           & \multicolumn{1}{l|}{\cellcolor[HTML]{ECF4FF}84.9}           & \cellcolor[HTML]{ECF4FF}Top-6           \\ \hline
\multicolumn{14}{|c|}{\textbf{Java Project}}                                                                                                                                                                                                                                                                                                                                                                                                                                                                                                                                                                                                                                                                                                                                                              \\ \hline
\multicolumn{1}{|l|}{LR}                                                                                  & \multicolumn{1}{l|}{\textbf{51.5}}  & \multicolumn{1}{l|}{\textbf{Top-6}} & \multicolumn{1}{l|}{32.9}           & \multicolumn{1}{l|}{Top-2}          & \multicolumn{1}{l|}{32.1}                 & \multicolumn{1}{l|}{\cellcolor[HTML]{ECF4FF}99.4}          & \multicolumn{1}{l|}{\cellcolor[HTML]{ECF4FF}Top-4}          & \multicolumn{1}{l|}{\cellcolor[HTML]{ECF4FF}\textbf{35.5}} & \multicolumn{1}{l|}{\cellcolor[HTML]{ECF4FF}\textbf{Top-13}} & \multicolumn{1}{l|}{\cellcolor[HTML]{ECF4FF}\textbf{11.4}}  & \multicolumn{1}{l|}{\cellcolor[HTML]{ECF4FF}\textbf{Top-27}} & \multicolumn{1}{l|}{\cellcolor[HTML]{ECF4FF}\textbf{11.8}}  & \cellcolor[HTML]{ECF4FF}\textbf{Top-27} \\ \hline
\multicolumn{1}{|l|}{DT}                                                                                  & \multicolumn{1}{l|}{\textbf{53.5}}  & \multicolumn{1}{l|}{\textbf{Top-2}} & \multicolumn{1}{l|}{45.2}           & \multicolumn{1}{l|}{Top-1}          & \multicolumn{1}{l|}{18.4}                 & \multicolumn{1}{l|}{\cellcolor[HTML]{ECF4FF}98.4}          & \multicolumn{1}{l|}{\cellcolor[HTML]{ECF4FF}Top-5}          & \multicolumn{1}{l|}{\cellcolor[HTML]{C0C0C0}\textbf{0.0}}  & \multicolumn{1}{l|}{\cellcolor[HTML]{C0C0C0}\textbf{Top-0}}  & \multicolumn{1}{l|}{\cellcolor[HTML]{ECF4FF}87.9}           & \multicolumn{1}{l|}{\cellcolor[HTML]{ECF4FF}Top-27}          & \multicolumn{1}{l|}{\cellcolor[HTML]{ECF4FF}87.9}           & \cellcolor[HTML]{ECF4FF}Top-27          \\ \hline
\multicolumn{1}{|l|}{RF}                                                                                  & \multicolumn{1}{l|}{\textbf{29.1}}  & \multicolumn{1}{l|}{\textbf{Top-2}} & \multicolumn{1}{l|}{27.3}           & \multicolumn{1}{l|}{Top-1}          & \multicolumn{1}{l|}{14.8}                 & \multicolumn{1}{l|}{\cellcolor[HTML]{ECF4FF}98.6}          & \multicolumn{1}{l|}{\cellcolor[HTML]{ECF4FF}Top-7}          & \multicolumn{1}{l|}{\cellcolor[HTML]{C0C0C0}\textbf{0.0}}  & \multicolumn{1}{l|}{\cellcolor[HTML]{C0C0C0}\textbf{Top-0}}  & \multicolumn{1}{l|}{\cellcolor[HTML]{ECF4FF}99.9}           & \multicolumn{1}{l|}{\cellcolor[HTML]{ECF4FF}Top-27}          & \multicolumn{1}{l|}{\cellcolor[HTML]{ECF4FF}100}            & \cellcolor[HTML]{ECF4FF}Top-27          \\ \hline
\multicolumn{1}{|l|}{MLP}                                                                                 & \multicolumn{1}{l|}{\textbf{53.5}}  & \multicolumn{1}{l|}{\textbf{Top-1}} & \multicolumn{1}{l|}{44.3}           & \multicolumn{1}{l|}{Top-1}          & \multicolumn{1}{l|}{26.3}                 & \multicolumn{1}{l|}{\cellcolor[HTML]{ECF4FF}99.6}          & \multicolumn{1}{l|}{\cellcolor[HTML]{ECF4FF}Top-6}          & \multicolumn{1}{l|}{\cellcolor[HTML]{ECF4FF}\textbf{26.1}} & \multicolumn{1}{l|}{\cellcolor[HTML]{ECF4FF}\textbf{Top-13}} & \multicolumn{1}{l|}{\cellcolor[HTML]{ECF4FF}\textbf{14.2}}  & \multicolumn{1}{l|}{\cellcolor[HTML]{ECF4FF}\textbf{Top-27}} & \multicolumn{1}{l|}{\cellcolor[HTML]{ECF4FF}\textbf{14.3}}  & \cellcolor[HTML]{ECF4FF}\textbf{Top-27} \\ \hline
\multicolumn{1}{|l|}{ADA}                                                                                 & \multicolumn{1}{l|}{\textbf{53.6}}  & \multicolumn{1}{l|}{\textbf{Top-3}} & \multicolumn{1}{l|}{22.4}           & \multicolumn{1}{l|}{Top-2}          & \multicolumn{1}{l|}{32.9}                 & \multicolumn{1}{l|}{\cellcolor[HTML]{C0C0C0}\textbf{0.0}}  & \multicolumn{1}{l|}{\cellcolor[HTML]{C0C0C0}\textbf{Top-0}} & \multicolumn{1}{l|}{\cellcolor[HTML]{C0C0C0}\textbf{0.0}}  & \multicolumn{1}{l|}{\cellcolor[HTML]{C0C0C0}\textbf{Top-0}}  & \multicolumn{1}{l|}{\cellcolor[HTML]{ECF4FF}\textbf{12.4}}  & \multicolumn{1}{l|}{\cellcolor[HTML]{ECF4FF}\textbf{Top-27}} & \multicolumn{1}{l|}{\cellcolor[HTML]{ECF4FF}\textbf{12.4}}  & \cellcolor[HTML]{ECF4FF}\textbf{Top-27} \\ \hline
\multicolumn{1}{|l|}{BAG}                                                                                 & \multicolumn{1}{l|}{\textbf{38.3}}  & \multicolumn{1}{l|}{\textbf{Top-2}} & \multicolumn{1}{l|}{35.8}           & \multicolumn{1}{l|}{Top-1}          & \multicolumn{1}{l|}{21.7}                 & \multicolumn{1}{l|}{\cellcolor[HTML]{ECF4FF}73.1}          & \multicolumn{1}{l|}{\cellcolor[HTML]{ECF4FF}Top-7}          & \multicolumn{1}{l|}{\cellcolor[HTML]{ECF4FF}\textbf{0.25}} & \multicolumn{1}{l|}{\cellcolor[HTML]{ECF4FF}\textbf{Top-3}}  & \multicolumn{1}{l|}{\cellcolor[HTML]{ECF4FF}99.7}           & \multicolumn{1}{l|}{\cellcolor[HTML]{ECF4FF}Top-27}          & \multicolumn{1}{l|}{\cellcolor[HTML]{ECF4FF}99.8}           & \cellcolor[HTML]{ECF4FF}Top-27          \\ \hline
\multicolumn{1}{|l|}{GBC}                                                                                 & \multicolumn{1}{l|}{\textbf{45.4}}  & \multicolumn{1}{l|}{\textbf{Top-2}} & \multicolumn{1}{l|}{38.6}           & \multicolumn{1}{l|}{Top-1}          & \multicolumn{1}{l|}{22.2}                 & \multicolumn{1}{l|}{\cellcolor[HTML]{ECF4FF}99.6}          & \multicolumn{1}{l|}{\cellcolor[HTML]{ECF4FF}Top-6}          & \multicolumn{1}{l|}{\cellcolor[HTML]{C0C0C0}\textbf{0.0}}  & \multicolumn{1}{l|}{\cellcolor[HTML]{C0C0C0}\textbf{Top-0}}  & \multicolumn{1}{l|}{\cellcolor[HTML]{ECF4FF}\textbf{11.5}}  & \multicolumn{1}{l|}{\cellcolor[HTML]{ECF4FF}\textbf{Top-27}} & \multicolumn{1}{l|}{\cellcolor[HTML]{ECF4FF}\textbf{11.4}}  & \cellcolor[HTML]{ECF4FF}\textbf{Top-27} \\ \hline
\multicolumn{14}{|c|}{\textbf{Postgres}}                                                                                                                                                                                                                                                                                                                                                                                                                                                                                                                                                                                                                                                                                                                                                                  \\ \hline
\multicolumn{1}{|l|}{LR}                                                                                  & \multicolumn{1}{l|}{\textbf{86.6}}  & \multicolumn{1}{l|}{\textbf{Top-1}} & \multicolumn{1}{l|}{83.4}           & \multicolumn{1}{l|}{Top-2}          & \multicolumn{1}{l|}{32.0}                 & \multicolumn{1}{l|}{\cellcolor[HTML]{ECF4FF}99.4}          & \multicolumn{1}{l|}{\cellcolor[HTML]{ECF4FF}Top-5}          & \multicolumn{1}{l|}{\cellcolor[HTML]{ECF4FF}99.5}          & \multicolumn{1}{l|}{\cellcolor[HTML]{ECF4FF}Top-9}           & \multicolumn{1}{l|}{\cellcolor[HTML]{ECF4FF}\textbf{19.9}}  & \multicolumn{1}{l|}{\cellcolor[HTML]{ECF4FF}\textbf{Top-12}} & \multicolumn{1}{l|}{\cellcolor[HTML]{ECF4FF}\textbf{19.9}}  & \cellcolor[HTML]{ECF4FF}\textbf{Top-12} \\ \hline
\multicolumn{1}{|l|}{DT}                                                                                  & \multicolumn{1}{l|}{80.3}           & \multicolumn{1}{l|}{Top-2}          & \multicolumn{1}{l|}{\textbf{82.8}}  & \multicolumn{1}{l|}{\textbf{Top-2}} & \multicolumn{1}{l|}{25.0}                 & \multicolumn{1}{l|}{\cellcolor[HTML]{ECF4FF}100}           & \multicolumn{1}{l|}{\cellcolor[HTML]{ECF4FF}Top-4}          & \multicolumn{1}{l|}{\cellcolor[HTML]{ECF4FF}88.7}          & \multicolumn{1}{l|}{\cellcolor[HTML]{ECF4FF}Top-3}           & \multicolumn{1}{l|}{\cellcolor[HTML]{ECF4FF}\textbf{35.9}}  & \multicolumn{1}{l|}{\cellcolor[HTML]{ECF4FF}\textbf{Top-12}} & \multicolumn{1}{l|}{\cellcolor[HTML]{ECF4FF}\textbf{33.2}}  & \cellcolor[HTML]{ECF4FF}\textbf{Top-12} \\ \hline
\multicolumn{1}{|l|}{RF}                                                                                  & \multicolumn{1}{l|}{70.9}           & \multicolumn{1}{l|}{Top-2}          & \multicolumn{1}{l|}{\textbf{71.6}}  & \multicolumn{1}{l|}{\textbf{Top-3}} & \multicolumn{1}{l|}{20.7}                 & \multicolumn{1}{l|}{\cellcolor[HTML]{ECF4FF}99.6}          & \multicolumn{1}{l|}{\cellcolor[HTML]{ECF4FF}Top-5}          & \multicolumn{1}{l|}{\cellcolor[HTML]{ECF4FF}91.6}          & \multicolumn{1}{l|}{\cellcolor[HTML]{ECF4FF}Top-4}           & \multicolumn{1}{l|}{\cellcolor[HTML]{ECF4FF}\textbf{45.5}}  & \multicolumn{1}{l|}{\cellcolor[HTML]{ECF4FF}\textbf{Top-12}} & \multicolumn{1}{l|}{\cellcolor[HTML]{ECF4FF}\textbf{45.3}}  & \cellcolor[HTML]{ECF4FF}\textbf{Top-12} \\ \hline
\multicolumn{1}{|l|}{MLP}                                                                                 & \multicolumn{1}{l|}{\textbf{67.2}}  & \multicolumn{1}{l|}{\textbf{Top-1}} & \multicolumn{1}{l|}{63.9}           & \multicolumn{1}{l|}{Top-3}          & \multicolumn{1}{l|}{36.6}                 & \multicolumn{1}{l|}{\cellcolor[HTML]{ECF4FF}93.5}          & \multicolumn{1}{l|}{\cellcolor[HTML]{ECF4FF}Top-6}          & \multicolumn{1}{l|}{\cellcolor[HTML]{ECF4FF}91.1}          & \multicolumn{1}{l|}{\cellcolor[HTML]{ECF4FF}Top-12}          & \multicolumn{1}{l|}{\cellcolor[HTML]{ECF4FF}\textbf{31.9}}  & \multicolumn{1}{l|}{\cellcolor[HTML]{ECF4FF}\textbf{Top-12}} & \multicolumn{1}{l|}{\cellcolor[HTML]{ECF4FF}\textbf{30.8}}  & \cellcolor[HTML]{ECF4FF}\textbf{Top-12} \\ \hline
\multicolumn{1}{|l|}{ADA}                                                                                 & \multicolumn{1}{l|}{\textbf{77.3}}  & \multicolumn{1}{l|}{\textbf{Top-1}} & \multicolumn{1}{l|}{64.4}           & \multicolumn{1}{l|}{Top-1}          & \multicolumn{1}{l|}{21.7}                 & \multicolumn{1}{l|}{\cellcolor[HTML]{ECF4FF}\textbf{5.6}}  & \multicolumn{1}{l|}{\cellcolor[HTML]{ECF4FF}\textbf{Top-2}} & \multicolumn{1}{l|}{\cellcolor[HTML]{ECF4FF}\textbf{66.8}} & \multicolumn{1}{l|}{\cellcolor[HTML]{ECF4FF}\textbf{Top-3}}  & \multicolumn{1}{l|}{\cellcolor[HTML]{ECF4FF}\textbf{19.9}}  & \multicolumn{1}{l|}{\cellcolor[HTML]{ECF4FF}\textbf{Top-12}} & \multicolumn{1}{l|}{\cellcolor[HTML]{ECF4FF}\textbf{19.9}}  & \cellcolor[HTML]{ECF4FF}\textbf{Top-12} \\ \hline
\multicolumn{1}{|l|}{BAG}                                                                                 & \multicolumn{1}{l|}{66.7}           & \multicolumn{1}{l|}{Top-2}          & \multicolumn{1}{l|}{\textbf{72.3}}  & \multicolumn{1}{l|}{\textbf{Top-2}} & \multicolumn{1}{l|}{11.5}                 & \multicolumn{1}{l|}{\cellcolor[HTML]{ECF4FF}100}           & \multicolumn{1}{l|}{\cellcolor[HTML]{ECF4FF}Top-4}          & \multicolumn{1}{l|}{\cellcolor[HTML]{ECF4FF}87.2}          & \multicolumn{1}{l|}{\cellcolor[HTML]{ECF4FF}Top-7}           & \multicolumn{1}{l|}{\cellcolor[HTML]{ECF4FF}\textbf{29.4}}  & \multicolumn{1}{l|}{\cellcolor[HTML]{ECF4FF}\textbf{Top-12}} & \multicolumn{1}{l|}{\cellcolor[HTML]{ECF4FF}\textbf{29.4}}  & \cellcolor[HTML]{ECF4FF}\textbf{Top-12} \\ \hline
\multicolumn{1}{|l|}{GBC}                                                                                 & \multicolumn{1}{l|}{57.3}           & \multicolumn{1}{l|}{Top-2}          & \multicolumn{1}{l|}{\textbf{63.7}}  & \multicolumn{1}{l|}{\textbf{Top-2}} & \multicolumn{1}{l|}{30.7}                 & \multicolumn{1}{l|}{\cellcolor[HTML]{ECF4FF}100}           & \multicolumn{1}{l|}{\cellcolor[HTML]{ECF4FF}Top-5}          & \multicolumn{1}{l|}{\cellcolor[HTML]{ECF4FF}88.2}          & \multicolumn{1}{l|}{\cellcolor[HTML]{ECF4FF}Top-4}           & \multicolumn{1}{l|}{\cellcolor[HTML]{ECF4FF}88.2}           & \multicolumn{1}{l|}{\cellcolor[HTML]{ECF4FF}Top-12}          & \multicolumn{1}{l|}{\cellcolor[HTML]{ECF4FF}90.1}           & \cellcolor[HTML]{ECF4FF}Top-12          \\ \hline
\multicolumn{14}{|c|}{\textbf{CLCDSA}}                                                                                                                                                                                                                                                                                                                                                                                                                                                                                                                                                                                                                                                                                                                                                                    \\ \hline
\multicolumn{1}{|l|}{LR}                                                                                  & \multicolumn{1}{l|}{34.5}           & \multicolumn{1}{l|}{Top-1}          & \multicolumn{1}{l|}{35.5}           & \multicolumn{1}{l|}{Top-1}          & \multicolumn{1}{l|}{\textbf{55.2}}        & \multicolumn{1}{l|}{\cellcolor[HTML]{ECF4FF}99.0}          & \multicolumn{1}{l|}{\cellcolor[HTML]{ECF4FF}Top-6}          & \multicolumn{1}{l|}{\cellcolor[HTML]{ECF4FF}\textbf{9.7}}  & \multicolumn{1}{l|}{\cellcolor[HTML]{ECF4FF}\textbf{Top-13}} & \multicolumn{1}{l|}{\cellcolor[HTML]{ECF4FF}100}            & \multicolumn{1}{l|}{\cellcolor[HTML]{ECF4FF}Top-18}          & \multicolumn{1}{l|}{\cellcolor[HTML]{ECF4FF}100}            & \cellcolor[HTML]{ECF4FF}Top-18          \\ \hline
\multicolumn{1}{|l|}{DT}                                                                                  & \multicolumn{1}{l|}{\textbf{53.9}}  & \multicolumn{1}{l|}{\textbf{Top-2}} & \multicolumn{1}{l|}{40.9}           & \multicolumn{1}{l|}{Top-2}          & \multicolumn{1}{l|}{13.4}                 & \multicolumn{1}{l|}{\cellcolor[HTML]{ECF4FF}95.5}          & \multicolumn{1}{l|}{\cellcolor[HTML]{ECF4FF}Top-5}          & \multicolumn{1}{l|}{\cellcolor[HTML]{C0C0C0}\textbf{0.0}}  & \multicolumn{1}{l|}{\cellcolor[HTML]{C0C0C0}\textbf{Top-0}}  & \multicolumn{1}{l|}{\cellcolor[HTML]{ECF4FF}74.8}           & \multicolumn{1}{l|}{\cellcolor[HTML]{ECF4FF}Top-18}          & \multicolumn{1}{l|}{\cellcolor[HTML]{ECF4FF}74.5}           & \cellcolor[HTML]{ECF4FF}Top-18          \\ \hline
\multicolumn{1}{|l|}{RF}                                                                                  & \multicolumn{1}{l|}{\textbf{46.1}}  & \multicolumn{1}{l|}{\textbf{Top-2}} & \multicolumn{1}{l|}{36.5}           & \multicolumn{1}{l|}{Top-2}          & \multicolumn{1}{l|}{15.3}                 & \multicolumn{1}{l|}{\cellcolor[HTML]{ECF4FF}96.2}          & \multicolumn{1}{l|}{\cellcolor[HTML]{ECF4FF}Top-8}          & \multicolumn{1}{l|}{\cellcolor[HTML]{C0C0C0}\textbf{0.0}}  & \multicolumn{1}{l|}{\cellcolor[HTML]{C0C0C0}\textbf{Top-0}}  & \multicolumn{1}{l|}{\cellcolor[HTML]{ECF4FF}64.1}           & \multicolumn{1}{l|}{\cellcolor[HTML]{ECF4FF}Top-18}          & \multicolumn{1}{l|}{\cellcolor[HTML]{ECF4FF}63.5}           & \cellcolor[HTML]{ECF4FF}Top-18          \\ \hline
\multicolumn{1}{|l|}{MLP}                                                                                 & \multicolumn{1}{l|}{\textbf{61.4}}  & \multicolumn{1}{l|}{\textbf{Top-2}} & \multicolumn{1}{l|}{49.8}           & \multicolumn{1}{l|}{Top-3}          & \multicolumn{1}{l|}{36.8}                 & \multicolumn{1}{l|}{\cellcolor[HTML]{ECF4FF}96.3}          & \multicolumn{1}{l|}{\cellcolor[HTML]{ECF4FF}Top-6}          & \multicolumn{1}{l|}{\cellcolor[HTML]{ECF4FF}15.3}          & \multicolumn{1}{l|}{\cellcolor[HTML]{ECF4FF}Top-13}          & \multicolumn{1}{l|}{\cellcolor[HTML]{ECF4FF}100}            & \multicolumn{1}{l|}{\cellcolor[HTML]{ECF4FF}Top-18}          & \multicolumn{1}{l|}{\cellcolor[HTML]{ECF4FF}100}            & \cellcolor[HTML]{ECF4FF}Top-18          \\ \hline
\multicolumn{1}{|l|}{ADA}                                                                                 & \multicolumn{1}{l|}{\textbf{51.1}}  & \multicolumn{1}{l|}{\textbf{Top-2}} & \multicolumn{1}{l|}{42.1}           & \multicolumn{1}{l|}{Top-2}          & \multicolumn{1}{l|}{17.3}                 & \multicolumn{1}{l|}{\cellcolor[HTML]{C0C0C0}\textbf{0.0}}  & \multicolumn{1}{l|}{\cellcolor[HTML]{C0C0C0}\textbf{Top-0}} & \multicolumn{1}{l|}{\cellcolor[HTML]{C0C0C0}\textbf{0.0}}  & \multicolumn{1}{l|}{\cellcolor[HTML]{C0C0C0}\textbf{Top-0}}  & \multicolumn{1}{l|}{\cellcolor[HTML]{ECF4FF}96.2}           & \multicolumn{1}{l|}{\cellcolor[HTML]{ECF4FF}Top-18}          & \multicolumn{1}{l|}{\cellcolor[HTML]{ECF4FF}96.1}           & \cellcolor[HTML]{ECF4FF}Top-18          \\ \hline
\multicolumn{1}{|l|}{BAG}                                                                                 & \multicolumn{1}{l|}{28.1}           & \multicolumn{1}{l|}{Top-3}          & \multicolumn{1}{l|}{\textbf{40.2}}  & \multicolumn{1}{l|}{\textbf{Top-2}} & \multicolumn{1}{l|}{16.1}                 & \multicolumn{1}{l|}{\cellcolor[HTML]{ECF4FF}82.9}          & \multicolumn{1}{l|}{\cellcolor[HTML]{ECF4FF}Top-5}          & \multicolumn{1}{l|}{\cellcolor[HTML]{ECF4FF}0.63}          & \multicolumn{1}{l|}{\cellcolor[HTML]{ECF4FF}Top-7}           & \multicolumn{1}{l|}{\cellcolor[HTML]{ECF4FF}84.2}           & \multicolumn{1}{l|}{\cellcolor[HTML]{ECF4FF}Top-18}          & \multicolumn{1}{l|}{\cellcolor[HTML]{ECF4FF}85.2}           & \cellcolor[HTML]{ECF4FF}Top-18          \\ \hline
\multicolumn{1}{|l|}{GBC}                                                                                 & \multicolumn{1}{l|}{\textbf{34.3}}  & \multicolumn{1}{l|}{\textbf{Top-2}} & \multicolumn{1}{l|}{33.3}           & \multicolumn{1}{l|}{Top-2}          & \multicolumn{1}{l|}{14.9}                 & \multicolumn{1}{l|}{\cellcolor[HTML]{ECF4FF}98.5}          & \multicolumn{1}{l|}{\cellcolor[HTML]{ECF4FF}Top-6}          & \multicolumn{1}{l|}{\cellcolor[HTML]{C0C0C0}\textbf{0.0}}  & \multicolumn{1}{l|}{\cellcolor[HTML]{C0C0C0}\textbf{Top-0}}  & \multicolumn{1}{l|}{\cellcolor[HTML]{ECF4FF}88.3}           & \multicolumn{1}{l|}{\cellcolor[HTML]{ECF4FF}Top-18}          & \multicolumn{1}{l|}{\cellcolor[HTML]{ECF4FF}87.8}           & \cellcolor[HTML]{ECF4FF}Top-18          \\ \hline
\multicolumn{14}{|c|}{\textbf{Code Review}}                                                                                                                                                                                                                                                                                                                                                                                                                                                                                                                                                                                                                                                                                                                                                               \\ \hline
\multicolumn{1}{|l|}{LR}                                                                                  & \multicolumn{1}{l|}{\textbf{53.9}}  & \multicolumn{1}{l|}{\textbf{Top-1}} & \multicolumn{1}{l|}{48.9}           & \multicolumn{1}{l|}{Top-1}          & \multicolumn{1}{l|}{41.9}                 & \multicolumn{1}{l|}{\cellcolor[HTML]{ECF4FF}99.3}          & \multicolumn{1}{l|}{\cellcolor[HTML]{ECF4FF}Top-3}          & \multicolumn{1}{l|}{\cellcolor[HTML]{ECF4FF}100}           & \multicolumn{1}{l|}{\cellcolor[HTML]{ECF4FF}Top-10}          & \multicolumn{1}{l|}{\cellcolor[HTML]{ECF4FF}100}            & \multicolumn{1}{l|}{\cellcolor[HTML]{ECF4FF}Top-15}          & \multicolumn{1}{l|}{\cellcolor[HTML]{ECF4FF}100}            & \cellcolor[HTML]{ECF4FF}Top-15          \\ \hline
\multicolumn{1}{|l|}{DT}                                                                                  & \multicolumn{1}{l|}{\textbf{38.0}}  & \multicolumn{1}{l|}{\textbf{Top-2}} & \multicolumn{1}{l|}{35.3}           & \multicolumn{1}{l|}{Top-3}          & \multicolumn{1}{l|}{26.9}                 & \multicolumn{1}{l|}{\cellcolor[HTML]{ECF4FF}99.6}          & \multicolumn{1}{l|}{\cellcolor[HTML]{ECF4FF}Top-3}          & \multicolumn{1}{l|}{\cellcolor[HTML]{ECF4FF}50.0}          & \multicolumn{1}{l|}{\cellcolor[HTML]{ECF4FF}Top-2}           & \multicolumn{1}{l|}{\cellcolor[HTML]{ECF4FF}68.0}           & \multicolumn{1}{l|}{\cellcolor[HTML]{ECF4FF}Top-15}          & \multicolumn{1}{l|}{\cellcolor[HTML]{ECF4FF}66.0}           & \cellcolor[HTML]{ECF4FF}Top-15          \\ \hline
\multicolumn{1}{|l|}{RF}                                                                                  & \multicolumn{1}{l|}{\textbf{39.3}}  & \multicolumn{1}{l|}{\textbf{Top-3}} & \multicolumn{1}{l|}{32.9}           & \multicolumn{1}{l|}{Top-2}          & \multicolumn{1}{l|}{21.1}                 & \multicolumn{1}{l|}{\cellcolor[HTML]{ECF4FF}99.3}          & \multicolumn{1}{l|}{\cellcolor[HTML]{ECF4FF}Top-4}          & \multicolumn{1}{l|}{\cellcolor[HTML]{ECF4FF}86.7}          & \multicolumn{1}{l|}{\cellcolor[HTML]{ECF4FF}Top-4}           & \multicolumn{1}{l|}{\cellcolor[HTML]{ECF4FF}66.5}           & \multicolumn{1}{l|}{\cellcolor[HTML]{ECF4FF}Top-15}          & \multicolumn{1}{l|}{\cellcolor[HTML]{ECF4FF}66.4}           & \cellcolor[HTML]{ECF4FF}Top-15          \\ \hline
\multicolumn{1}{|l|}{MLP}                                                                                 & \multicolumn{1}{l|}{\textbf{71.6}}  & \multicolumn{1}{l|}{\textbf{Top-1}} & \multicolumn{1}{l|}{57.5}           & \multicolumn{1}{l|}{Top-2}          & \multicolumn{1}{l|}{28.3}                 & \multicolumn{1}{l|}{\cellcolor[HTML]{ECF4FF}79.8}          & \multicolumn{1}{l|}{\cellcolor[HTML]{ECF4FF}Top-5}          & \multicolumn{1}{l|}{\cellcolor[HTML]{ECF4FF}89.5}          & \multicolumn{1}{l|}{\cellcolor[HTML]{ECF4FF}Top-13}          & \multicolumn{1}{l|}{\cellcolor[HTML]{ECF4FF}100}            & \multicolumn{1}{l|}{\cellcolor[HTML]{ECF4FF}Top-15}          & \multicolumn{1}{l|}{\cellcolor[HTML]{ECF4FF}100}            & \cellcolor[HTML]{ECF4FF}Top-15          \\ \hline
\multicolumn{1}{|l|}{ADA}                                                                                 & \multicolumn{1}{l|}{\textbf{62.9}}  & \multicolumn{1}{l|}{\textbf{Top-1}} & \multicolumn{1}{l|}{47.9}           & \multicolumn{1}{l|}{Top-2}          & \multicolumn{1}{l|}{38.1}                 & \multicolumn{1}{l|}{\cellcolor[HTML]{ECF4FF}\textbf{40.0}} & \multicolumn{1}{l|}{\cellcolor[HTML]{ECF4FF}\textbf{Top-3}} & \multicolumn{1}{l|}{\cellcolor[HTML]{ECF4FF}66.4}          & \multicolumn{1}{l|}{\cellcolor[HTML]{ECF4FF}Top-3}           & \multicolumn{1}{l|}{\cellcolor[HTML]{ECF4FF}\textbf{57.8}}  & \multicolumn{1}{l|}{\cellcolor[HTML]{ECF4FF}\textbf{Top-15}} & \multicolumn{1}{l|}{\cellcolor[HTML]{ECF4FF}\textbf{52.1}}  & \cellcolor[HTML]{ECF4FF}\textbf{Top-15} \\ \hline
\multicolumn{1}{|l|}{BAG}                                                                                 & \multicolumn{1}{l|}{\textbf{27.6}}  & \multicolumn{1}{l|}{\textbf{Top-2}} & \multicolumn{1}{l|}{23.5}           & \multicolumn{1}{l|}{Top-2}          & \multicolumn{1}{l|}{26.9}                 & \multicolumn{1}{l|}{\cellcolor[HTML]{ECF4FF}79.3}          & \multicolumn{1}{l|}{\cellcolor[HTML]{ECF4FF}Top-4}          & \multicolumn{1}{l|}{\cellcolor[HTML]{ECF4FF}80.7}          & \multicolumn{1}{l|}{\cellcolor[HTML]{ECF4FF}Top-8}           & \multicolumn{1}{l|}{\cellcolor[HTML]{ECF4FF}100}            & \multicolumn{1}{l|}{\cellcolor[HTML]{ECF4FF}Top-15}          & \multicolumn{1}{l|}{\cellcolor[HTML]{ECF4FF}100}            & \cellcolor[HTML]{ECF4FF}Top-15          \\ \hline
\multicolumn{1}{|l|}{GBC}                                                                                 & \multicolumn{1}{l|}{\textbf{44.4}}  & \multicolumn{1}{l|}{\textbf{Top-2}} & \multicolumn{1}{l|}{38.1}           & \multicolumn{1}{l|}{Top-3}          & \multicolumn{1}{l|}{26.9}                 & \multicolumn{1}{l|}{\cellcolor[HTML]{ECF4FF}100}           & \multicolumn{1}{l|}{\cellcolor[HTML]{ECF4FF}Top-4}          & \multicolumn{1}{l|}{\cellcolor[HTML]{ECF4FF}70.4}          & \multicolumn{1}{l|}{\cellcolor[HTML]{ECF4FF}Top-4}           & \multicolumn{1}{l|}{\cellcolor[HTML]{ECF4FF}92.9}           & \multicolumn{1}{l|}{\cellcolor[HTML]{ECF4FF}Top-15}          & \multicolumn{1}{l|}{\cellcolor[HTML]{ECF4FF}89.4}           & \cellcolor[HTML]{ECF4FF}Top-15          \\ \hline
\multicolumn{14}{|c|}{\textbf{BigCloneBench}}                                                                                                                                                                                                                                                                                                                                                                                                                                                                                                                                                                                                                                                                                                                                                             \\ \hline
\multicolumn{1}{|l|}{LR}                                                                                  & \multicolumn{1}{l|}{\textbf{68.0}} & \multicolumn{1}{l|}{\textbf{Top-1}} & \multicolumn{1}{l|}{56.2}          & \multicolumn{1}{l|}{Top-1}          & \multicolumn{1}{l|}{52.4}                & \multicolumn{1}{l|}{\cellcolor[HTML]{ECF4FF}100}           & \multicolumn{1}{l|}{\cellcolor[HTML]{ECF4FF}Top-3}          & \multicolumn{1}{l|}{\cellcolor[HTML]{ECF4FF}100}           & \multicolumn{1}{l|}{\cellcolor[HTML]{ECF4FF}Top-5}           & \multicolumn{1}{l|}{\cellcolor[HTML]{ECF4FF}99.9}          & \multicolumn{1}{l|}{\cellcolor[HTML]{ECF4FF}Top-6}           & \multicolumn{1}{l|}{\cellcolor[HTML]{ECF4FF}99.9}          & \cellcolor[HTML]{ECF4FF}Top-6           \\ \hline
\multicolumn{1}{|l|}{DT}                                                                                  & \multicolumn{1}{l|}{50.9}           & \multicolumn{1}{l|}{Top-1}          & \multicolumn{1}{l|}{\textbf{63.5}} & \multicolumn{1}{l|}{\textbf{Top-3}} & \multicolumn{1}{l|}{17.8}                & \multicolumn{1}{l|}{\cellcolor[HTML]{ECF4FF}100}           & \multicolumn{1}{l|}{\cellcolor[HTML]{ECF4FF}Top-3}          & \multicolumn{1}{l|}{\cellcolor[HTML]{ECF4FF}87.8}          & \multicolumn{1}{l|}{\cellcolor[HTML]{ECF4FF}Top-4}           & \multicolumn{1}{l|}{\cellcolor[HTML]{ECF4FF}\textbf{51.2}}  & \multicolumn{1}{l|}{\cellcolor[HTML]{ECF4FF}\textbf{Top-6}}  & \multicolumn{1}{l|}{\cellcolor[HTML]{ECF4FF}\textbf{51.2}}  & \cellcolor[HTML]{ECF4FF}\textbf{Top-6}  \\ \hline
\multicolumn{1}{|l|}{RF}                                                                                  & \multicolumn{1}{l|}{53.8}           & \multicolumn{1}{l|}{Top-2}          & \multicolumn{1}{l|}{\textbf{64.9}} & \multicolumn{1}{l|}{\textbf{Top-2}} & \multicolumn{1}{l|}{20.5}                & \multicolumn{1}{l|}{\cellcolor[HTML]{ECF4FF}99.9}         & \multicolumn{1}{l|}{\cellcolor[HTML]{ECF4FF}Top-5}          & \multicolumn{1}{l|}{\cellcolor[HTML]{ECF4FF}98.6}          & \multicolumn{1}{l|}{\cellcolor[HTML]{ECF4FF}Top-5}           & \multicolumn{1}{l|}{\cellcolor[HTML]{ECF4FF}\textbf{52.6}} & \multicolumn{1}{l|}{\cellcolor[HTML]{ECF4FF}\textbf{Top-6}}  & \multicolumn{1}{l|}{\cellcolor[HTML]{ECF4FF}\textbf{52.6}} & \cellcolor[HTML]{ECF4FF}\textbf{Top-6}  \\ \hline
\multicolumn{1}{|l|}{MLP}                                                                                 & \multicolumn{1}{l|}{\textbf{69.5}} & \multicolumn{1}{l|}{\textbf{Top-1}} & \multicolumn{1}{l|}{68.6}          & \multicolumn{1}{l|}{Top-2}          & \multicolumn{1}{l|}{44.4}                & \multicolumn{1}{l|}{\cellcolor[HTML]{ECF4FF}100}           & \multicolumn{1}{l|}{\cellcolor[HTML]{ECF4FF}Top-4}          & \multicolumn{1}{l|}{\cellcolor[HTML]{ECF4FF}100}           & \multicolumn{1}{l|}{\cellcolor[HTML]{ECF4FF}Top-6}           & \multicolumn{1}{l|}{\cellcolor[HTML]{ECF4FF}99.4}          & \multicolumn{1}{l|}{\cellcolor[HTML]{ECF4FF}Top-6}           & \multicolumn{1}{l|}{\cellcolor[HTML]{ECF4FF}99.0}          & \cellcolor[HTML]{ECF4FF}Top-6           \\ \hline
\multicolumn{1}{|l|}{ADA}                                                                                 & \multicolumn{1}{l|}{\textbf{69.0}}  & \multicolumn{1}{l|}{\textbf{Top-1}} & \multicolumn{1}{l|}{68.9}          & \multicolumn{1}{l|}{Top-2}          & \multicolumn{1}{l|}{37.3}                & \multicolumn{1}{l|}{\cellcolor[HTML]{ECF4FF}92.6}          & \multicolumn{1}{l|}{\cellcolor[HTML]{ECF4FF}Top-3}          & \multicolumn{1}{l|}{\cellcolor[HTML]{ECF4FF}99.6}          & \multicolumn{1}{l|}{\cellcolor[HTML]{ECF4FF}Top-5}           & \multicolumn{1}{l|}{\cellcolor[HTML]{ECF4FF}\textbf{53.3}} & \multicolumn{1}{l|}{\cellcolor[HTML]{ECF4FF}\textbf{Top-6}}  & \multicolumn{1}{l|}{\cellcolor[HTML]{ECF4FF}\textbf{53.3}} & \cellcolor[HTML]{ECF4FF}\textbf{Top-6}  \\ \hline
\multicolumn{1}{|l|}{BAG}                                                                                 & \multicolumn{1}{l|}{\textbf{64.7}} & \multicolumn{1}{l|}{\textbf{Top-1}} & \multicolumn{1}{l|}{63.0}          & \multicolumn{1}{l|}{Top-2}          & \multicolumn{1}{l|}{17.8}                & \multicolumn{1}{l|}{\cellcolor[HTML]{ECF4FF}100}           & \multicolumn{1}{l|}{\cellcolor[HTML]{ECF4FF}Top-5}          & \multicolumn{1}{l|}{\cellcolor[HTML]{ECF4FF}97.2}         & \multicolumn{1}{l|}{\cellcolor[HTML]{ECF4FF}Top-5}           & \multicolumn{1}{l|}{\cellcolor[HTML]{ECF4FF}\textbf{53.0}} & \multicolumn{1}{l|}{\cellcolor[HTML]{ECF4FF}\textbf{Top-6}}  & \multicolumn{1}{l|}{\cellcolor[HTML]{ECF4FF}\textbf{53.0}} & \cellcolor[HTML]{ECF4FF}\textbf{Top-6}  \\ \hline
\multicolumn{1}{|l|}{GBC}                                                                                 & \multicolumn{1}{l|}{\textbf{67.6}} & \multicolumn{1}{l|}{\textbf{Top-1}} & \multicolumn{1}{l|}{67.8}          & \multicolumn{1}{l|}{Top-2}          & \multicolumn{1}{l|}{18.7}                & \multicolumn{1}{l|}{\cellcolor[HTML]{ECF4FF}99.8}         & \multicolumn{1}{l|}{\cellcolor[HTML]{ECF4FF}Top-5}          & \multicolumn{1}{l|}{\cellcolor[HTML]{ECF4FF}95.5}         & \multicolumn{1}{l|}{\cellcolor[HTML]{ECF4FF}Top-5}           & \multicolumn{1}{l|}{\cellcolor[HTML]{ECF4FF}\textbf{52.2}} & \multicolumn{1}{l|}{\cellcolor[HTML]{ECF4FF}\textbf{Top-6}}  & \multicolumn{1}{l|}{\cellcolor[HTML]{ECF4FF}\textbf{52.2}} & \cellcolor[HTML]{ECF4FF}\textbf{Top-6}  \\ \hline
\end{tabular}
\end{adjustbox}
\label{attack_result}
\textbf{Note:} Bolded cells indicate cases where the attack techniques failed to produce any successful adversarial examples
\end{table*}

\textbf{Conducting Adversarial Attacks to assess the Robustness of ML Models:}
To address this research question, we hypothesize that altering the values of the top-$k$ important features identified through ML explainability techniques should result in a change in the predictions made by the ML models. If there is concrete experimental evidence supporting this hypothesis, it suggests that ML explainability techniques effectively generate adversarial examples capable of assessing the robustness of ML models.

Figures \ref{pyexp_a} and \ref{pyexp_b} illustrate how changing the important features identified through ML explainability affects the prediction probability of the ML model. The research question (RQ1) findings further demonstrate a significant difference in prediction probabilities between the original and transformed instances when changing the top-$k$ important feature values. Thus, a follow-up question arises: \textit{Is the prediction probability difference sufficient to reverse the decision of the ML model?} We promptly find the answer to this question in Figures \ref{pyexp_a} and \ref{pyexp_b}. These two figures collectively reveal that altering the top-$k$ important features identified by ML explainability flips the prediction of the ML model (changing the colour from \textit{green} to \textit{orange}). Now, we investigate whether this scenario applies to all the test datasets and ML models.

Table \ref{attack_result} presents the ASR metric and top-$k$ values both for the explanation-guided adversarial attacks and the selected state-of-the-art adversarial attacks on tabular data. The bolded cells highlighted in grey represent cases where the attack techniques failed to produce any successful adversarial examples. The ASR metric quantifies the degree to which the models' accuracy is compromised when specific features are altered to generate adversarial examples. For instance, in Table \ref{attack_result}, it is evident that the BAG model trained on the cross-project mobile apps dataset fails to accurately predict $76.4\%$ of instances that it correctly predicted before adversarial attacks when considering SHAP. Similarly, the LR model fails to predict $75.6\%$ and $56.3\%$ instances accurately after adversarial attacks when considering LIME and PyExplainer, respectively. The ASR metric value for the cross-project mobile apps dataset ranges between $23.3\%$ and $76.4\%$. Similarly, the range is 14.8\%--53.6\% for the Java project dataset, 11.5\%--86.6\% for the Postgres dataset, 13.4\%--61.4\% for the CLCDSA dataset, 17.87\%--69.0\% for the BCB dataset, and 26.9\%--71.6\% for the code review dataset, respectively. In summary, the ASR metric value ranges between 11.5\%--86.6\% for all the ML models, considering all the datasets.

From Table \ref{attack_result}, it is evident that ML models fail to accurately predict a large number of instances that they correctly predicted before adversarial attacks, even when altering just up to \textit{three} feature values. The only exception is the LR model trained on the Java project dataset, where we modified \textit{six} feature values to generate adversarial examples. In most cases, we only needed to alter the values of the top-$1$ or top-$2$ important features. Interestingly, the LR model trained on the Postgres dataset exhibited the highest ASR metric value (e.g., $86.6\%$), even when changing only \textit{one} feature. Therefore, in terms of imperceptibility defined in Section \ref{Adv Attack}, our proposed explanation-guided adversarial attack shows promising results. These experimental results underscore the effectiveness of the important features identified by ML explainability techniques in generating adversarial examples. Furthermore, our findings emphasize that ML models struggled to accurately predict up to $86.6\%$ of correctly predicted instances after undergoing adversarial attacks.


Following existing studies, we apply sampling techniques (e.g., SMOTE) only to the training data, leaving the test data unaffected. As a result, the test datasets remain highly imbalanced, which impacts the F1-scores of some models. For instance, after the train-test split, the Postgres test dataset includes 1,524 samples of the negative class (e.g., \textit{clean} commits) and 520 samples of the positive class (e.g., \textit{buggy} commits) when testing the LR model. Due to the high imbalance in the test dataset, we perform additional experiments using the undersampling technique to balance the test dataset. Table \ref{dataset_balanced} shows that after balancing the test dataset, we observe improved F1-scores along with nearly similar accuracy and AUC values.

\begin{table}[htbp]
\centering
\caption{Performance of the ML models on different datasets without adversarial attacks considering different evaluation metrics such as Accuracy (Acc), F1-score (F1), and AUC values before balancing (BB) and after balancing (AB) the test data}
\begin{tabular}{l|llllllllllll}
\hline
                                                                     & \multicolumn{12}{c}{\textbf{Dataset}}                                                                                                                                                                                                                                                                                                                                                                                                                                                                                                                                                                                                                                                                                                                                                                                                                                                                                                                                                                                                                                                                                                     \\ \cmidrule{2-13} 
                                                                     & \multicolumn{6}{c|}{\textbf{Java Project}}                                                                                                                                                                                                                                                                                                                                                                                                                                                                                                                                     & \multicolumn{6}{c}{\textbf{Postgres}}                                                                                                                                                                                                                                                                                                                                                                                                                                                                                                                    \\ \cmidrule{2-13} 
                                                                     & \multicolumn{2}{c|}{\textbf{Acc}}                                                                                                                                                        & \multicolumn{2}{c|}{\textbf{F-1}}                                                                                                                                                        & \multicolumn{2}{c|}{\textbf{AUC}}                                                                                                                                                        & \multicolumn{2}{c|}{\textbf{Acc}}                                                                                                                                                        & \multicolumn{2}{c|}{\textbf{F-1}}                                                                                                                                                        & \multicolumn{2}{c}{\textbf{AUC}}                                                                                                                                   \\ \cmidrule{2-13} 
\multirow{-4}{*}{\begin{tabular}[c]{@{}l@{}}ML\\ Model\end{tabular}} & \multicolumn{1}{l|}{\begin{tabular}[c]{@{}l@{}}BB\end{tabular}} & \multicolumn{1}{l|}{\cellcolor[HTML]{DAE8FC}\begin{tabular}[c]{@{}l@{}}AB\end{tabular}} & \multicolumn{1}{l|}{\begin{tabular}[c]{@{}l@{}}BB\end{tabular}} & \multicolumn{1}{l|}{\cellcolor[HTML]{DAE8FC}\begin{tabular}[c]{@{}l@{}}AB\end{tabular}} & \multicolumn{1}{l|}{\begin{tabular}[c]{@{}l@{}}BB\end{tabular}} & \multicolumn{1}{l|}{\cellcolor[HTML]{DAE8FC}\begin{tabular}[c]{@{}l@{}}AB\end{tabular}} & \multicolumn{1}{l|}{\begin{tabular}[c]{@{}l@{}}BB\end{tabular}} & \multicolumn{1}{l|}{\cellcolor[HTML]{DAE8FC}\begin{tabular}[c]{@{}l@{}}AB\end{tabular}} & \multicolumn{1}{l|}{\begin{tabular}[c]{@{}l@{}}BB\end{tabular}} & \multicolumn{1}{l|}{\cellcolor[HTML]{DAE8FC}\begin{tabular}[c]{@{}l@{}}AB\end{tabular}} & \multicolumn{1}{l|}{\begin{tabular}[c]{@{}l@{}}BB\end{tabular}} & \cellcolor[HTML]{DAE8FC}\begin{tabular}[c]{@{}l@{}}AB\end{tabular} \\ \hline
LR                                                                   & \multicolumn{1}{l|}{0.75}                                                       & \multicolumn{1}{l|}{\cellcolor[HTML]{DAE8FC}0.69}                                                      & \multicolumn{1}{l|}{0.37}                                                       & \multicolumn{1}{l|}{\cellcolor[HTML]{DAE8FC}0.67}                                                      & \multicolumn{1}{l|}{0.76}                                                       & \multicolumn{1}{l|}{\cellcolor[HTML]{DAE8FC}0.76}                                                      & \multicolumn{1}{l|}{0.75}                                                       & \multicolumn{1}{l|}{\cellcolor[HTML]{DAE8FC}0.72}                                                      & \multicolumn{1}{l|}{0.57}                                                       & \multicolumn{1}{l|}{\cellcolor[HTML]{DAE8FC}0.68}                                                      & \multicolumn{1}{l|}{0.76}                                                       & \cellcolor[HTML]{DAE8FC}0.77                                                      \\ \hline
DT                                                                   & \multicolumn{1}{l|}{0.80}                                                       & \multicolumn{1}{l|}{\cellcolor[HTML]{DAE8FC}0.74}                                                      & \multicolumn{1}{l|}{0.49}                                                       & \multicolumn{1}{l|}{\cellcolor[HTML]{DAE8FC}0.72}                                                      & \multicolumn{1}{l|}{0.80}                                                       & \multicolumn{1}{l|}{\cellcolor[HTML]{DAE8FC}0.79}                                                      & \multicolumn{1}{l|}{0.78}                                                       & \multicolumn{1}{l|}{\cellcolor[HTML]{DAE8FC}0.72}                                                      & \multicolumn{1}{l|}{0.59}                                                       & \multicolumn{1}{l|}{\cellcolor[HTML]{DAE8FC}0.67}                                                      & \multicolumn{1}{l|}{0.79}                                                       & \cellcolor[HTML]{DAE8FC}0.79                                                      \\ \hline
RF                                                                   & \multicolumn{1}{l|}{0.88}                                                       & \multicolumn{1}{l|}{\cellcolor[HTML]{DAE8FC}0.76}                                                      & \multicolumn{1}{l|}{0.55}                                                       & \multicolumn{1}{l|}{\cellcolor[HTML]{DAE8FC}0.70}                                                      & \multicolumn{1}{l|}{0.85}                                                       & \multicolumn{1}{l|}{\cellcolor[HTML]{DAE8FC}0.87}                                                      & \multicolumn{1}{l|}{0.80}                                                       & \multicolumn{1}{l|}{\cellcolor[HTML]{DAE8FC}0.73}                                                      & \multicolumn{1}{l|}{0.59}                                                       & \multicolumn{1}{l|}{\cellcolor[HTML]{DAE8FC}0.67}                                                      & \multicolumn{1}{l|}{0.82}                                                       & \cellcolor[HTML]{DAE8FC}0.82                                                      \\ \hline
MLP                                                                  & \multicolumn{1}{l|}{0.80}                                                       & \multicolumn{1}{l|}{\cellcolor[HTML]{DAE8FC}0.73}                                                      & \multicolumn{1}{l|}{0.48}                                                       & \multicolumn{1}{l|}{\cellcolor[HTML]{DAE8FC}0.70}                                                      & \multicolumn{1}{l|}{0.81}                                                       & \multicolumn{1}{l|}{\cellcolor[HTML]{DAE8FC}0.80}                                                      & \multicolumn{1}{l|}{0.71}                                                       & \multicolumn{1}{l|}{\cellcolor[HTML]{DAE8FC}0.66}                                                      & \multicolumn{1}{l|}{0.51}                                                       & \multicolumn{1}{l|}{\cellcolor[HTML]{DAE8FC}0.63}                                                      & \multicolumn{1}{l|}{0.73}                                                       & \cellcolor[HTML]{DAE8FC}0.74                                                      \\ \hline
ADA                                                                  & \multicolumn{1}{l|}{0.79}                                                       & \multicolumn{1}{l|}{\cellcolor[HTML]{DAE8FC}0.76}                                                      & \multicolumn{1}{l|}{0.47}                                                       & \multicolumn{1}{l|}{\cellcolor[HTML]{DAE8FC}0.75}                                                      & \multicolumn{1}{l|}{0.81}                                                       & \multicolumn{1}{l|}{\cellcolor[HTML]{DAE8FC}0.81}                                                      & \multicolumn{1}{l|}{0.78}                                                       & \multicolumn{1}{l|}{\cellcolor[HTML]{DAE8FC}0.72}                                                      & \multicolumn{1}{l|}{0.58}                                                       & \multicolumn{1}{l|}{\cellcolor[HTML]{DAE8FC}0.68}                                                      & \multicolumn{1}{l|}{0.79}                                                       & \cellcolor[HTML]{DAE8FC}0.79                                                      \\ \hline
BAG                                                                  & \multicolumn{1}{l|}{0.78}                                                       & \multicolumn{1}{l|}{\cellcolor[HTML]{DAE8FC}0.75}                                                      & \multicolumn{1}{l|}{0.50}                                                       & \multicolumn{1}{l|}{\cellcolor[HTML]{DAE8FC}0.75}                                                      & \multicolumn{1}{l|}{0.83}                                                       & \multicolumn{1}{l|}{\cellcolor[HTML]{DAE8FC}0.82}                                                      & \multicolumn{1}{l|}{0.78}                                                       & \multicolumn{1}{l|}{\cellcolor[HTML]{DAE8FC}0.75}                                                      & \multicolumn{1}{l|}{0.61}                                                       & \multicolumn{1}{l|}{\cellcolor[HTML]{DAE8FC}0.73}                                                      & \multicolumn{1}{l|}{0.82}                                                       & \cellcolor[HTML]{DAE8FC}0.82                                                      \\ \hline
GBC                                                                  & \multicolumn{1}{l|}{0.81}                                                       & \multicolumn{1}{l|}{\cellcolor[HTML]{DAE8FC}0.72}                                                      & \multicolumn{1}{l|}{0.48}                                                       & \multicolumn{1}{l|}{\cellcolor[HTML]{DAE8FC}0.68}                                                      & \multicolumn{1}{l|}{0.80}                                                       & \multicolumn{1}{l|}{\cellcolor[HTML]{DAE8FC}0.79}                                                      & \multicolumn{1}{l|}{0.78}                                                       & \multicolumn{1}{l|}{\cellcolor[HTML]{DAE8FC}0.72}                                                      & \multicolumn{1}{l|}{0.57}                                                       & \multicolumn{1}{l|}{\cellcolor[HTML]{DAE8FC}0.67}                                                      & \multicolumn{1}{l|}{0.80}                                                       & \cellcolor[HTML]{DAE8FC}0.81                                                      \\ \hline
\end{tabular}
\label{dataset_balanced}
\end{table}

Table \ref{attack_result_balanced} presents the ASR metric and top-$k$ values for both the explanation-guided and selected state-of-the-art adversarial attacks using the balanced test data. From this table, it is evident that by changing only up to three feature values, the accuracy of the ML models drops by up to $85.89\%$. Interestingly, when changing only one feature value, the LR model fails to predict $85.89\%$ of instances after undergoing adversarial attacks. Therefore, we still get a similar conclusion after balancing the test dataset.

\begin{center}
\begin{tcolorbox}[
    enhanced,
    attach boxed title to top left={yshift=-3mm,yshifttext=-1mm}, 
    colback=mycolor_box,                 
    colframe=black,                
    colbacktitle= mycolor_title,            
    coltitle=black,                
    title=Result RQ2-a,            
    fonttitle=\bfseries,           
    boxed title style={size=small},
    width=0.9\textwidth            
]
    When altering the top-$k$ feature values to generate adversarial examples and testing ML models on such examples, the accuracy of the ML models under attack can be compromised by up to \textbf{86.6\%}.
\end{tcolorbox}
\label{adv_result}
\end{center}

\textbf{Comparison with the baselines:}
We compare our approach with four state-of-the-art techniques: \textit{Zoo} \cite{chen2017zoo}, \textit{Boundary attack} \cite{brendel2017decision}, \textit{PermuteAttack} \cite{hashemi2020permuteattack}, and \textit{HopSkipJump} \cite{chen2020hopskipjumpattack}, in terms of ASR metric values and imperceptibility (e.g., minimal $\ell_0$ perturbation) across various ML models. It is important to note that we consider the best result (highlighted in bold text) from the three explanation-guided attacks for comparison with the baselines. Table \ref{attack_result} demonstrates that our technique outperforms \textit{HopSkipJump} and \textit{Boundary} attacks significantly in terms of imperceptibility. In many cases (highlighted in bold text), we observe that our approach outperforms \textit{HopSkipJump} and \textit{Boundary} attacks both in ASR metric values and imperceptibility. Additionally, \textit{HopSkipJump} and \textit{Boundary} attacks modify all features of test instances, thereby contradicting the imperceptibility property of the adversarial examples defined in Section \ref{Adv Attack}. Shifting to the \textit{Zoo} attack, in the majority of cases (highlighted in bold text), our approach surpasses the \textit{Zoo} attack in terms of ASR metric values and imperceptibility. Furthermore, the \textit{Zoo} attack fails to generate a single adversarial example for a few models (highlighted in grey cells). Moreover, regarding imperceptibility, our approach exhibits superior performance over the \textit{Zoo} attack.

\begin{table*}[htbp]
\centering
\caption{Attack Success Rate (ASR) metric values and the Top-$k$ modified features (e.g., a minimal $\ell_0$ perturbation) when we apply explanation-guided adversarial attacks and selected baseline attacks on the ML models for the balanced Java project and Postgres datasets}
\begin{adjustbox}{max width=\textwidth}
\begin{tabular}{|llllllllllllll|}
\hline
\multicolumn{1}{|l|}{}                                                                     & \multicolumn{13}{c|}{\textbf{Java Project}}                                                                                                                                                                                                                                                                                                                                                                                                                                                                                                                                                                                                                                                \\ \cmidrule{2-14} 
\multicolumn{1}{|l|}{}                                                                     & \multicolumn{2}{c|}{\textbf{SHAP}}                                        & \multicolumn{2}{c|}{\textbf{LIME}}                                        & \multicolumn{1}{c|}{\textbf{PyExplainer}} & \multicolumn{2}{c|}{\textbf{PermuteAttack}}                                                                             & \multicolumn{2}{c|}{\textbf{Zoo}}                                                                                       & \multicolumn{2}{c|}{\textbf{Boundary}}                                                                                     & \multicolumn{2}{c|}{\textbf{HopSkipJump}}                                                             \\ \cmidrule{2-14} 
\multicolumn{1}{|l|}{\multirow{-3}{*}{\begin{tabular}[c]{@{}l@{}}ML\\ Model\end{tabular}}} & \multicolumn{1}{l|}{ASR}            & \multicolumn{1}{l|}{Top-$k$}        & \multicolumn{1}{l|}{ASR}            & \multicolumn{1}{l|}{Top-$k$}        & \multicolumn{1}{l|}{}                     & \multicolumn{1}{l|}{\cellcolor[HTML]{DAE8FC}ASR}          & \multicolumn{1}{l|}{\cellcolor[HTML]{DAE8FC}Top-$k$}        & \multicolumn{1}{l|}{\cellcolor[HTML]{DAE8FC}ASR}          & \multicolumn{1}{l|}{\cellcolor[HTML]{DAE8FC}Top-$k$}        & \multicolumn{1}{l|}{\cellcolor[HTML]{DAE8FC}ASR}            & \multicolumn{1}{l|}{\cellcolor[HTML]{DAE8FC}Top-$k$}         & \multicolumn{1}{l|}{\cellcolor[HTML]{DAE8FC}ASR}            & \cellcolor[HTML]{DAE8FC}Top-$k$         \\ \hline
\multicolumn{1}{|l|}{LR}                                                                   & \multicolumn{1}{l|}{40.98}          & \multicolumn{1}{l|}{Top-3}          & \multicolumn{1}{l|}{\textbf{48.78}}          & \multicolumn{1}{l|}{\textbf{Top-3}}          & \multicolumn{1}{l|}{25.06}                & \multicolumn{1}{l|}{\cellcolor[HTML]{DAE8FC}97.98}        & \multicolumn{1}{l|}{\cellcolor[HTML]{DAE8FC}Top-7}          & \multicolumn{1}{l|}{\cellcolor[HTML]{DAE8FC}23.38}        & \multicolumn{1}{l|}{\cellcolor[HTML]{DAE8FC}Top-13}         & \multicolumn{1}{l|}{\cellcolor[HTML]{DAE8FC}45.43}          & \multicolumn{1}{l|}{\cellcolor[HTML]{DAE8FC}Top-27}          & \multicolumn{1}{l|}{\cellcolor[HTML]{DAE8FC}44.98}          & \cellcolor[HTML]{DAE8FC}Top-27          \\ \hline
\multicolumn{1}{|l|}{DT}                                                                   & \multicolumn{1}{l|}{\textbf{60.71}}          & \multicolumn{1}{l|}{\textbf{Top-2}}          & \multicolumn{1}{l|}{47.49}          & \multicolumn{1}{l|}{Top-1}          & \multicolumn{1}{l|}{20.59}                & \multicolumn{1}{l|}{\cellcolor[HTML]{DAE8FC}97.95}        & \multicolumn{1}{l|}{\cellcolor[HTML]{DAE8FC}Top-5}          & \multicolumn{1}{l|}{\cellcolor[HTML]{9B9B9B}\textbf{0.0}} & \multicolumn{1}{l|}{\cellcolor[HTML]{9B9B9B}\textbf{Top-0}} & \multicolumn{1}{l|}{\cellcolor[HTML]{DAE8FC}55.12}          & \multicolumn{1}{l|}{\cellcolor[HTML]{DAE8FC}Top-27}          & \multicolumn{1}{l|}{\cellcolor[HTML]{DAE8FC}55.12}          & \cellcolor[HTML]{DAE8FC}Top-27          \\ \hline
\multicolumn{1}{|l|}{RF}                                                                   & \multicolumn{1}{l|}{\textbf{44.58}}          & \multicolumn{1}{l|}{\textbf{Top-2}}          & \multicolumn{1}{l|}{39.58}          & \multicolumn{1}{l|}{Top-3}          & \multicolumn{1}{l|}{9.07}                 & \multicolumn{1}{l|}{\cellcolor[HTML]{DAE8FC}98.75}        & \multicolumn{1}{l|}{\cellcolor[HTML]{DAE8FC}Top-6}          & \multicolumn{1}{l|}{\cellcolor[HTML]{9B9B9B}\textbf{0.0}} & \multicolumn{1}{l|}{\cellcolor[HTML]{9B9B9B}\textbf{Top-0}} & \multicolumn{1}{l|}{\cellcolor[HTML]{DAE8FC}100}            & \multicolumn{1}{l|}{\cellcolor[HTML]{DAE8FC}Top-27}          & \multicolumn{1}{l|}{\cellcolor[HTML]{DAE8FC}100}            & \cellcolor[HTML]{DAE8FC}Top-27          \\ \hline
\multicolumn{1}{|l|}{MLP}                                                                  & \multicolumn{1}{l|}{\textbf{50.09}}          & \multicolumn{1}{l|}{\textbf{Top-2}}          & \multicolumn{1}{l|}{38.4}           & \multicolumn{1}{l|}{Top-1}          & \multicolumn{1}{l|}{22.26}                & \multicolumn{1}{l|}{\cellcolor[HTML]{DAE8FC}99.27}        & \multicolumn{1}{l|}{\cellcolor[HTML]{DAE8FC}Top-7}          & \multicolumn{1}{l|}{\cellcolor[HTML]{DAE8FC}16.67}        & \multicolumn{1}{l|}{\cellcolor[HTML]{DAE8FC}Top-12}         & \multicolumn{1}{l|}{\cellcolor[HTML]{DAE8FC}45.82}          & \multicolumn{1}{l|}{\cellcolor[HTML]{DAE8FC}Top-27}          & \multicolumn{1}{l|}{\cellcolor[HTML]{DAE8FC}45.08}          & \cellcolor[HTML]{DAE8FC}Top-27          \\ \hline
\multicolumn{1}{|l|}{ADA}                                                                  & \multicolumn{1}{l|}{\textbf{46.62}}          & \multicolumn{1}{l|}{\textbf{Top-3}}          & \multicolumn{1}{l|}{41.59}          & \multicolumn{1}{l|}{Top-3}          & \multicolumn{1}{l|}{19.11}                & \multicolumn{1}{l|}{\cellcolor[HTML]{9B9B9B}\textbf{0.0}} & \multicolumn{1}{l|}{\cellcolor[HTML]{9B9B9B}\textbf{Top-0}} & \multicolumn{1}{l|}{\cellcolor[HTML]{9B9B9B}\textbf{0.0}} & \multicolumn{1}{l|}{\cellcolor[HTML]{9B9B9B}\textbf{Top-0}} & \multicolumn{1}{l|}{\cellcolor[HTML]{DAE8FC}46.61}          & \multicolumn{1}{l|}{\cellcolor[HTML]{DAE8FC}Top-27}          & \multicolumn{1}{l|}{\cellcolor[HTML]{DAE8FC}46.42}          & \cellcolor[HTML]{DAE8FC}Top-27          \\ \hline
\multicolumn{1}{|l|}{BAG}                                                                  & \multicolumn{1}{l|}{\textbf{51.96}} & \multicolumn{1}{l|}{\textbf{Top-2}} & \multicolumn{1}{l|}{44.21}          & \multicolumn{1}{l|}{Top-2}          & \multicolumn{1}{l|}{17.86}                & \multicolumn{1}{l|}{\cellcolor[HTML]{DAE8FC}56.19}        & \multicolumn{1}{l|}{\cellcolor[HTML]{DAE8FC}Top-6}          & \multicolumn{1}{l|}{\cellcolor[HTML]{DAE8FC}0.02}         & \multicolumn{1}{l|}{\cellcolor[HTML]{DAE8FC}Top-1}          & \multicolumn{1}{l|}{\cellcolor[HTML]{DAE8FC}99.63}          & \multicolumn{1}{l|}{\cellcolor[HTML]{DAE8FC}Top-27}          & \multicolumn{1}{l|}{\cellcolor[HTML]{DAE8FC}99.63}          & \cellcolor[HTML]{DAE8FC}Top-27          \\ \hline
\multicolumn{1}{|l|}{GBC}                                                                  & \multicolumn{1}{l|}{\textbf{55.97}}          & \multicolumn{1}{l|}{\textbf{Top-3}}          & \multicolumn{1}{l|}{46.27}          & \multicolumn{1}{l|}{Top-1}          & \multicolumn{1}{l|}{14.37}                & \multicolumn{1}{l|}{\cellcolor[HTML]{DAE8FC}99.25}        & \multicolumn{1}{l|}{\cellcolor[HTML]{DAE8FC}Top-6}          & \multicolumn{1}{l|}{\cellcolor[HTML]{9B9B9B}\textbf{0.0}} & \multicolumn{1}{l|}{\cellcolor[HTML]{9B9B9B}\textbf{Top-0}} & \multicolumn{1}{l|}{\cellcolor[HTML]{DAE8FC}41.6}           & \multicolumn{1}{l|}{\cellcolor[HTML]{DAE8FC}Top-27}          & \multicolumn{1}{l|}{\cellcolor[HTML]{DAE8FC}41.6}           & \cellcolor[HTML]{DAE8FC}Top-27          \\ \hline
\multicolumn{14}{|c|}{\textbf{Postgres}}                                                                                                                                                                                                                                                                                                                                                                                                                                                                                                                                                                                                                                                                                                                                                \\ \hline
\multicolumn{1}{|l|}{LR}                                                                   & \multicolumn{1}{l|}{\textbf{85.89}} & \multicolumn{1}{l|}{\textbf{Top-1}} & \multicolumn{1}{l|}{83.06}          & \multicolumn{1}{l|}{Top-3}          & \multicolumn{1}{l|}{34.54}                & \multicolumn{1}{l|}{\cellcolor[HTML]{DAE8FC}99.46}        & \multicolumn{1}{l|}{\cellcolor[HTML]{DAE8FC}Top-6}          & \multicolumn{1}{l|}{\cellcolor[HTML]{DAE8FC}99.32}        & \multicolumn{1}{l|}{\cellcolor[HTML]{DAE8FC}Top-8}          & \multicolumn{1}{l|}{\cellcolor[HTML]{DAE8FC}\textbf{41.8}}  & \multicolumn{1}{l|}{\cellcolor[HTML]{DAE8FC}\textbf{Top-12}} & \multicolumn{1}{l|}{\cellcolor[HTML]{DAE8FC}\textbf{41.8}}  & \cellcolor[HTML]{DAE8FC}\textbf{Top-12} \\ \hline
\multicolumn{1}{|l|}{DT}                                                                   & \multicolumn{1}{l|}{\textbf{85.63}} & \multicolumn{1}{l|}{\textbf{Top-2}} & \multicolumn{1}{l|}{82.92}          & \multicolumn{1}{l|}{Top-2}          & \multicolumn{1}{l|}{6.9}                  & \multicolumn{1}{l|}{\cellcolor[HTML]{DAE8FC}85.87}        & \multicolumn{1}{l|}{\cellcolor[HTML]{DAE8FC}Top-4}          & \multicolumn{1}{l|}{\cellcolor[HTML]{DAE8FC}85.87}        & \multicolumn{1}{l|}{\cellcolor[HTML]{DAE8FC}Top-12}         & \multicolumn{1}{l|}{\cellcolor[HTML]{DAE8FC}\textbf{66.54}} & \multicolumn{1}{l|}{\cellcolor[HTML]{DAE8FC}\textbf{Top-12}} & \multicolumn{1}{l|}{\cellcolor[HTML]{DAE8FC}\textbf{68.55}} & \cellcolor[HTML]{DAE8FC}\textbf{Top-12} \\ \hline
\multicolumn{1}{|l|}{RF}                                                                   & \multicolumn{1}{l|}{65.45}          & \multicolumn{1}{l|}{Top-2}          & \multicolumn{1}{l|}{\textbf{72.91}} & \multicolumn{1}{l|}{\textbf{Top-2}} & \multicolumn{1}{l|}{21.36}                  & \multicolumn{1}{l|}{\cellcolor[HTML]{DAE8FC}100}          & \multicolumn{1}{l|}{\cellcolor[HTML]{DAE8FC}Top-4}          & \multicolumn{1}{l|}{\cellcolor[HTML]{DAE8FC}91.99}        & \multicolumn{1}{l|}{\cellcolor[HTML]{DAE8FC}Top-4}          & \multicolumn{1}{l|}{\cellcolor[HTML]{DAE8FC}86.08}          & \multicolumn{1}{l|}{\cellcolor[HTML]{DAE8FC}Top-12}          & \multicolumn{1}{l|}{\cellcolor[HTML]{DAE8FC}84.77}          & \cellcolor[HTML]{DAE8FC}Top-12          \\ \hline
\multicolumn{1}{|l|}{MLP}                                                                  & \multicolumn{1}{l|}{56.99}          & \multicolumn{1}{l|}{Top-1}          & \multicolumn{1}{l|}{\textbf{59.45}} & \multicolumn{1}{l|}{\textbf{Top-2}} & \multicolumn{1}{l|}{22.4}                 & \multicolumn{1}{l|}{\cellcolor[HTML]{DAE8FC}87.94}        & \multicolumn{1}{l|}{\cellcolor[HTML]{DAE8FC}Top-6}          & \multicolumn{1}{l|}{\cellcolor[HTML]{DAE8FC}88.08}        & \multicolumn{1}{l|}{\cellcolor[HTML]{DAE8FC}Top-11}         & \multicolumn{1}{l|}{\cellcolor[HTML]{DAE8FC}\textbf{48.9}}  & \multicolumn{1}{l|}{\cellcolor[HTML]{DAE8FC}\textbf{Top-12}} & \multicolumn{1}{l|}{\cellcolor[HTML]{DAE8FC}\textbf{48.76}} & \cellcolor[HTML]{DAE8FC}\textbf{Top-12} \\ \hline
\multicolumn{1}{|l|}{ADA}                                                                  & \multicolumn{1}{l|}{\textbf{84.09}} & \multicolumn{1}{l|}{\textbf{Top-3}} & \multicolumn{1}{l|}{56.98}          & \multicolumn{1}{l|}{Top-1}          & \multicolumn{1}{l|}{19.61}                & \multicolumn{1}{l|}{\cellcolor[HTML]{DAE8FC}8.34}         & \multicolumn{1}{l|}{\cellcolor[HTML]{DAE8FC}Top-3}          & \multicolumn{1}{l|}{\cellcolor[HTML]{DAE8FC}56.97}        & \multicolumn{1}{l|}{\cellcolor[HTML]{DAE8FC}Top-4}          & \multicolumn{1}{l|}{\cellcolor[HTML]{DAE8FC}\textbf{40.93}} & \multicolumn{1}{l|}{\cellcolor[HTML]{DAE8FC}\textbf{Top-12}} & \multicolumn{1}{l|}{\cellcolor[HTML]{DAE8FC}\textbf{40.93}} & \cellcolor[HTML]{DAE8FC}\textbf{Top-12} \\ \hline
\multicolumn{1}{|l|}{BAG}                                                                  & \multicolumn{1}{l|}{\textbf{83.59}} & \multicolumn{1}{l|}{\textbf{Top-2}} & \multicolumn{1}{l|}{79.56}          & \multicolumn{1}{l|}{Top-3}          & \multicolumn{1}{l|}{36.98}                & \multicolumn{1}{l|}{\cellcolor[HTML]{DAE8FC}\textbf{100}} & \multicolumn{1}{l|}{\cellcolor[HTML]{DAE8FC}Top-4}          & \multicolumn{1}{l|}{\cellcolor[HTML]{DAE8FC}86.71}        & \multicolumn{1}{l|}{\cellcolor[HTML]{DAE8FC}Top-7}          & \multicolumn{1}{l|}{\cellcolor[HTML]{DAE8FC}\textbf{50.13}} & \multicolumn{1}{l|}{\cellcolor[HTML]{DAE8FC}\textbf{Top-12}} & \multicolumn{1}{l|}{\cellcolor[HTML]{DAE8FC}\textbf{49.47}} & \cellcolor[HTML]{DAE8FC}\textbf{Top-12} \\ \hline
\multicolumn{1}{|l|}{GBC}                                                                  & \multicolumn{1}{l|}{70.99}          & \multicolumn{1}{l|}{Top-2}          & \multicolumn{1}{l|}{\textbf{73.54}} & \multicolumn{1}{l|}{\textbf{Top-2}} & \multicolumn{1}{l|}{19.75}                & \multicolumn{1}{l|}{\cellcolor[HTML]{DAE8FC}100}          & \multicolumn{1}{l|}{\cellcolor[HTML]{DAE8FC}Top-4}          & \multicolumn{1}{l|}{\cellcolor[HTML]{DAE8FC}85.69}        & \multicolumn{1}{l|}{\cellcolor[HTML]{DAE8FC}Top-4}          & \multicolumn{1}{l|}{\cellcolor[HTML]{DAE8FC}93.24}          & \multicolumn{1}{l|}{\cellcolor[HTML]{DAE8FC}Top-12}          & \multicolumn{1}{l|}{\cellcolor[HTML]{DAE8FC}92.05}          & \cellcolor[HTML]{DAE8FC}Top-12          \\ \hline
\end{tabular}
\end{adjustbox}
\label{attack_result_balanced}
\textbf{Note:} Bolded cells indicate cases where the attack techniques failed to produce any successful adversarial examples
\end{table*}



Moving to the \textit{PermuteAttack}, our approach shows promising results over \textit{PermuteAttack} in terms of imperceptibility. Except for a few cases (highlighted in bold text), \textit{PermuteAttack} exhibits superior performance regarding ASR metric values. However, for a few ML models, \textit{PermuteAttack} fails to generate a single adversarial example. Moreover, our manual investigation found that, since \textit{PermuteAttack} employs the \textit{Genetic} algorithm and randomly selected features and their values for generating adversarial examples, it produces inconsistent outcomes for the same instance across multiple executions in a row. Therefore, our approach outperforms \textit{PermuteAttack} regarding consistency and imperceptibility.

Table \ref{attack_result_balanced} presents the ASR metric and top-$k$ values for both the explanation-guided and selected state-of-the-art adversarial attacks using the balanced test data. The bolded cells highlighted in grey represent cases where the attack techniques failed to produce any successful adversarial examples. From Table \ref{attack_result_balanced}, it is evident that our explanation-guided adversarial attack approach outperforms the \textit{ZOO}, \textit{HopSkipJump}, and \textit{Boundary} attack techniques in terms of the ASR metric and imperceptibility. Although the \textit{PermuteAttack} performs well in terms of the ASR metric, our approach achieves a better balance between the ASR metric and imperceptibility. Additionally, the \textit{PermuteAttack} fails to generate a single adversarial example for the ADA model trained on the Java project dataset and performs poorly on the same model trained on the Postgres dataset. Therefore, we reach a similar conclusion even after balancing the test dataset.

While the tabular results provide detailed comparisons, we also employ win-rate analysis to more effectively summarize the overall performance of our attack approach relative to baseline methods. Win-rate analysis offers a straightforward yet powerful means of comparing two approaches across multiple experimental settings (e.g., different datasets or models). We evaluate whether the proposed method outperforms, matches, or underperforms for each setting relative to the baseline. The win-rate is computed as the proportion of settings where the proposed method outperforms the baseline, defined as:

\begin{equation}
\text{Win Rate}_{A \text{ vs } B} = \frac{\text{Number of Wins of } A}{\text{Total Number of Comparisons}} \times 100\%
\label{WinRate}
\end{equation}

Where, a \textbf{win} is counted if method $A$ (e.g., our proposed approach) achieves better performance than method $B$ (e.g., baselines) according to the evaluation metric of interest (e.g., ASR).

\begin{figure}[htbp]
\centering     
\subfigure[ASR on Unbalanced Data]{\label{ASRUB}\includegraphics[width=6.25cm, height=4.25cm]{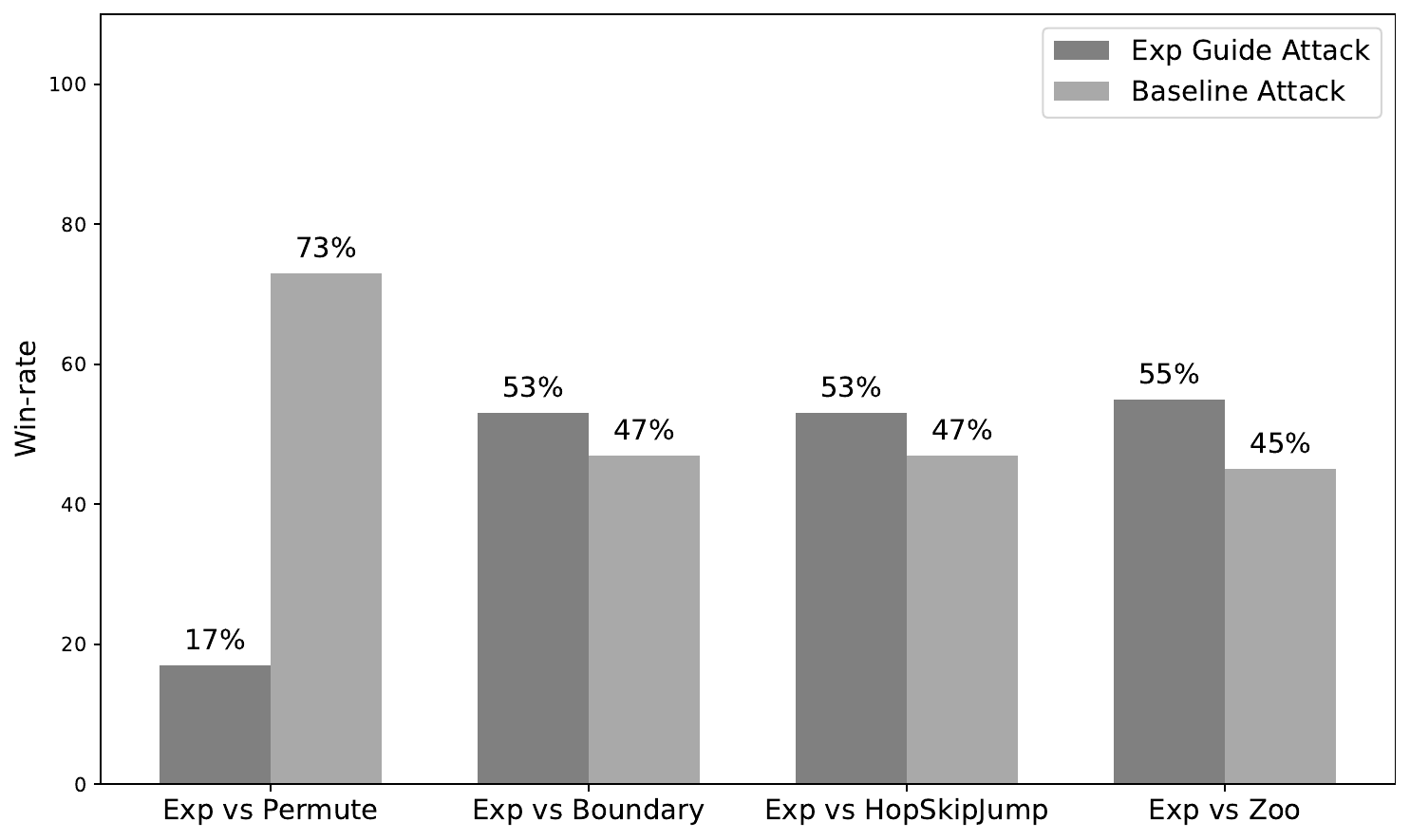}}
\subfigure[ASR on Balanced Data]{\label{ASRB}\includegraphics[width=6.25cm, height=4.25cm]{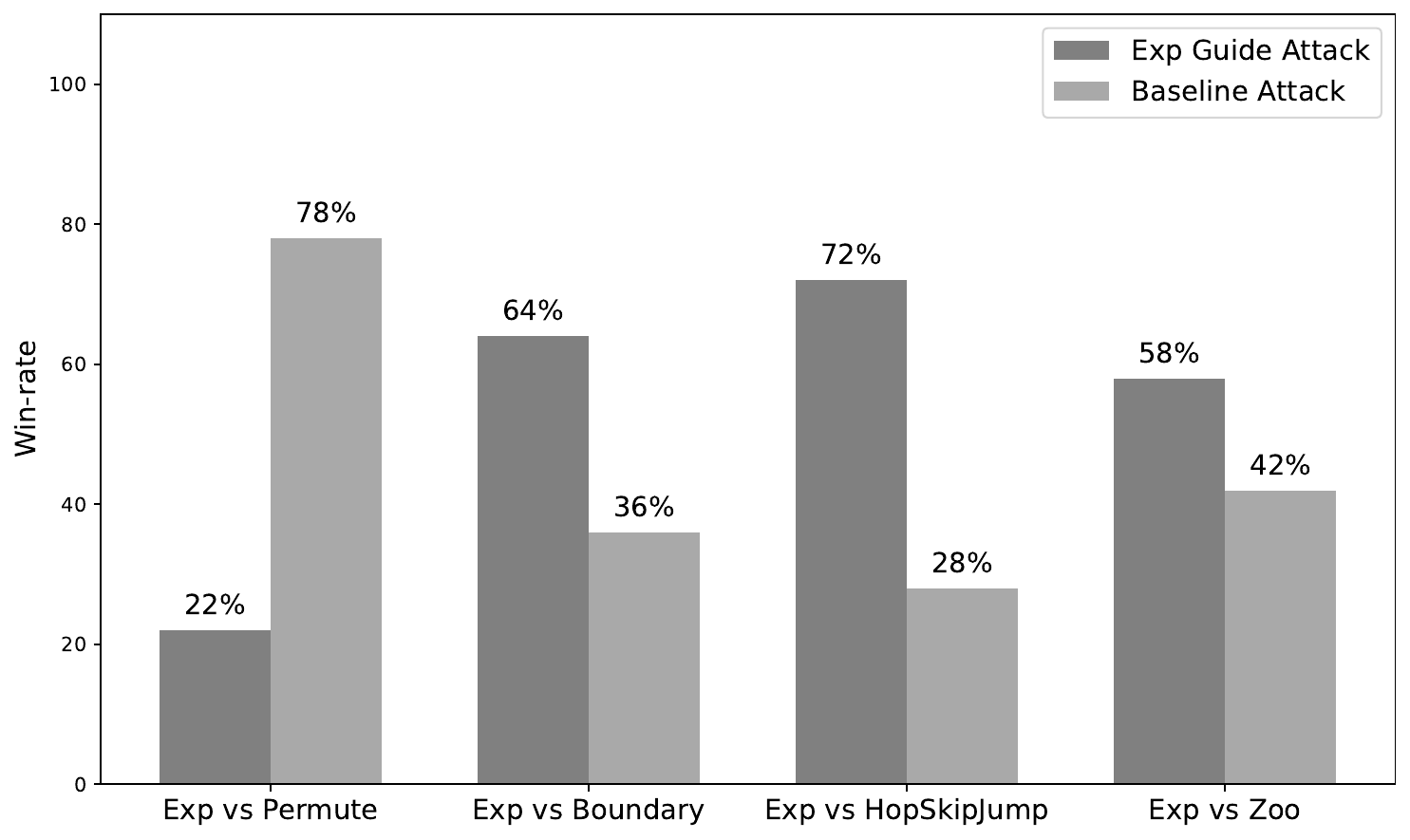}} 
\subfigure[Imperceptibility on Unbalanced Data]{\label{IMPUB}\includegraphics[width=6.25cm, height=4.25cm]{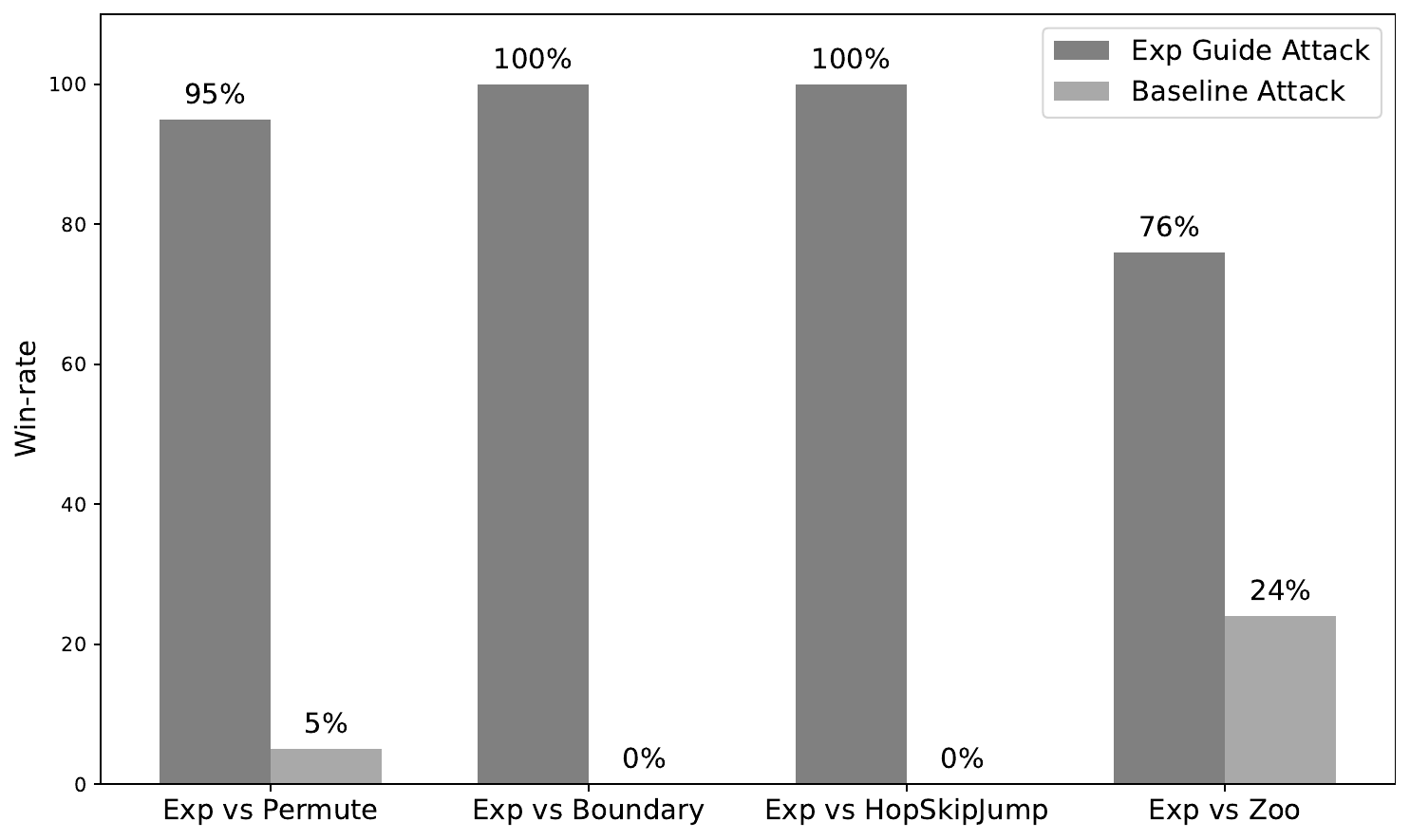}} 
\subfigure[Imperceptibility on Balanced Data]{\label{IMPB}\includegraphics[width=6.25cm, height=4.25cm]{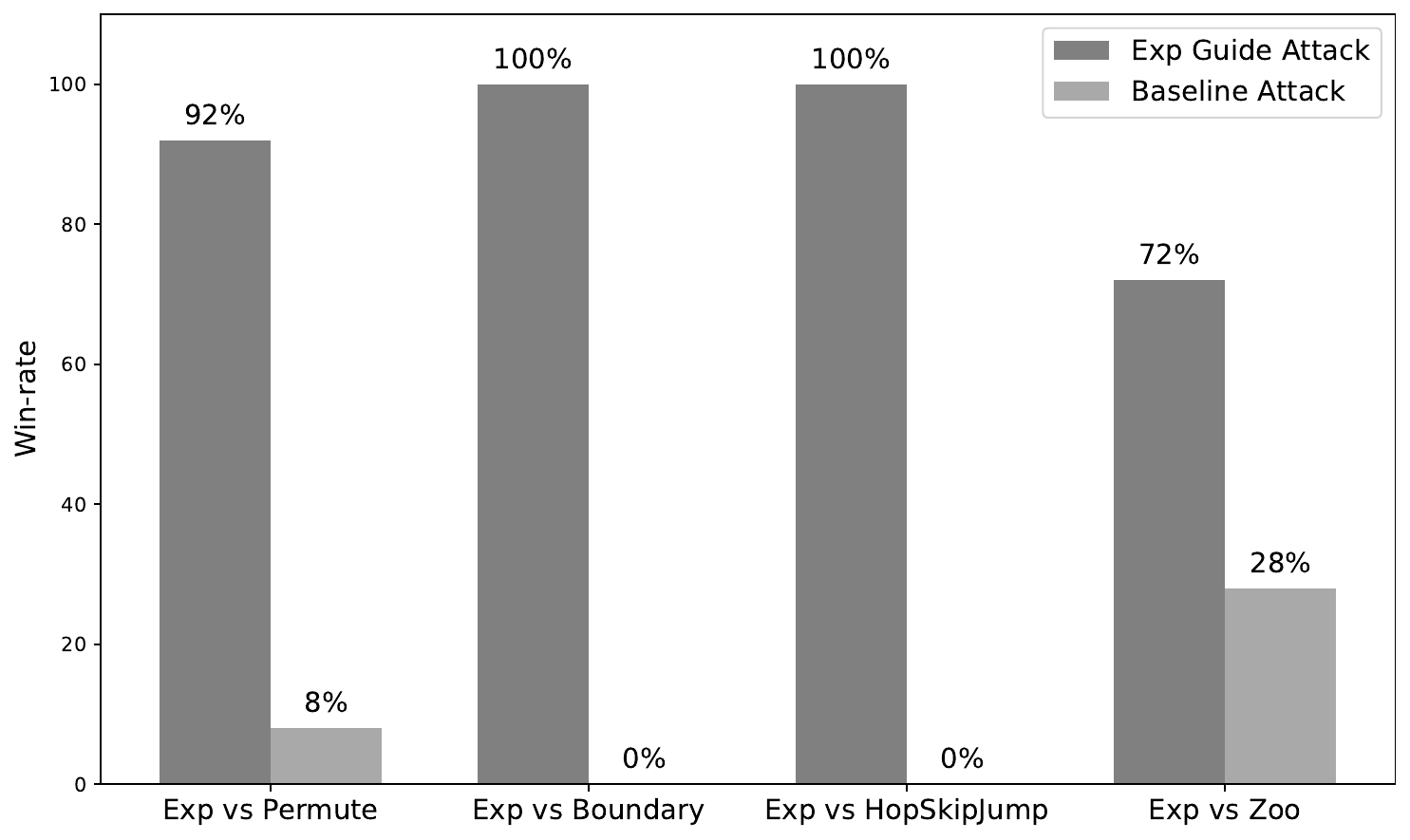}}
\vspace{-0.5em}

\caption{Comparison of the explanation-guided attack with four baselines based on ASR and imperceptibility metrics. Each group illustrates the performance difference between the proposed method and each baseline method.}
\label{RQ2WinRate}
\end{figure}

Figure \ref{RQ2WinRate} compares the win-rates between the explanation-guided attack and four baseline adversarial attacks— PermuteAttack, Zoo, Boundary, and HopSkipJump. In each pairwise comparison, the explanation-guided attack consistently achieves a higher win-rate than the corresponding baseline methods regarding the \textit{imperceptibility} metric. However, although the \textit{PermuteAttack} performs better than our approach, considering the ASR metric, our approach outperforms the other three baseline attacks. Overall, the win-rate analysis suggests that the explanation-guided attack is generally more effective in generating successful adversarial examples than the classical baseline attacks. The consistent improvement across multiple baselines highlights the robustness and reliability of the proposed attack strategy.

Finally, we extend our experiments by (1) randomly changing the feature values and (2) altering the least important features ranked at the bottom of the list. We refer to these methods as \textbf{makeshift tools}: (1) \textbf{BL}, where we change the number of features equal to the top-$k$ from the bottom of the feature importance rank, and (2) \textbf{BR}, where we randomly select and change the number of features equal to the top-$k$. It is important to note that for random changes, we exclude the top-$k$ features from the feature importance rank from consideration. This is our design choice, and we deliberately do this because there is a possibility that the most important feature could be randomly selected to conduct adversarial attacks. Thus, the ASR metric value for \textbf{BR} might be high. Finally, we apply the same feature value transformation strategy as mentioned in Section \ref{Adv Attack} for these makeshift tools.

\begin{figure}[htbp]
\centering     
\subfigure[CLCDSA + SHAP]{\label{clcdsa_sh_RQ2}\includegraphics[width=4.25cm, height=2.8cm]{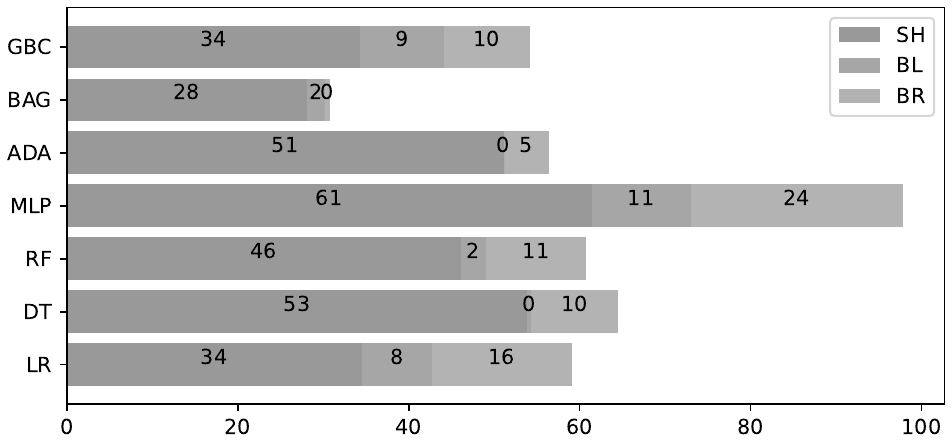}}
\subfigure[CLCDSA + LIME]{\label{clcdsa_lm_RQ2}\includegraphics[width=4.25cm, height=2.8cm]{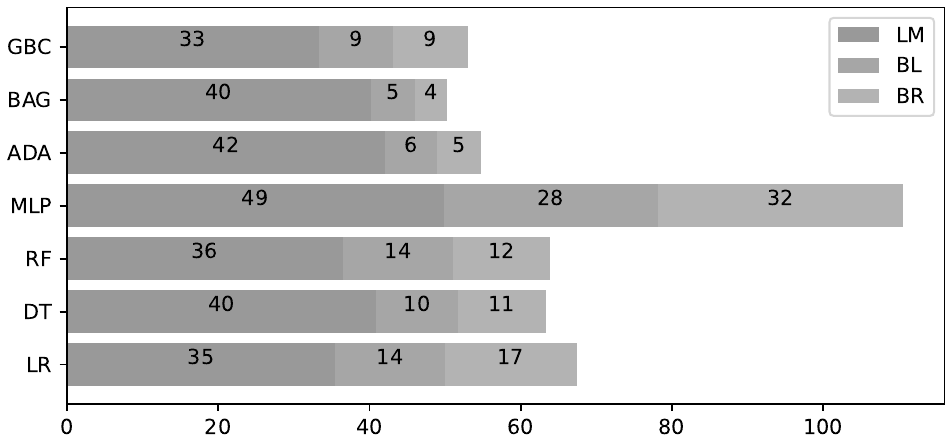}} 
\subfigure[CLCDSA + PyExplainer]{\label{clcdsa_pe_RQ2}\includegraphics[width=4.25cm, height=2.8cm]{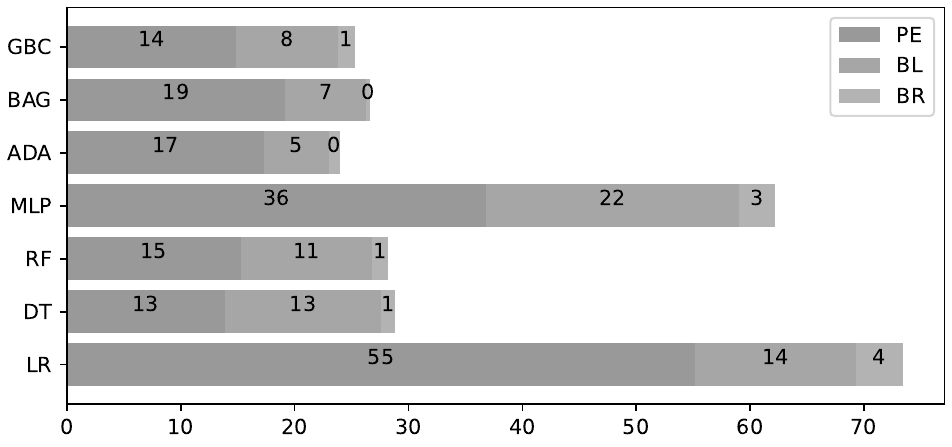}}  \\
\vspace{-.75em}
\subfigure[Code Review + SHAP]{\label{cr_sh_RQ2}\includegraphics[width=4.25cm, height=2.8cm]{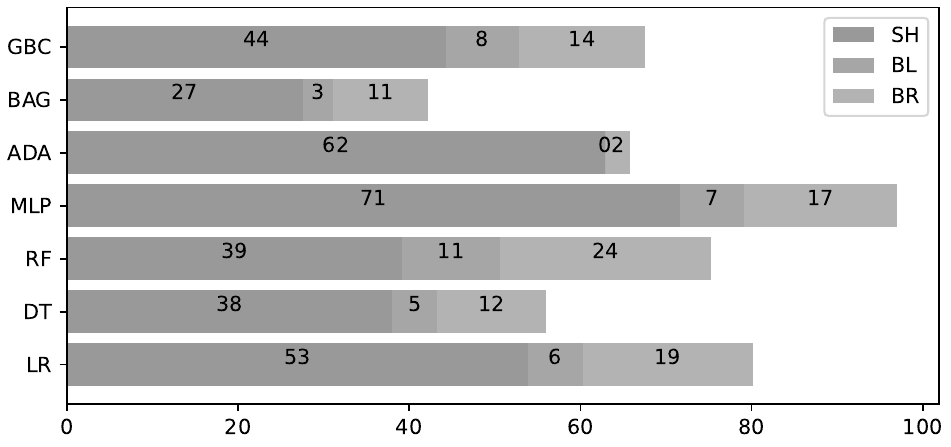}}
\subfigure[Code Review + LIME]{\label{cr_lm_RQ2}\includegraphics[width=4.25cm, height=2.8cm]{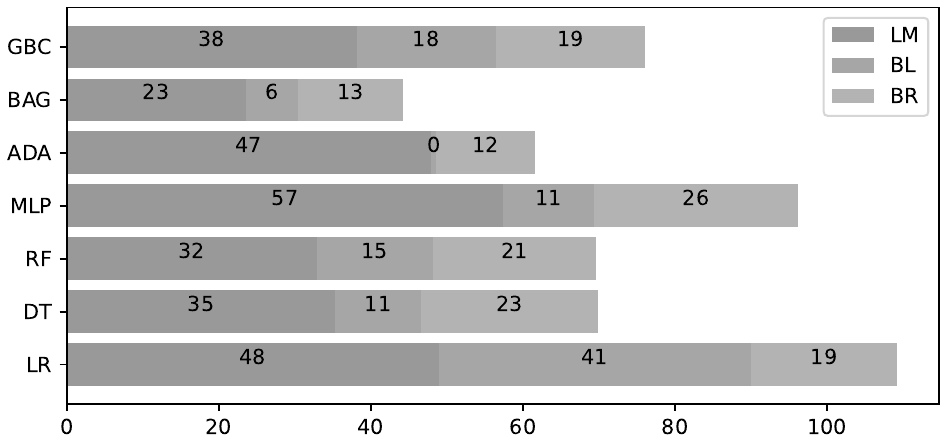}} 
\subfigure[Code Review + PyExplainer]{\label{cr_pe_RQ2}\includegraphics[width=4.25cm, height=2.8cm]{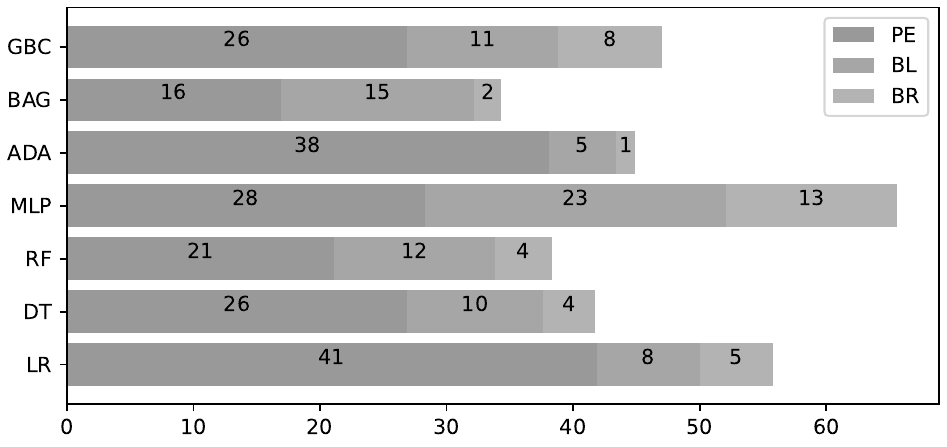}} \\
\vspace{-0.75em}
\subfigure[Postgres + SHAP]{\label{desk_sh_RQ2}\includegraphics[width=4.25cm, height=2.8cm]{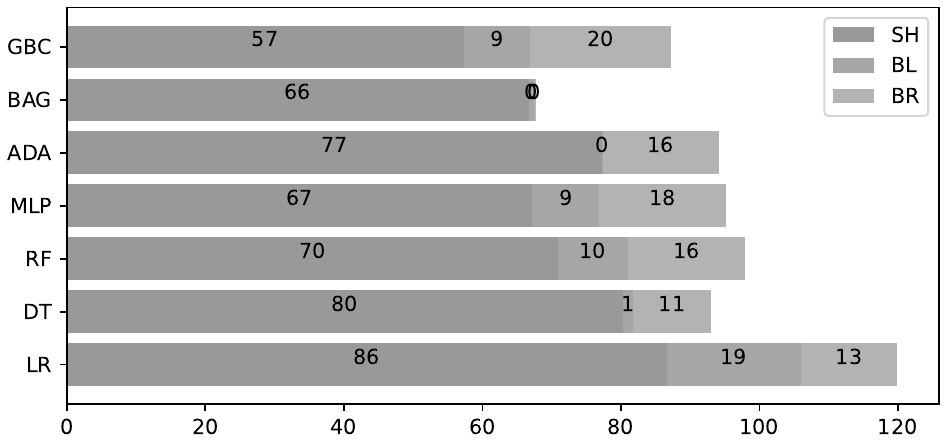}}
\subfigure[Postgres + LIME]{\label{desk_lm_RQ2}\includegraphics[width=4.25cm, height=2.8cm]{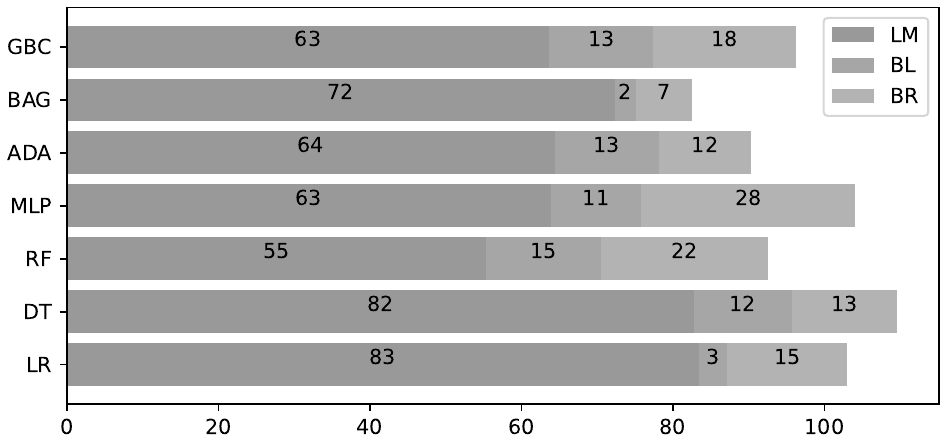}} 
\subfigure[Postgres + PyExplainer]{\label{desk_pe_RQ2}\includegraphics[width=4.25cm, height=2.8cm]{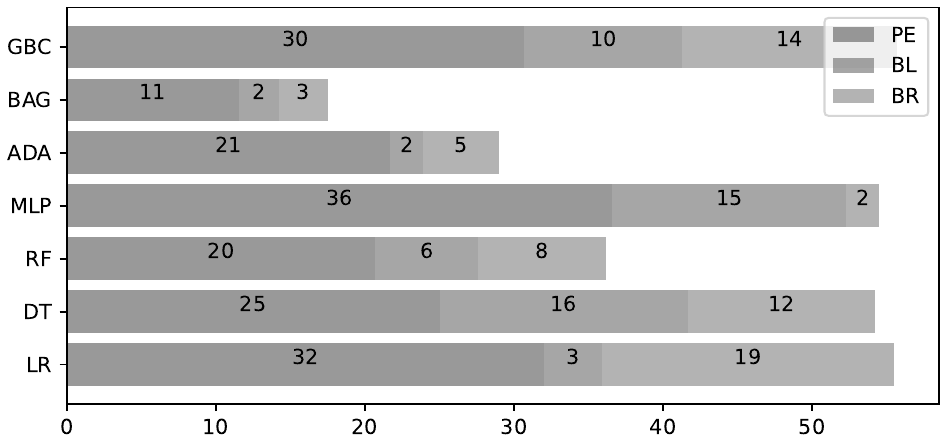}} \\
\subfigure[BCB + SHAP]{\label{bcb_sh_RQ2}\includegraphics[width=4.25cm, height=2.8cm]{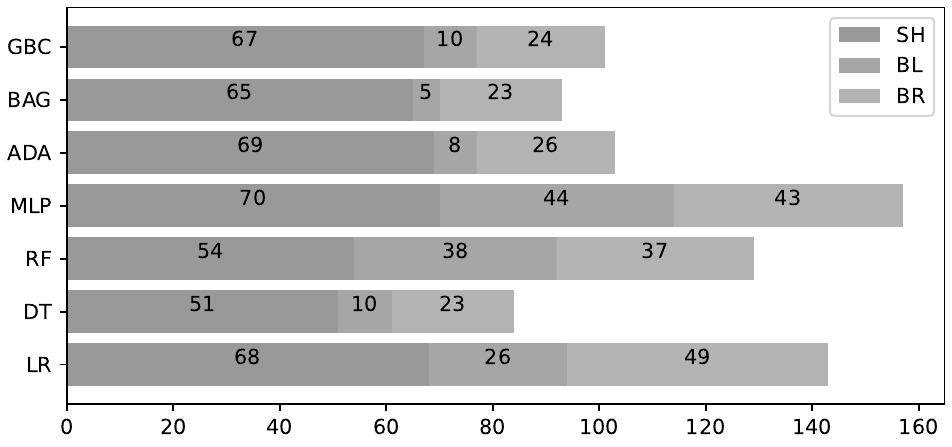}}
\subfigure[BCB + LIME]{\label{bcb_lm_RQ2}\includegraphics[width=4.25cm, height=2.8cm]{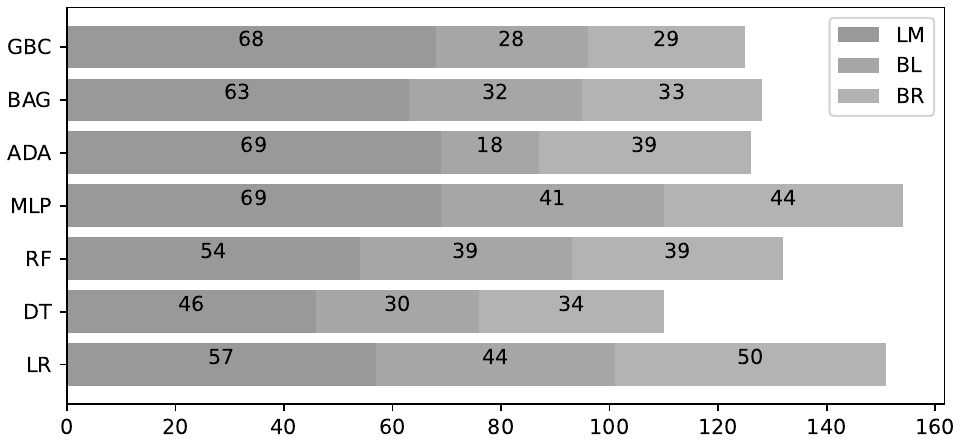}} 
\subfigure[BCB + PyExplainer]{\label{bcb_pe_RQ2}\includegraphics[width=4.25cm, height=2.8cm]{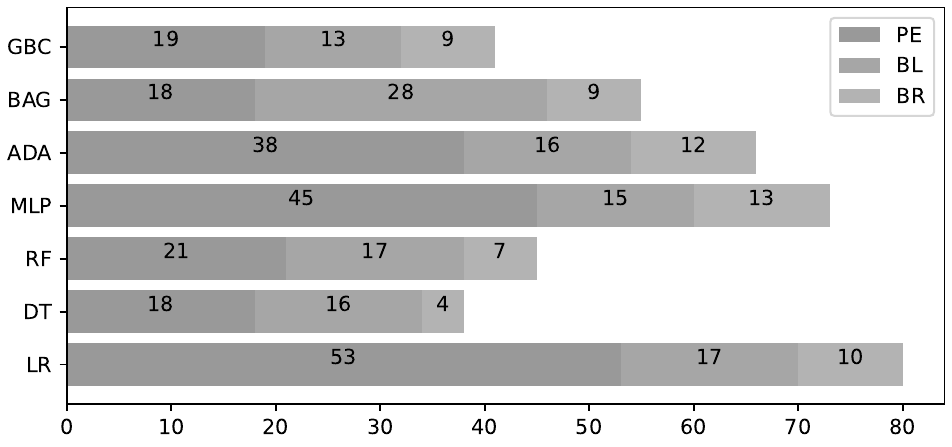}}

\caption{Comparison of our approach with the makeshift tools regarding ASR metric. The first row compares our approach with the makeshift tools when we change the top-k important features identified by SHAP (SH, Fig. \ref{clcdsa_sh_RQ2}), LIME (LM. Fig. \ref{clcdsa_lm_RQ2}), and PyExplainer (PE, Fig. \ref{clcdsa_pe_RQ2}) for the CLCDSA dataset. Similarly, the second, third, and fourth rows compare the code review, Postgres, and BCB datasets, respectively. Note: ASR metric values are rounded for better visualization.}
\label{RQ2_Fig}
\end{figure}

Figure \ref{RQ2_Fig} demonstrates that our technique outperforms makeshift tools regarding adversarial attacks on ML models, as measured by the ASR metric. For example, in our adversarial attacks on the MLP model trained on the CLCDSA dataset, changing the top-$k$ important feature values identified by SHAP results in a maximum ASR value of $61\%$. In contrast, makeshift tools BL and BR lead to ASR values of only $11\%$ and $24\%$, respectively. We observe a similar result for the MLP trained on the CLCDSA dataset for LIME explainability. For the LR model trained on the CLCDSA dataset, altering the important feature values identified by PyExplainer results in a maximum ASR value of $55\%$. In comparison, makeshift tools BL and BR lead to ASR values of only $14\%$ and $4\%$, respectively.

A closer examination of Figure \ref{RQ2_Fig} reveals a substantial difference between our approach and the makeshift tools regarding the ASR metric for adversarial attacks when using SHAP explainability. Similar results are observed for LIME and PyExplainer, with only a few exceptions. For instance, the difference between our approach and BL is only $4$ for the RF model trained on the CLCDSA dataset with LIME, and it's $0$ for the DT model trained on the CLCDSA dataset with PyExplainer. Similar results are observed for the code review datasets. In contrast, a significant difference in the ASR metric is observed for the Postgres and BCB datasets between our approach and the makeshift tools.

\begin{figure}[htbp]
\centering     
\subfigure[Java + SHAP]{\label{clcdsa_sh_RQ2_bal}\includegraphics[width=4.25cm, height=2.8cm]{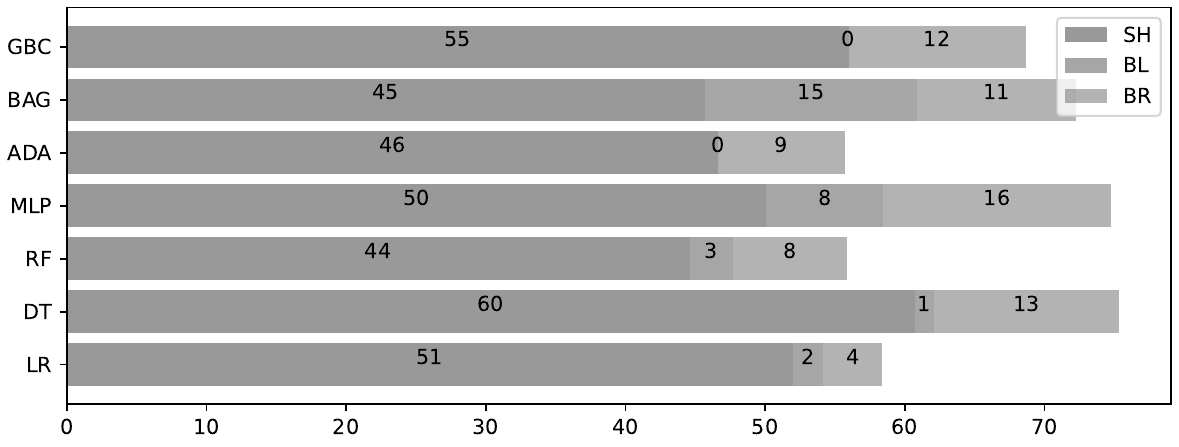}}
\subfigure[Java + LIME]{\label{clcdsa_lm_RQ2_bal}\includegraphics[width=4.25cm, height=2.8cm]{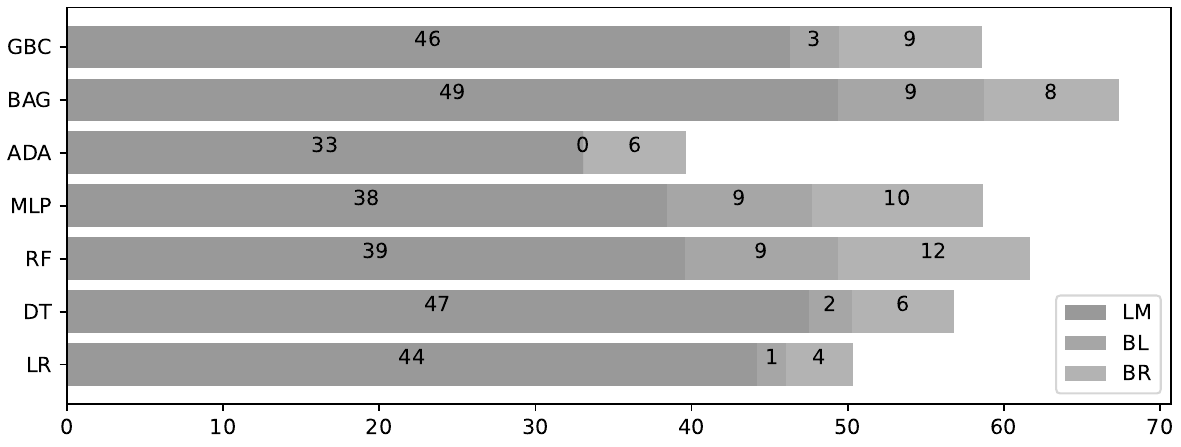}} 
\subfigure[Java + PyExplainer]{\label{clcdsa_pe_RQ2_bal}\includegraphics[width=4.25cm, height=2.8cm]{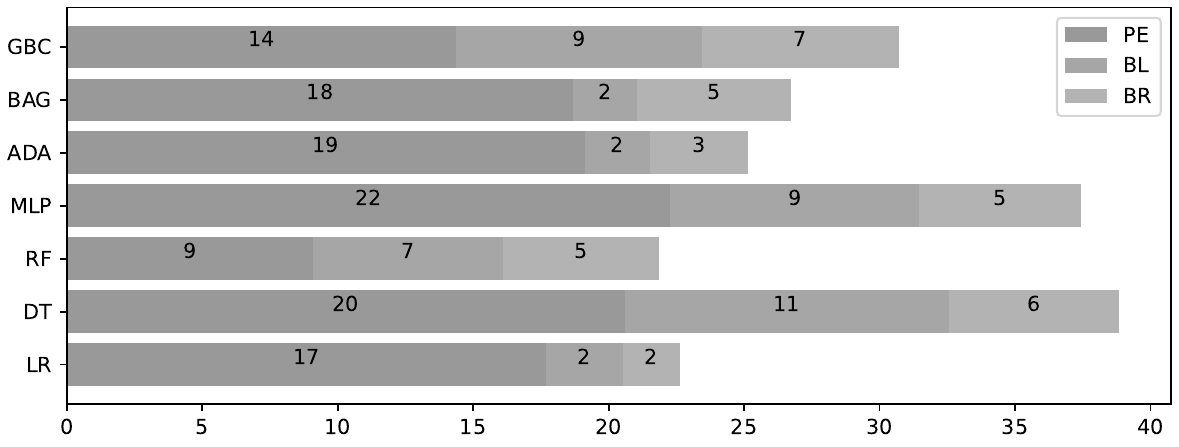}}  \\
\subfigure[Postgres + SHAP]{\label{desk_sh_RQ2_bal}\includegraphics[width=4.25cm, height=2.8cm]{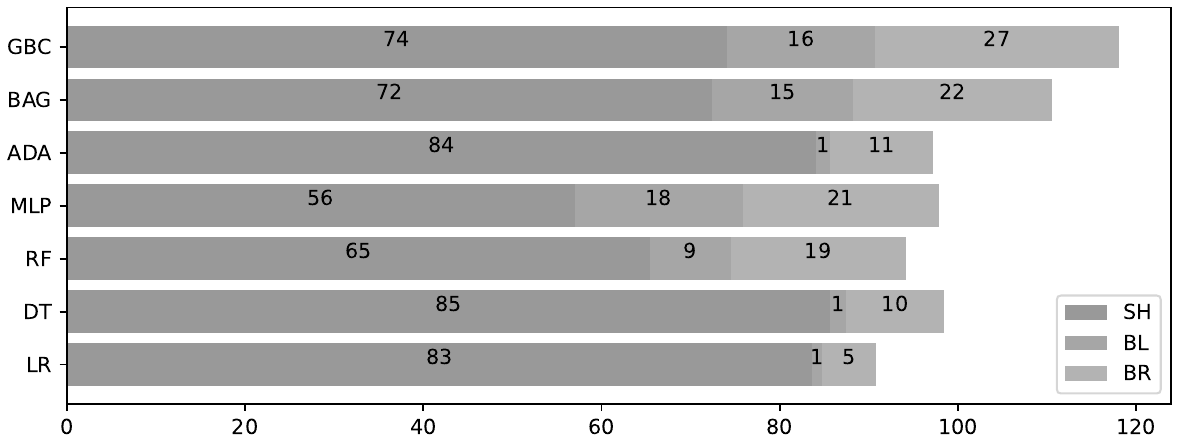}}
\subfigure[Postgres + LIME]{\label{desk_lm_RQ2_bal}\includegraphics[width=4.25cm, height=2.8cm]{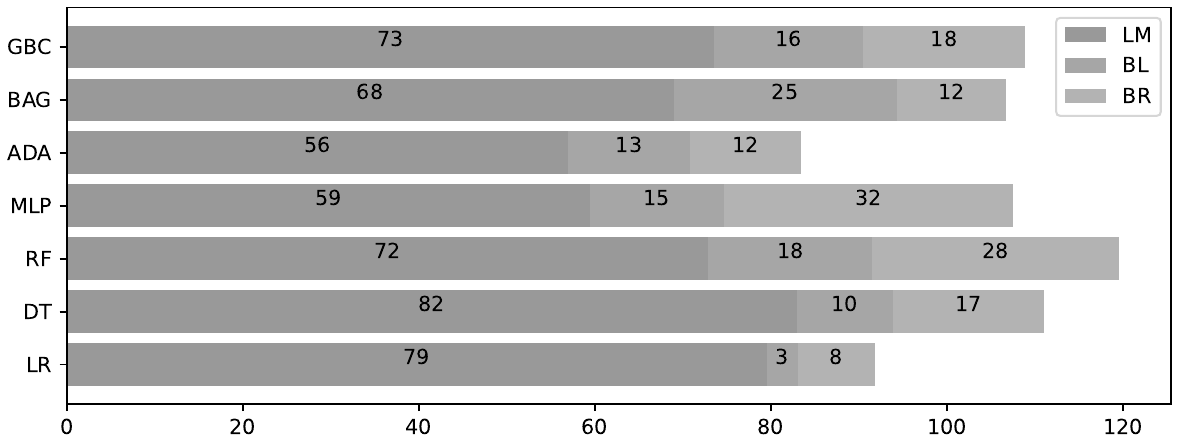}} 
\subfigure[Postgres + PyExplainer]{\label{desk_pe_RQ2_bal}\includegraphics[width=4.25cm, height=2.8cm]{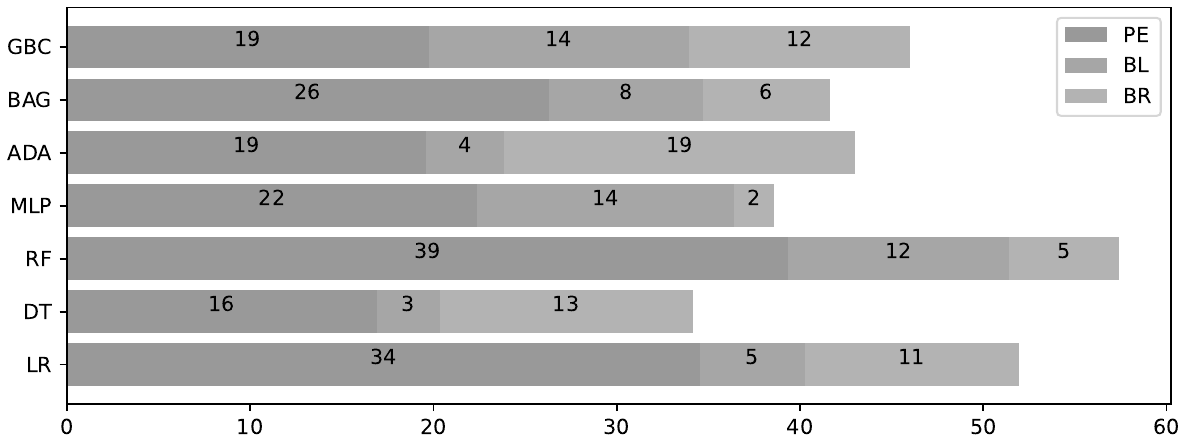}}

\caption{Comparison of our approach with the makeshift tools regarding ASR metric for the balanced test data. The first and second rows compare our approach with the makeshift tools when we change the top-k important features identified by SHAP, LIME, and PyExplainer for the Java project and Postgres datasets. Note: ASR metric values are rounded for better visualization.}
\label{RQ2_Fig_bal}
\end{figure}

Similar to RQ2(a), we conducted additional experiments to compare our approach with makeshift tools regarding the ASR metric using the balanced test datasets. Figure \ref{RQ2_Fig_bal} clearly shows that our explanation-guided adversarial attack approach outperforms the makeshift tools by a large margin. Therefore, our approach surpasses the baselines and makeshift tools in terms of the ASR metric and imperceptibility when performing adversarial attacks on the ML models in software analytics tasks.

\begin{center}
\begin{tcolorbox}[
    enhanced,
    attach boxed title to top left={yshift=-3mm,yshifttext=-1mm}, 
    colback=mycolor_box,                 
    colframe=black,                
    colbacktitle= mycolor_title,            
    coltitle=black,                
    title=Result RQ2-b,            
    fonttitle=\bfseries,           
    boxed title style={size=small},
    width=0.9\textwidth            
]
    Our approach demonstrates promising results compared to the baselines and makeshift tools in terms of ASR metric and imperceptibility when conducting adversarial attacks on machine learning models in software analytics tasks.
\end{tcolorbox}
\end{center}

A significant advantage of explanation-guided adversarial attacks is that we do not need to access the model's parameters to generate adversarial examples. Based on our experimental results, the key observations are:

\begin{enumerate}
    \item Modifying just one or two features can significantly affect the accuracy of ML models in software analytics tasks. Therefore, researchers should prioritize the development of more robust ML models and consider implementing countermeasures to mitigate such attacks.

    \item We find that changing the top-$k$ important features identified by SHAP performs better compared to LIME and PyExplainer in generating adversarial examples. The second best performer is LIME, while PyExplainer's performance is not as good as SHAP and LIME. Thus, our explanation-guided adversarial attacks could be a valuable tool to assess the effectiveness of the explanations offered by different explainability techniques. 
    
\end{enumerate}

\section{Threats to Validity}
\label{threat}
This section briefly describes the internal, construct and external threats related to our study.

\subsection{Internal Threat}
\label{internal}
The first internal threat is the accuracy of ML models. However, we countered this threat by choosing the hyperparameter settings described in previous studies \cite{catolino2019cross, pornprasit2021pyexplainer}. We used SMOTE and Autospearman and applied grid search, random search, and Bayesian optimization to find the best hyperparameter combinations for each ML model. Thus, our selected ML models have achieved good accuracy. We addressed the potential threat concerning the quality of important features identified by ML explainability techniques by choosing two widely recognized model-agnostic methods, SHAP and LIME. Additionally, we used PyExplainer \cite{pornprasit2021pyexplainer}, explicitly designed to explain JIT defect models. Therefore, the important features found by these three techniques are good enough to generate adversarial examples. Another internal threat could be selecting the top-$k$ features to generate adversarial examples. However, Section \ref{Adv Attack} describes how we mitigated this threat. Our experiments show that LIME and PyExplainer sometimes generate different feature importance rankings across multiple executions of the same instance. This variability in experimental outcomes is not a limitation of our study.

\subsection{Construct Threat}
\label{construct}

One potential threat could be how changes to top-$k$ features might affect source code. However, we addressed this concern by considering the characteristics of black-box adversarial attacks performed on extracted features stored in tabular format. Similar to studies in the financial domain \cite{cartella2021adversarial, mathov2020not, du2023extensive} and in the software analytics \cite{zhang2021advdoor}, we assume access to both training and test datasets for black-box attacks. We constructed surrogate ML models to carry out adversarial attacks, following the approaches outlined in \cite{cartella2021adversarial, mathov2020not}. In addition, similar to studies in computer vision \cite{xu2020towards, inkawhich2019feature, simonetto2021unified}, we attack the feature-space (e.g., metrics or properties of source code in tabular format) while generating adversarial examples. Since we only have access to the extracted features in tabular format (e.g., training and testing data) rather than the source code itself, and considering that these extracted features serve as the direct input to ML models during training, we explicitly state that observing the impact of altered features on the source code falls outside the scope of this study.

Another potential concern is that the preprocessing of the dataset and feature engineering, along with the validation of the model performance and the accuracy of the ML model, could affect the effectiveness of our approach in generating adversarial examples. However, we addressed this concern by comparing our approach with four state-of-the-art adversarial attack techniques on the same trained ML models and datasets. Our approach demonstrates promising results compared to baselines and makeshift tools discussed in Section \ref{exp_res}. Therefore, as long as we compare our approach with the baselines using the same trained ML models and datasets, this concern does not affect the effectiveness of our approach.

\subsection{External Threat}
\label{external}
The generalizability of our experimental results might threaten the external validity. However, we resolved this threat by choosing seven classical ML models extensively studied in different software analytics tasks, three ML explainability techniques and six distinct datasets used in the previous studies \cite{feng2024machine, wu2022detecting, hu2022treecen, catolino2019cross, roy2022don, pornprasit2021pyexplainer, chen2017adversarial}. Thus, we considered $6 \times 3 \times 7 = 126$ experimental combinations for the analysis. Therefore, our extensive experimental findings are possibly sufficiently robust to generalize our study.

\section{Related Work}
\label{RW}
There has been a growing interest in adversarial learning, which encompasses adversarial training, attacks, countermeasures, and generating adversarial examples. This section briefly discusses existing works focusing on adversarial attacks on ML models in software analytics tasks.


Szegedy et al. \cite{szegedy2013intriguing} first introduced the concept of adversarial attack for image classifiers. To the best of our knowledge, we have not found any research on adversarial attacks targeting classical ML models trained on extracted features stored in tabular format within software analytics tasks, such as JIT defect prediction, clone detection, and classification of useful code review comments. Several studies have been done on adversarial attacks targeting ML models trained on tabular data in other domains \cite{ballet2019imperceptible,hashemi2020permuteattack,grosse2017statistical,papernot2016transferability, mathov2020not,esmaeilpour2022rcc,kireev2022adversarial, melo2023adversarial,hyeong2022empirical,alshantti2023castgan,simonetto2023constrained,pleiter2023tabdoor,hu2021tablegan}. Ballet et al. \cite{ballet2019imperceptible} were the first to introduce systematic adversarial attacks on fully connected neural networks trained on tabular financial data. They altered feature values and modified less significant features based on human judgment to craft imperceptible adversarial examples. Similar approaches have been proposed by Grosse et al. \cite{grosse2017statistical}, Hashemi et al. \cite{hashemi2020permuteattack}, Levy et al. \cite{mathov2020not}, and Papernot et al. \cite{papernot2016transferability}. Additionally, Cartella et al. \cite{cartella2021adversarial} adapted state-of-the-art adversarial attack techniques such as the Zoo attack \cite{chen2017zoo}, boundary attack \cite{brendel2017decision}, and HopSkipJump attack \cite{chen2020hopskipjumpattack} from computer vision to the financial domain using tabular data. A different study by Levy et al. \cite{mathov2020not} introduced an approach for transforming a surrogate model while preserving the characteristics of the original model, enabling the application of established generation techniques. They demonstrated that a slight human imperceptible perturbation of the input image could change the prediction of the image classifiers. All of the aforementioned studies considered various constraints when generating adversarial examples, none of which are applicable to tabular data in software analytics tasks.

A plethora of research has been conducted on adversarial attacks targeting various software analytics tasks. Liu et al. \cite{liu2021practical}, and Erwin et al. \cite{quiring2019misleading} proposed a practical black-box attack on source code authorship identification classifiers based on a set of semantically equivalent program transformations. Nguyen et al. \cite{nguyen2021adversarial} showed that state-of-the-art API recommender systems are vulnerable to adversarial attacks if attackers corrupt the training corpus by injecting malicious data. Chen et al. \cite{chen2017adversarial} developed the EvenAttack model considering the contributions of the features of instances to the malware detection problem. Liu et al. \cite{liu2019atmpa} proposed a novel method (ATMPA) considering the gradient descent and L-norm optimization method. An attacker can use their technique to generate adversarial examples by introducing tiny perturbations to the input data. Grosse et al. and Suciu et al. \cite{suciu2019exploring} demonstrated how adversarial examples fool the malware detection classifiers with slightly modified input data. However, none of the methods proposed for adversarial attacks on ML models in software analytics consider the importance of the features identified by ML explainability techniques. Our work aims to manipulate the important features identified by ML explainability techniques to generate adversarial examples. Moreover, adversarial attacks in the feature-space have been studied in computer vision \cite{xu2020towards, inkawhich2019feature, simonetto2021unified}. However, this area remains largely unexplored in software analytics tasks. Therefore, our work aims to advance research in this direction.

Severi et al. \cite{severi2021explanation} applied ML explainability to select important features and values for malware classifier adversarial attacks. Unlike their work, our approach is designed specifically for ML inference, not training. Furthermore, our methodology diverges in selecting and modifying the top-$k$ important features while generating adversarial examples. Amich et al. \cite{amich2021explanation, amich2022eg} employ ML explainability to enhance and diagnose evasion attacks on ML models. Their study focuses on ML models trained on image classifier datasets like MNIST and CIFAR-10, with a different approach to perturbation and adversarial example generation than ours. Zhang et al. \cite{zhang2024harmonizing} proposed a novel ensemble-based adversarial attack approach that focuses on balancing two key aspects: 1) transferability and 2) imperceptibility, based on model interpretability for image data. Sun et al. \cite{sun2023mate} proposed an explainability-guided, model-agnostic testing framework to assess the robustness of malware detectors through feature-space manipulation. The framework uses \textit{Accured Malicious Magnitude (AMM)} to identify the fragile features for manipulation while generating adversarial examples. The primary distinction between our work and that of Sun et al. \cite{sun2023mate} is that we introduced the concept of the \textit{Reverse Elbow Method} for identifying fragile features (e.g., top-$k$ important features) for feature-space manipulation. Another important distinction is that their adversarial attack technique is a white-box method, assuming the attacker has complete knowledge of the model architecture and training dataset. In contrast, our explanation-guided adversarial attack technique does not have access to any information about the model's architecture. Additionally, we applied a multi-objective optimization function, as defined in Section \ref{Adv Attack}, to generate adversarial examples.



Numerous studies have investigated adversarial attacks targeting deep learning models and large language models (LLMs) trained on source code for various software analytics tasks such as clone detection, source code authorship attribution, vulnerability detection, code comment generation, and code summarization \cite{yefet2020adversarial, srikant2021generating, zhang2020generating, tian2021generating, zhou2022adversarial, bielik2020adversarial, springer2020strata, li2022ropgen, li2021deeppayload, tian2023code, zhou2024evolutionary, sun2023codemark, wan2022you, zhang2021advdoor, wong2022deceiving, liu2024alanca, jia2023clawsat, zhu2023robust, he2023large}. Yefet et al. \cite{yefet2020adversarial} proposed a white-box attack technique, DAMP, that changes the identifier's name in the code snippet based on the gradient information of the victim model. Srikant et al. \cite{srikant2021generating} generated adversarial examples for models of code using optimized obfuscation. Zhang et al. \cite{zhang2020generating} proposed a black-box attack technique MHM based on Metropolis-Hasting sampling-based variable renaming approach. Tian et al. \cite{tian2021generating} introduced QMDP (Q-Learning-based Markov decision process) to generate adversarial examples for models of code based on the semantic equivalent program transformations. Their experimental results demonstrated that QMDP generates adversarial examples and enhances the robustness of the source code classification models by over 44\%. 

Pour et al. \cite{pour2021search} developed a search-based adversarial test generation framework to measure the robustness of the neural source code embedding methods (i.e., Code2vec, Code2seq, and CodeBERT). Their experimental results show that the generated adversarial examples can, on average, decrease the performance of these embedding methods from 5.41\% to 9.58\%. Applis et al. \cite{applis2021assessing} proposed a testing framework: LAMPION, to assess the robustness of the ML-based program analysis tools adopting metamorphic program transformations. Zhou et al. \cite{zhou2022adversarial} studied the robustness of the code comment generation tasks in adversarial settings. They proposed ACCENT to generate adversarial examples by substituting identifiers in code snippets to generate syntactically correct and semantically similar ones to original code snippets. Recently, several works have focused on assessing the robustness of pre-trained code models under adversarial attacks \cite{yang2022natural, du2023extensive, zeng2022extensive, jha2023codeattack}. Yang et al. \cite{yang2022natural} performed a natural attack on the pre-trained code models. They transformed the code snippet, preserving the original inputs' operational and natural semantics. Bielik et al. \cite{bielik2020adversarial}, and Springer et al. \cite{springer2020strata} both evaluated the robustness of models of code in adversarial settings. Schuster et al. \cite{schuster2021you} demonstrated that carefully designed adversarial files attached to the training corpus make the code completion model vulnerable.

All the studies above pertain to adversarial attacks on DL models or LLMs trained solely on source code for software analytics tasks. However, the attack methods developed for DL models or LLMs are not directly applicable to classical ML models due to differences in the characteristics of the datasets used for training. Additionally, many attacks on DL models rely on renaming identifiers to generate successful adversarial examples, which does not affect the extracted features from the source code. For instance, in the CLCDSA dataset, the feature \textit{`Number of Variables Declared'} remains unchanged regardless of identifier renaming. Finally, while the above adversarial attack techniques target input-space manipulation for generating adversarial examples, our approach focuses on feature-space manipulation for this purpose.


\section{Conclusion}
\label{end}
In recent years, ML models have achieved widespread success in software analytics tasks such as JIT defect prediction, clone detection, code comment generation, code completion, API recommendation, malware detection, and code authorship attribution. However, ML models are vulnerable to adversarial attacks when the test input is changed with minimal perturbations, which may lead to substantial monetary losses in the software development and maintenance process. Many techniques have been proposed in the literature to assess the robustness of ML models under adversarial attacks. However, in this paper, we investigated how ML explainability leads to generating adversarial examples to assess the robustness of ML models against attacks on the feature-space. Our experimental results demonstrate a positive correlation between ML explainability and adversarial attacks. Modifying up to the top-$3$ most influential features identified by ML explainability techniques can create adversarial examples that can be used to assess the robustness of ML models. Our experimental results underscore the importance of developing robust ML models and the countermeasures against explanation-guided adversarial attacks on ML models in software analytics tasks.  


\section*{Data Availability}
Our code and the corresponding dataset are publicly available to enhance further research\footnote{\href{https://zenodo.org/doi/10.5281/zenodo.7865487}{Replication-package}}.

\section*{Funding}
This research is supported in part by the Natural Sciences and Engineering Research Council of Canada (NSERC) Discovery Grants program, the Canada Foundation for Innovation's John R. Evans Leaders Fund (CFI-JELF), and by the industry-stream NSERC CREATE in Software Analytics Research (SOAR).


\section*{Declaration of generative AI and AI-assisted technologies in the writing process}
During the preparation of this work, the author(s) used Grammarly \footnote{https://app.grammarly.com/} and ChatGPT\footnote{https://chat.openai.com/} to find grammatical mistakes and improve sentence clarity/ presentation. After using these tools/services, the author(s) reviewed and edited the content as needed and take(s) full responsibility for the content of the publication.

\section*{Conflict of Interest}
The authors declare that they have no competing interests, as defined by Springer, nor any other interests that might be perceived to influence the results and/or discussion reported in this paper.


\bibliography{sn-bibliography}

\end{document}